\documentclass{cernyrep}
\usepackage{texnames}
\usepackage[T1]{fontenc}
\begin{document}
\newcommand{\divop}{\nabla\cdot}
\newcommand{\curlop}{\nabla\times}
\newcommand{\gradop}{\nabla}
\title{Theory of Electromagnetic Fields}
 
\author{Andrzej Wolski}

\institute{University of Liverpool, and the Cockcroft Institute, UK}

\maketitle 

\begin{abstract}
We discuss the theory of electromagnetic fields, with an emphasis on aspects relevant
to radiofrequency systems in particle accelerators.  We begin by reviewing Maxwell's
equations and their physical significance.  We show that in free space, there are solutions
to Maxwell's equations representing the propagation of electromagnetic fields as waves.
We introduce electromagnetic potentials, and show how they can be used to simplify
the calculation of the fields in the presence of sources.  We derive Poynting's theorem,
which leads to expressions for the energy density and energy flux in an electromagnetic
field.  We discuss the properties of electromagnetic waves in cavities, waveguides
and transmission lines.
\end{abstract}
 
\section{Maxwell's equations}
 
Maxwell's equations may be written in differential form as follows:
\begin{eqnarray}
\divop \vec{D} & = & \rho, \label{eq:maxwell1} \\
\divop \vec{B} & = & 0, \label{eq:maxwell2} \\
\curlop \vec{H} & = & \vec{J} + \frac{\partial \vec{D}}{\partial t}, \label{eq:maxwell4} \\
\curlop \vec{E} & = & -\frac{\partial \vec{B}}{\partial t}. \label{eq:maxwell3}
\end{eqnarray}
The fields $\vec{B}$ (magnetic flux density) and $\vec{E}$ (electric
field strength) determine the force on a particle of charge $q$ travelling with
velocity $\vec{v}$ (the Lorentz force equation):
\begin{equation}
\vec{F} = q\left( \vec{E} + \vec{v} \times \vec{B} \right). \nonumber 
\end{equation}
The electric displacement $\vec{D}$ and magnetic intensity $\vec{H}$ are
related to the electric field and magnetic flux density by the \emph{constitutive relations}:
\begin{eqnarray}
\vec{D} & = & \varepsilon \vec{E}, \nonumber \\
\vec{B} & = & \mu \vec{H}. \nonumber
\end{eqnarray}
The electric permittivity $\varepsilon$ and magnetic permeability $\mu$
depend on the medium within which the fields exist.  The values of these
quantities in vacuum are fundamental physical constants.  In SI units:
\begin{eqnarray}
\mu_0 & = & 4\pi \times 10^{-7}\,\textrm{Hm}^{-1}, \nonumber \\
\varepsilon_0 & = & \frac{1}{\mu_0 c^2}, \nonumber
\end{eqnarray}
where $c$ is the speed of light in vacuum.  The permittivity and permeability
of a material characterize the response of that material to electric and
magnetic fields.  In simplified models, they are often regarded as constants
for a given material; however, in reality the permittivity and permeability
can have a complicated dependence on the fields that are present.  Note that
the \emph{relative permittivity} $\varepsilon_r$ and the
\emph{relative permeability} $\mu_r$ are frequently used.  These are
dimensionless quantities, defined by:
\begin{equation}
\varepsilon_r = \frac{\varepsilon}{\varepsilon_0}, \quad
\mu_r = \frac{\mu}{\mu_0}.
\end{equation}
That is, the relative permittivity is the permittivity of a material relative
to the permittivity of free space, and similarly for the relative permeability.

The quantities $\rho$ and $\vec{J}$ are respectively the electric charge
density (charge per unit volume) and electric current density
($\vec{J} \cdot \vec{n}$ is the charge crossing unit area perpendicular
to unit vector $\vec{n}$ per unit time).  Equations (\ref{eq:maxwell2})
and (\ref{eq:maxwell3}) are independent of $\rho$ and $\vec{J}$, and are
generally referred to as the ``homogeneous'' equations; the other two
equations, (\ref{eq:maxwell1}) and (\ref{eq:maxwell4}) are dependent
on $\rho$ and $\vec{J}$, and are generally referred to as the
``inhomogeneous'' equations.  The charge density and current density
may be regarded as \emph{sources} of electromagnetic fields.  When the
charge density and current density are specified (as functions of space,
and, generally, time), one can integrate Maxwell's equations
(\ref{eq:maxwell1})--(\ref{eq:maxwell4}) to find possible electric and
magnetic fields in the system.  Usually, however, the solution one finds
by integration is not unique: for example, as we shall see, there are many
possible field patterns that may exist in a cavity (or waveguide) of given
geometry.

\begin{figure}[t]
\centering
\includegraphics[width=0.7\linewidth]{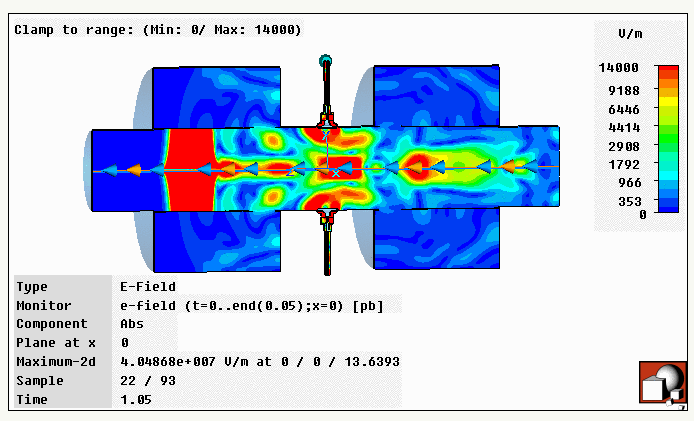}
\caption{Snapshot of a numerical solution to Maxwell's equations for a bunch of electrons
moving through a beam position monitor in an accelerator vacuum chamber.
The colours show the strength of the electric field.  The bunch is moving from right
to left: the location of the bunch corresponds to the large region of high field intensity
towards the left hand side.  (Image courtesy of M.\,Korostelev.)
\label{fig:bpmwakefield}}
\end{figure}

Most realistic situations are sufficiently complicated that solutions
to Maxwell's equations cannot be obtained analytically.  A variety of
computer codes exist to provide solutions numerically, once the charges,
currents, and properties of the materials present are all specified,
see for example References \cite{bib:vectorfields, bib:cst, bib:radia}.
Solving for the fields in realistic systems (with three spatial dimensions,
and a dependence on time) often
requires a considerable amount of computing power; some sophisticated
techniques have been developed for solving Maxwell's equations
numerically with good efficiency \cite{bib:russenschuck}.  An example of
a numerical solution to Maxwell's equations in the context of a particle
accelerator is shown in Fig.~\ref{fig:bpmwakefield}.  We do not
consider such techniques here, but focus instead on the analytical
solutions that may be obtained in idealized situations.  Although the
solutions in such cases may not be sufficiently accurate to complete
the design of real accelerator components, the analytical solutions do
provide a useful basis for describing the fields in (for example) real
RF cavities and waveguides.

An important feature of Maxwell's equations is that, for systems containing
materials with constant permittivity and permeability (i.e. permittivity and
permeability that are independent of the fields present), the equations are
\emph{linear} in the fields and sources.  That is, each term in the equations 
involves a field or a source to (at most) the first power, and products of
fields or sources do not appear.  As a consequence, the \emph{principle of
superposition} applies: if $\vec{E}_1, \vec{B}_1$ and $\vec{E}_2, \vec{B}_2$
are solutions of Maxwell's equations with given boundary conditions, then
$\vec{E}_T = \vec{E}_1 + \vec{E}_2$ and 
$\vec{B}_T = \vec{B}_1 + \vec{B}_2$ will also be solutions of Maxwell's
equations, with the same boundary conditions.
This means that it is possible to represent complicated fields as
superpositions of simpler fields.  An important and widely used
analysis technique for electromagnetic systems, including RF cavities
and waveguides, is to find a set of solutions to Maxwell's equations from
which more complete and complicated solutions may be constructed.
The members of the set are known as \emph{modes}; the modes can
generally be labelled using \emph{mode indices}.
For example, plane electromagnetic waves in free space may be labelled
using the three components of the wave vector that describes the direction
and wavelength of the wave.  Important properties of the electromagnetic
fields, such as the frequency of oscillation, can often be expressed in terms
of the mode indices.

Solutions to Maxwell's equations lead to a rich diversity of
phenomena, including the fields around charges and currents in certain
basic configurations, and the generation, transmission and absorption of
electromagnetic radiation.  Many existing texts cover these phenomena
in detail; for example, Grant and Phillips \cite{bib:grantandphillips}, or the
authoritative text by Jackson \cite{bib:jackson}.  We consider these aspects
rather briefly, with an emphasis on those features of the theory that are
important for understanding the properties of RF components in accelerators.

\section{Integral theorems and the physical interpretation of Maxwell's equations
\label{sec:integraltheorems}}

\subsection{Gauss' theorem and Coulomb's law}

Guass' theorem states that for any smooth vector field $\vec{a}$:
\begin{equation}
\int_V \divop \vec{a} \,dV = \oint_{\partial V} \vec{a} \cdot d\vec{S}, \nonumber
\end{equation}
where $V$ is a volume bounded by the closed surface $\partial V$.  Note that
the area element $d\vec{S}$ is oriented to point \emph{out} of $V$.

Gauss' theorem is helpful for obtaining physical interpretations of two
of Maxwell's equations, (\ref{eq:maxwell1}) and (\ref{eq:maxwell2}).
First, applying Gauss' theorem to (\ref{eq:maxwell1}) gives:
\begin{equation}
\int_V \divop \vec{D} \,dV = \oint_{\partial V} \vec{D} \cdot d\vec{S} = q,
\label{eq:coulomb1}
\end{equation}
where $q = \int_V \rho \, dV$ is the total charge enclosed by $\partial V$.

\begin{figure}[t]
\centering
\includegraphics[width=0.4\linewidth]{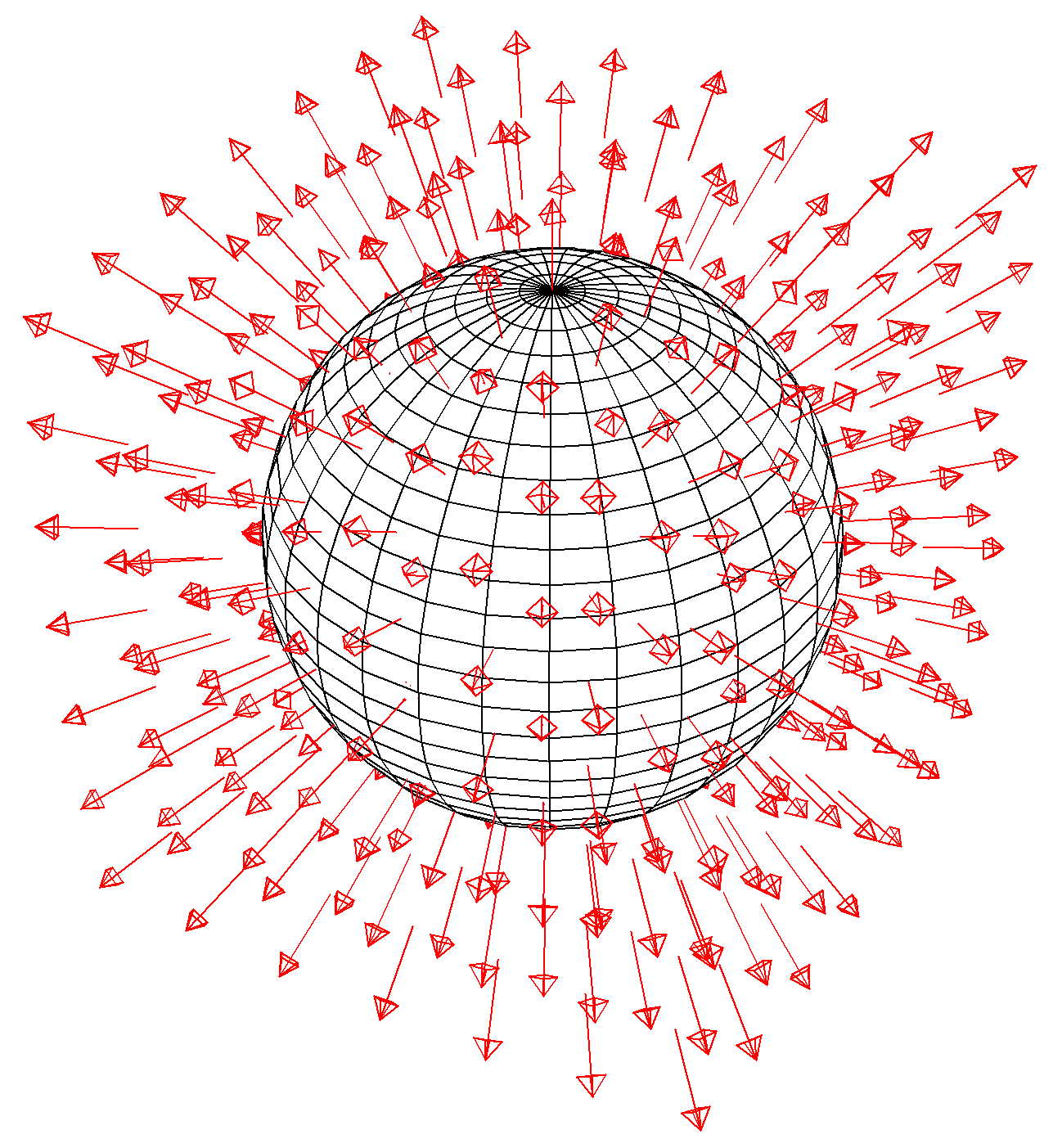}
\caption{Electric field lines from a point charge $q$.  The field lines are
everywhere perpendicular to a spherical surface centered on the charge.
\label{fig:coulombslaw}}
\end{figure}

Suppose that we have a single isolated point charge in an homogeneous, isotropic
medium with constant permittivity $\varepsilon$.  In this case, it
is interesting to take $\partial V$ to be a sphere of radius $r$.  By symmetry,
the magnitude of the electric field must be the same at all points on $\partial V$,
and must be normal to the surface at each point.  Then, we can perform the
surface integral in (\ref{eq:coulomb1}):
\begin{equation}
\oint_{\partial V} \vec{D} \cdot d\vec{S} = 4\pi r^2 D. \nonumber
\end{equation}
This is illustrated in Fig.~\ref{fig:coulombslaw}: the outer circle represents a
cross-section of a sphere ($\partial V$) enclosing volume $V$, with the charge $q$
at its centre.  The red arrows in Fig.~\ref{fig:coulombslaw} represent the
electric field lines, which are everywhere perpendicular to the surface $\partial V$.
Since $\vec{D} = \varepsilon \vec{E}$, we
find Coulomb's law for the magnitude of the electric field around a point
charge:
\begin{equation}
E = \frac{q}{4\pi \varepsilon r^2}. \nonumber
\end{equation}

Applied to Maxwell's equation (\ref{eq:maxwell2}), Gauss' theorem leads to:
\begin{equation}
\int_V \divop \vec{B}\, dV = \oint_{\partial V} \vec{B}\cdot d\vec{S} = 0. \nonumber
\end{equation}
In other words, the magnetic flux integrated over any closed surface must
equal zero -- at least, until we discover magnetic monopoles.  Lines of
magnetic flux \emph{always} occur in closed loops; lines of electric field
may occur in closed loops, but in the presence of electric charges will have start
(and end) points on the electric charges.

\subsection{Stokes' theorem, Amp\`ere's law, and Faraday's law}

\begin{figure}[t]
\centering
\includegraphics[width=0.35\linewidth]{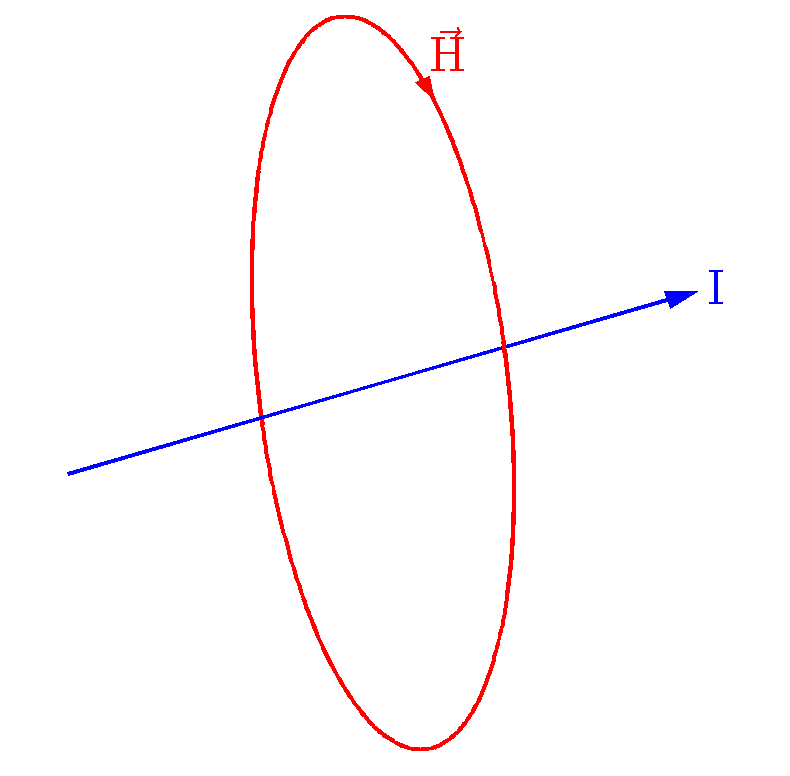}
\caption{Magnetic field lines around a long straight wire carrying a current $I$.
\label{fig:ampereslaw}}
\end{figure}

Stokes' theorem states that for any smooth vector field $\vec{a}$:
\begin{equation}
\int_S \curlop \vec{a} \cdot d\vec{S} = \oint_{\partial S} \vec{a} \cdot d\vec{l},
\label{eq:stokestheorem}
\end{equation}
where the closed loop $\partial S$ bounds the surface $S$.  Applied to Maxwell's
equation (\ref{eq:maxwell4}), Stokes' theorem leads to:
\begin{equation}
\oint_{\partial S} \vec{H} \cdot d\vec{l} = \int_S \vec{J} \cdot d\vec{S},
\label{eq:ampere1}
\end{equation}
which is Amp\`ere's law.  From Amp\`ere's law, we can derive an expression for
the strength of the magnetic field around a long, straight wire carrying
current $I$.  The magnetic field must have rotational symmetry around the wire.
There are two possibilities: a radial field, or a field consisting of closed
concentric loops centred on the wire (or some superposition of these fields).
A radial field would violate Maxwell's equation (\ref{eq:maxwell2}).  Therefore,
the field must consist of closed concentric loops; and by considering a
circular loop of radius $r$, we can perform the integral in Eq.~(\ref{eq:ampere1}):
\begin{equation}
2\pi r H = I, \nonumber
\end{equation}
where $I$ is the total current carried by the wire.  In this case, the line integral
is performed around a loop $\partial S$ centered on the wire, and in a plane
perpendicular to the wire: essentially, this corresponds to one of the magnetic
field lines, see Fig.~\ref{fig:ampereslaw}.  The total current passing through
the surface $S$ bounded by the loop $\partial S$ is simply the total current $I$.

In an homogeneous, isotropic
medium with constant permeability $\mu$, $\vec{B} = \mu_0 \vec{H}$, and we obtain
the expression for the magnetic flux density at distance $r$ from the wire:
\begin{equation}
B = \frac{\mu I}{2\pi r}. \label{eq:ampere2}
\end{equation}

Finally, applying Stokes' theorem to the homogeneous Maxwell's equation
(\ref{eq:maxwell3}), we find:
\begin{equation}
\oint_{\partial S} \vec{E} \cdot d\vec{l} = -\frac{\partial}{\partial t}
\int_S \vec{B} \cdot d\vec{S}. \label{eq:faraday1}
\end{equation}
Defining the electromotive force $\mathscr{E}$ as the integral of the
electric field around a closed loop, and the magnetic flux $\Phi$ as the
integral of the magnetic flux density over the surface bounded by the loop,
Eq.~(\ref{eq:faraday1})
gives:
\begin{equation}
\mathscr{E} = -\frac{\partial \Phi}{\partial t},
\end{equation}
which is Faraday's law of electromagnetic induction.

Maxwell's equations (\ref{eq:maxwell4}) and (\ref{eq:maxwell3}) are significant for RF systems:
they tell us that a time dependent electric field will induce a magnetic field; and a time dependent
magnetic field will induce an electric field.  Consequently, the fields in RF cavities and waveguides
always consist of both electric and magnetic fields.

\section{Electromagnetic waves in free space}

In free space (i.e. in the absence of any charges or currents) Maxwell's
equations have a trivial solution in which all the fields vanish.  However, there
are also non-trivial solutions with considerable practical importance.
In general, it is difficult to write down solutions to Maxwell's equations, because
two of the equations involve both the electric and magnetic fields.  However, by
taking additional derivatives, it is possible to write equations for the fields that
involve only either the electric or the magnetic field.  This makes it easier to write
down solutions: however, the drawback is that instead of first-order differential
equations, the new equations are second-order in the derivatives.  There is no
guarantee that a solution to the second-order equations will also satisfy the
first-order equations, and it is necessary to impose additional constraints to
ensure that the first-order equations are satisfied.  Fortunately, it turns out
that this is not difficult to do, and taking additional derivatives is a useful
technique for simplifying the analytical solution of Maxwell's equations in
simple cases.

\subsection{Wave equation for the electric field}
In free space, Maxwell's equations (\ref{eq:maxwell1}) -- (\ref{eq:maxwell3})
take the form:
\begin{eqnarray}
\divop \vec{E} & = & 0, \label{eq:maxwell1fs} \\
\divop \vec{B} & = & 0, \label{eq:maxwell2fs} \\
\curlop \vec{B} & = & \frac{1}{c^2} \frac{\partial \vec{E}}{\partial t}, \label{eq:maxwell4fs} \\
\curlop \vec{E} & = & -\frac{\partial \vec{B}}{\partial t}, \label{eq:maxwell3fs}
\end{eqnarray}
where we have defined:
\begin{equation}
\frac{1}{c^2} = \mu_0 \varepsilon_0.
\label{speedoflight}
\end{equation}
Our goal is to find a form of the equations in which the fields $\vec{E}$ and
$\vec{B}$ appear separately, and not together in the same equation.
As a first step, we take the curl of both sides of Eq.~(\ref{eq:maxwell3fs}),
and interchange the order of differentiation
on the right hand side (which we are allowed to do, since the space and
time coordinates are independent).  We obtain:
\begin{equation}
\curlop \curlop \vec{E} = -\frac{\partial}{\partial t} \curlop \vec{B}.
\label{curlcurle}
\end{equation}
Substituting for $\curlop \vec{B}$ from Eq.~(\ref{eq:maxwell4fs}),
this becomes:
\begin{equation}
\curlop \curlop \vec{E} = -\frac{1}{c^2} \frac{\partial^2 \vec{E}}{\partial t^2}.
\end{equation}
This second-order differential equation involves only the electric field, $\vec{E}$, so we
have achieved our aim of decoupling the field equations.  However, it is possible to make
a further simplification, using a mathematical identity. For any differentiable vector field $\vec{a}$:
\begin{equation}
\curlop \curlop \vec{a} \equiv \gradop (\divop \vec{a}) - \nabla^2  \vec{a}.
\label{curlcurlidentity}
\end{equation}
Using the identity (\ref{curlcurlidentity}), and also making use of Eq.~(\ref{eq:maxwell1fs}),
we obtain finally:
\begin{equation}
\nabla^2  \vec{E} - \frac{1}{c^2} \frac{\partial^2 \vec{E}}{\partial t^2} = 0.
\label{waveequatione}
\end{equation}
Eq.~(\ref{waveequatione}) is the wave equation in three spatial dimensions.  Note that
each component of the electric field independently satisfies the wave equation.  The
solution, representing a plane wave propagating in the direction of the vector $\vec{k}$,
may be written in the form:
\begin{equation}
\vec{E} = \vec{E}_0 \cos \! \left( \vec{k} \cdot \vec{r}  - \omega t + \phi_0 \right),
\label{waveeqnesolution}
\end{equation}
where: $\vec{E}_0$ is a constant vector; $\phi_0$ is a constant phase; $\omega$ and
$\vec{k}$ are constants related to the frequency $f$ and wavelength $\lambda$ of the
wave by:
\begin{eqnarray}
\omega & = & 2\pi f, \\
\lambda & = & \frac{2\pi}{|\vec{k}|}.
\end{eqnarray}
If we substitute Eq.~(\ref{waveeqnesolution}) into the wave equation (\ref{waveequatione}),
we find that it provides a valid solution as long as the angular frequency $\omega$
and wave vector $\vec{k}$ satisfy the \emph{dispersion relation}:
\begin{equation}
\frac{\omega}{| \vec{k} |} = c.
\label{freespacedispersionrelation}
\end{equation}
If we inspect Eq.~(\ref{waveeqnesolution}), we see that a particle travelling in the direction
of $\vec{k}$ has to move at a speed $\omega / |\vec{k}|$ in order to remain at the
same phase in the wave: thus, the quantity $c$ is the \emph{phase velocity} of the wave.
This quantity $c$ is, of course, the speed of light in a vacuum; and the identification of
light with an electromagnetic wave (with the phase velocity related to the electric permittivity
and magnetic permeability by Eq.~(\ref{speedoflight})) was one of the great
achievements of 19$^\textrm{th}$ century physics.

\subsection{Wave equation for the magnetic field}
So far, we have only considered the electric field.  But Maxwell's equation (\ref{eq:maxwell4})
tells us that an electric field that varies with time must have a magnetic field associated with it.
Therefore, we should look for a (non-trivial) solution for the magnetic field in free space.
Starting with Eq.~(\ref{eq:maxwell4fs}),
and following the same procedure as above, we find that the magnetic field also
satisfies the wave equation:
\begin{equation}
\nabla^2  \vec{B} - \frac{1}{c^2} \frac{\partial^2 \vec{B}}{\partial t^2} = 0,
\end{equation}
with a similar solution:
\begin{equation}
\vec{B} = \vec{B}_0 \cos \! \left( \vec{k} \cdot \vec{r}  - \omega t + \phi_0 \right).
\label{waveeqnbsolution}
\end{equation}
Here, we have written the same constants $\omega$, $\vec{k}$ and $\phi_0$ as we used
for the electric field, though we do not so far know they have to be the same.  We shall show
in the following section that these constants do indeed need to be the same for both the
electric field and the magnetic field.

\subsection{Relations between electric and magnetic fields in a plane wave in free space}
As we commented above, although taking additional derivatives of Maxwell's equations allows
us to decouple the equations for the electric and magnetic fields, we must impose additional
constraints on the solutions, to ensure that the first-order equations are satisfied.  In
particular, substituting the expressions for the fields (\ref{waveeqnesolution}) and
(\ref{waveeqnbsolution}) into Eqs.~(\ref{eq:maxwell1fs}) and (\ref{eq:maxwell2fs})
respectively, and noting that the latter equations must be satisfied at all points in space
and at all times, we obtain:
\begin{eqnarray}
\vec{k} \cdot \vec{E}_0 & = & 0, \\
\vec{k} \cdot \vec{B}_0 & = & 0.
\end{eqnarray} 
Since $\vec{k}$ represents the direction of propagation of the wave, we see that the
electric and magnetic fields must at all times and all places be perpendicular to the
direction in which the wave is travelling.  This is a feature that does not appear if we
only consider the second-order equations.

\begin{figure}[t]
\centering
\includegraphics[width=0.6\linewidth]{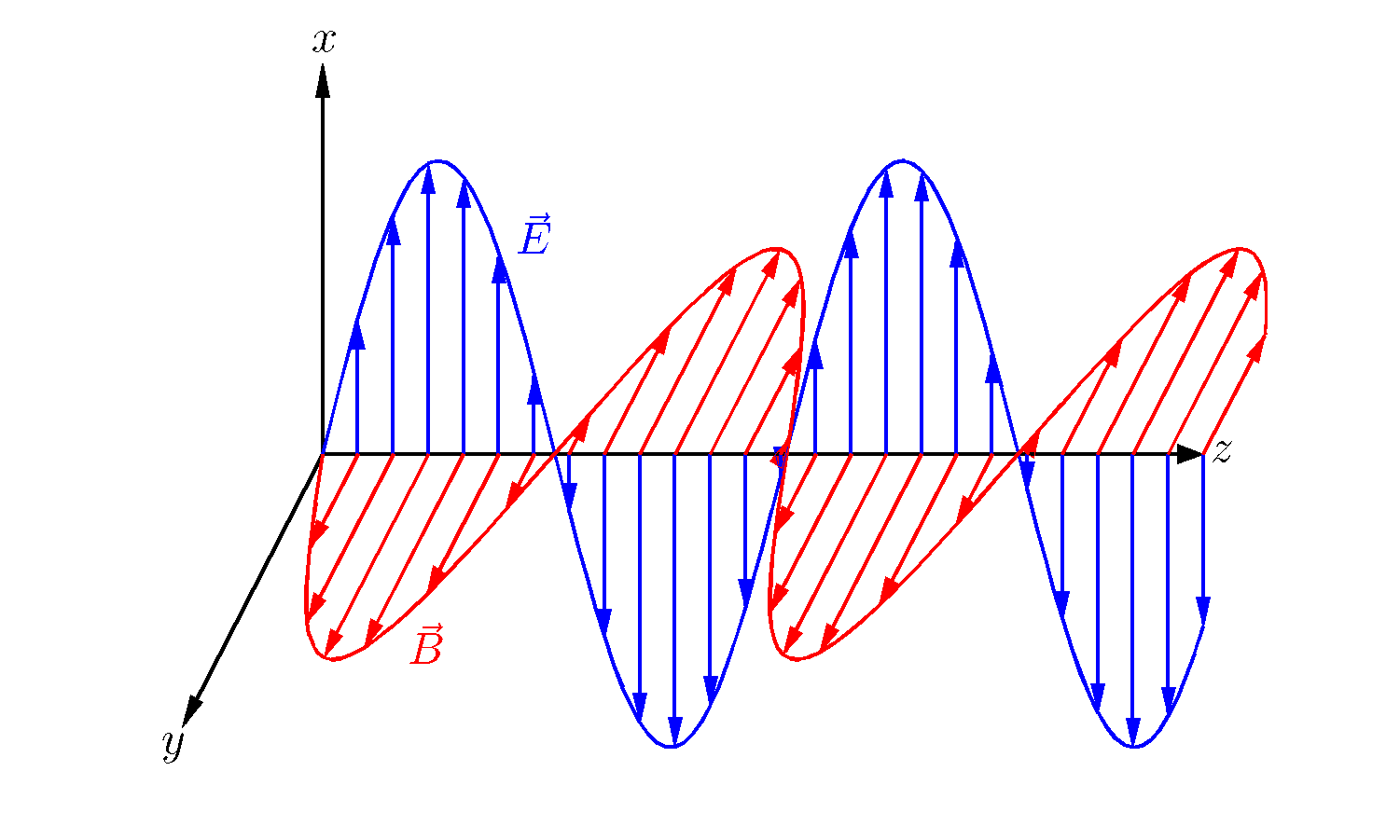}
\caption{Electric and magnetic fields in a plane electromagnetic wave in free space.
The wave vector $\vec{k}$ is in the direction of the $+z$ axis.
\label{fig:planewaveebfields}}
\end{figure}

Finally, substituting the expressions for the fields (\ref{waveeqnesolution}) and
(\ref{waveeqnbsolution}) into Eqs.~(\ref{eq:maxwell3fs}) and (\ref{eq:maxwell4fs})
respectively, and again noting that the latter equations must be satisfied at all points in space
and at all times, we see first that the quantities $\omega$, $\vec{k}$ and $\phi_0$
appearing in (\ref{waveeqnesolution}) and (\ref{waveeqnbsolution}) must be the same
in each case.  Also, we have the following relations between the magnitudes and
directions of the fields:
\begin{eqnarray}
\vec{k} \times \vec{E}_0 & = & \omega \vec{B}_0, \\
\vec{k} \times \vec{B}_0 & = & -\omega \vec{E}_0.
\end{eqnarray} 
If we choose a coordinate system so that $\vec{E}_0$ is parallel to the $x$ axis and
$\vec{B}_0$ is parallel to the $y$ axis, then $\vec{k}$ must be parallel to the $z$ axis:
note that the vector product $\vec{E} \times \vec{B}$ is in the same direction as
the direction of propagation of the wave -- see Fig.\,\ref{fig:planewaveebfields}.
The magnitudes of the electric and
magnetic fields are related by:
\begin{equation}
\frac{| \vec{E} |}{| \vec{B} |} = c.
\label{amplituderelation}
\end{equation}

Note that the wave vector $\vec{k}$ can be chosen arbitrarily: there are infinitely
many ``modes'' in which an electromagnetic wave propagating in free space may
appear; and the most general solution will be a sum over all modes.  When the
mode is specified (by giving the components of $\vec{k}$), the frequency is
determined from the dispersion relation (\ref{freespacedispersionrelation}).
However, the amplitude and phase are not determined (although the electric
and magnetic fields must have the same phase, and their amplitudes must be
related by Eq.~(\ref{amplituderelation})).

Finally, note that all the results derived in this section are strictly true only for
electromagnetic fields in a vacuum.  The generalisation to fields in uniform, homogenous,
linear (i.e. constant permeability $\mu$ and permittivity $\varepsilon$) nonconducting
media is straightforward.  However, new features appear for waves in conductors,
on boundaries, or in nonlinear media.

\subsection{Complex notation for electromagnetic waves}
We finish this section by introducing the complex notation for free waves.
Note that the electric field given by equation (\ref{waveeqnesolution}) can also be
written as:
\begin{equation}
\vec{E} = \textrm{Re} \, \vec{E}_0 e^{i \phi_0} e^{i( \vec{k} \cdot \vec{r} - \omega t )}.
\end{equation}
To avoid continually writing a constant phase factor when dealing with solutions
to the wave equation, we replace the real (constant) vector $\vec{E}_0$ by the
complex (constant) vector $\vec{E}^\prime_0 = \vec{E}_0 e^{i\phi_0}$.  Also,
we note that since all the equations describing the fields are linear, and that any
two solutions can be linearly superposed to construct a third solution, the complex
vectors:
\begin{eqnarray}
\vec{E}^\prime & = & \vec{E}^\prime_0 e^{i( \vec{k} \cdot \vec{r} - \omega t )}, \\
\vec{B}^\prime & = & \vec{B}^\prime_0 e^{i( \vec{k} \cdot \vec{r} - \omega t )}
\end{eqnarray}
provide \emph{mathematically} valid solutions to Maxwell's equations in free space,
with the same relationships between the various quantities (frequency, wave vector,
amplitudes, phase) as the solutions given in Eqs.~(\ref{waveeqnesolution}) and (\ref{waveeqnbsolution}).  Therefore, as long as we deal with linear equations,
we can carry out all the algebraic manipulation using \emph{complex} field vectors,
where it is implicit that the physical quantities are obtained by taking the real parts of
the complex vectors.  However, when using the complex notation, particular care is
needed when taking the product of two complex vectors: to be safe, one should always
take the real part \emph{before} multiplying two complex quantities, the real parts
of which represent physical quantities.  Products of the electromagnetic field vectors
occur in expressions for the energy density and energy flux in an electromagnetic
field, as we shall see below.

\section{Electromagnetic waves in conductors}

Electromagnetic waves in free space are characterized by an amplitude that remains
constant in space and time.  This is also true for waves travelling through any
isotropic, homogeneous, linear, non-conducting medium, which we may refer to as
an ``ideal'' dielectric.  The fact that real materials
contain electric charges that can respond to electromagnetic fields means that
the vacuum is really the only ideal dielectric.  Some real materials
(for example, many gases, and materials such as glass) have properties that
approximate those of an ideal dielectric, at least over certain frequency ranges:
such materials are transparent.  However, we know that many materials are
not transparent: even a thin sheet of a good conductor such as aluminium or
copper, for example, can provide an effective barrier for electromagnetic radiation
over a wide range of frequencies.

To understand the shielding effect of good conductors is relatively straightforward.
Essentially, we follow the same procedure to derive the wave equations for the
electromagnetic fields as we did for the case of a vacuum, but we include
additional terms to represent the conductivity of the medium.  These additional
terms have the consequence that the amplitude of the wave decays as the wave
propagates through the medium.  The rate of decay of the wave is characterised
by the skin depth, which depends (amongst other things) on the conductivity of the
medium.

Let us consider an ohmic conductor.  An ohmic conductor is defined by the
relationship between the current density $\vec{J}$ at a point in the conductor, and
the electric field $\vec{E}$ existing at the same point in the conductor:
\begin{equation}
\vec{J} = \sigma \vec{E},
\end{equation}
where $\sigma$ is a constant, the \emph{conductivity} of the material.

In an uncharged ohmic conductor, Maxwell's equations (\ref{eq:maxwell1}) -- (\ref{eq:maxwell3})
take the form:
\begin{eqnarray}
\divop \vec{E} & = & 0, \label{eq:maxwell1cond} \\
\divop \vec{B} & = & 0, \label{eq:maxwell2cond} \\
\curlop \vec{B} & = & \mu \sigma \vec{E} + \mu \varepsilon \frac{\partial \vec{E}}{\partial t}, \label{eq:maxwell4cond} \\
\curlop \vec{E} & = & -\frac{\partial \vec{B}}{\partial t}, \label{eq:maxwell3cond}
\end{eqnarray}
where $\mu$ is the (absolute) permeability of the medium, and $\varepsilon$ is the
(absolute) permittivity.  Notice the appearance of the additional term on the right hand
side of Eq.~(\ref{eq:maxwell4cond}), compared to Eq.~(\ref{eq:maxwell4fs}).

Following the same procedure as led to Eq.~(\ref{waveequatione}), we derive the following
equation for the electric field in a conducting medium:
\begin{equation}
\nabla^2 \vec{E} - \mu \sigma \frac{\partial \vec{E}}{\partial t} - \mu \varepsilon \frac{\partial^2 \vec{E}}{\partial t^2}= 0.
\label{waveeqnecond}
\end{equation}
This is again a wave equation, but with a term that includes a first-order time derivative.
In the equation for a simple harmonic oscillator, such a term would represent a ``frictional''
force that leads to dissipation of the energy in the oscillator.  There is a similar effect here;
to see this, let us try a solution of the same form as for a wave in free space.  The results
we are seeking can be obtained more directly if we used the complex notation:
\begin{equation}
\vec{E} = \vec{E}_0 e^{i(\vec{k} \cdot\vec{r} - \omega t)}.
\label{wavesolncond}
\end{equation}
Substituting into the wave equation (\ref{waveeqnecond}), we obtain the dispersion relation:
\begin{equation}
-\vec{k}^2 +i \mu \sigma \omega + \mu \varepsilon \omega^2 = 0.
\label{dispersionrelationcond}
\end{equation}
Let us assume that the frequency $\omega$ is a real number. Then, to find a solution
to Eq.~(\ref{dispersionrelationcond}), we have to allow the wave vector $\vec{k}$ to be
complex.  Let us write the real and imaginary parts as $\vec{\alpha}$ and $\vec{\beta}$
respectively:
\begin{equation}
\vec{k} = \vec{\alpha} + i\vec{\beta}.
\label{krealimag}
\end{equation}
Substituting (\ref{krealimag}) into (\ref{dispersionrelationcond}) and equating
real and imaginary parts, we find (after some algebra) that:
\begin{eqnarray}
| \vec{\alpha} | & = &  \omega \sqrt{\mu \varepsilon}
\left( \frac{1}{2} + \frac{1}{2} \sqrt{1 + \frac{\sigma^2}{\omega^2 \varepsilon^2}} \right)^\frac{1}{2}, \\
| \vec{\beta} | & = & \frac{\omega \mu \sigma}{2 |\vec{\alpha}|}.
\end{eqnarray}
To understand the physical significance of $\vec{\alpha}$ and $\vec{\beta}$, we write the
solution (\ref{wavesolncond}) to the wave equation as:
\begin{equation}
\vec{E} = \vec{E}_0 e^{-\vec{\beta} \cdot \vec{r}} e^{i(\vec{\alpha} \cdot\vec{r} - \omega t)}.
\end{equation}
We see that there is still a wave-like oscillation of the electric field, but there is now also an
exponential decay of the amplitude.  The wavelength is determined by the \emph{real part} of
the wave vector:
\begin{equation}
\lambda = \frac{2\pi}{| \vec{\alpha} |}.
\end{equation}
The imaginary part of the wave vector gives the distance  $\delta$ over which the amplitude of the
wave falls by a factor $1/e$, known as the \emph{skin depth}:
\begin{equation}
\delta = \frac{1}{| \vec{\beta} |}.
\end{equation}

\begin{figure}[t]
\centering
\includegraphics[width=0.6\linewidth]{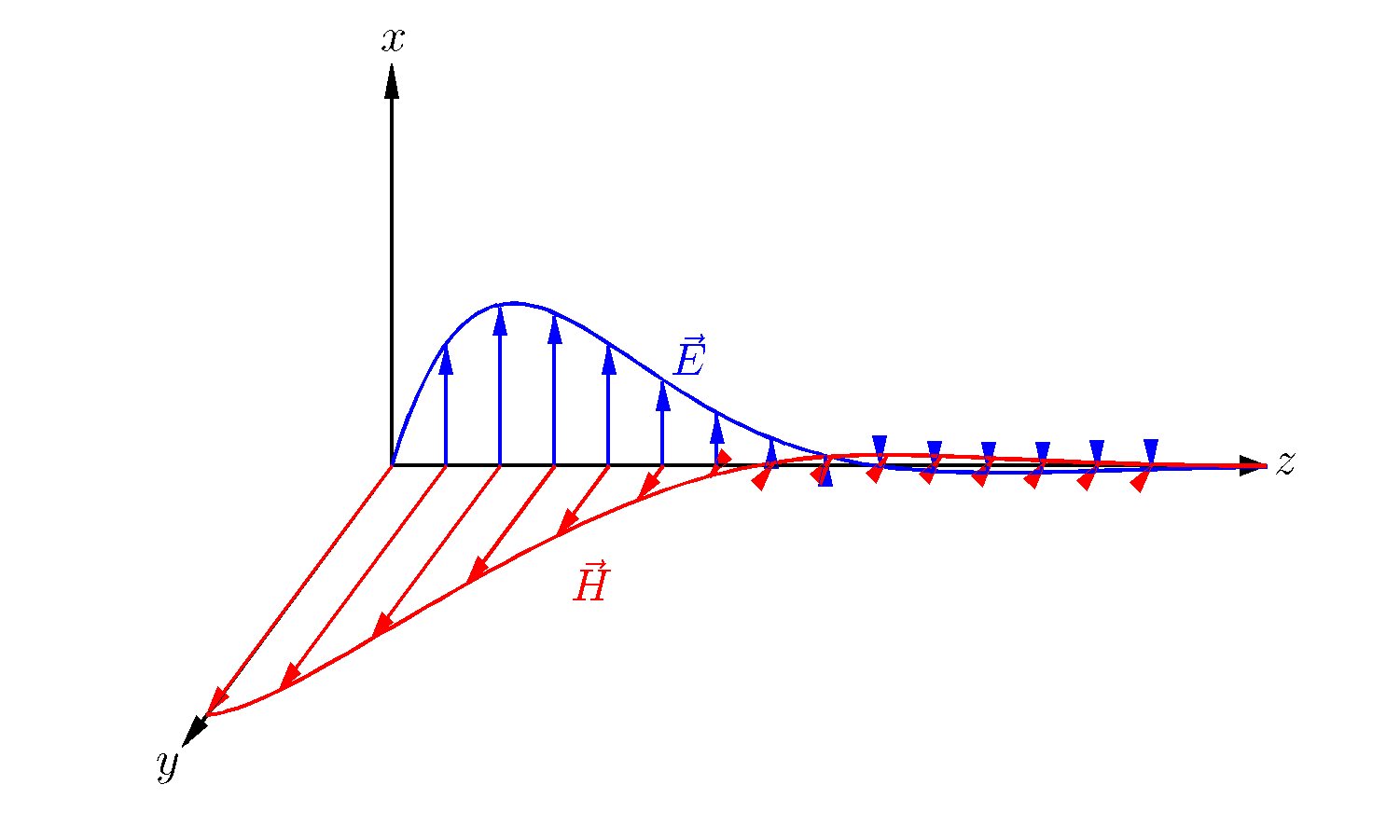}
\caption{Electric and magnetic fields in a plane electromagnetic wave in a conductor.
The wave vector is in the direction of the $+z$ axis.
\label{fig:planewaveebfieldsconductor}}
\end{figure}

Accompanying the electric field, there must be a magnetic field:
\begin{equation}
\vec{B} = \vec{B}_0 \, e^{i (\vec{k}\cdot \vec{r} - \omega t)}.
\end{equation}
From Maxwell's equation (\ref{eq:maxwell3}), the amplitudes of the electric and magnetic
fields must be related by:
\begin{equation}
\vec{k} \times \vec{E}_0 = \omega \vec{B}_0.
\label{fieldamplitudesconductor}
\end{equation}
The electric and magnetic fields are perpendicular to each other, and to the wave vector:
this is the same situation as occurred for a plane wave in free space.  However, 
since $\vec{k}$ is complex for a wave in a conductor, there is a phase difference between the
electric and magnetic fields, given by the complex phase of $\vec{k}$.  The fields in a plane
wave in a conductor are illustrated in Fig.\,\ref{fig:planewaveebfieldsconductor}.

The dispersion relation (\ref{dispersionrelationcond}) gives a rather complicated algebraic
relationship between the frequency and the wave vector, in which the electromagnetic properties
of the medium (permittivity, permeability and conductivity) all appear.  However, in many cases
it is possible to write much simpler expressions that provide good approximations.  First, there
is the ``poor conductor'' regime:
\begin{equation}
\textrm{if }
\sigma \ll \omega \varepsilon,
\textrm{ then }
| \vec{\alpha} | \approx \omega \sqrt{\mu \varepsilon}, \quad
| \vec{\beta} | \approx \frac{\sigma}{2} \sqrt{\frac{\mu}{\varepsilon}}.
\end{equation}
The wavelength is related to the frequency in the way that we would expect for a dielectric.

Next there is the ``good conductor'' regime:
\begin{equation}
\textrm{if }
\sigma \gg \omega \varepsilon,
\textrm{ then }
| \vec{\alpha} | \approx \sqrt{\frac{\omega \sigma \mu}{2}}, \quad
| \vec{\beta} | \approx | \vec{\alpha} |.
\label{wavevectorgoodconductor}
\end{equation}
Here the situation is very different.  The wavelength depends directly on the conductivity:
for a good conductor, the wavelength is very much shorter than it would be for a wave at
the same frequency in free space.  The real and imaginary parts of the wave vector are
approximately equal: this means that there is a significant reduction in the amplitude of
the wave even over one wavelength.  Also, the electric and magnetic fields are
approximately $\pi/4$ out of phase.

The reduction in amplitude of a wave as it travels through a conductor is not difficult to
understand.  The electric charges in the conductor move in response to the electric field
in the wave.  The motion of the charges constitutes an electric current in the conductor,
which results in ohmic losses: ultimately, the energy in the wave is dissipated as heat in
the conductor.  Note that whether or not a given material can be described as a ``good
conductor'' depends on the frequency of the wave (and permittivity of the material): at a
high enough frequency, any material will become a poor conductor.

\section{Energy in electromagnetic fields}

Waves are generally associated with the propagation of energy: the question then arises
as to whether this is the case with electromagnetic waves, and, if so, how much energy is
carried by a wave of a given amplitude.  To address this question, we first need to find
general expressions for the energy density and energy flux in an electromagnetic field.
The appropriate expressions follow from Poynting's theorem, which may be derived from
Maxwell's equations.

\subsection{Poynting's theorem}

First, we take the scalar product of Maxwell's equation (\ref{eq:maxwell3}) with the magnetic
intensity $\vec{H}$ on both sides, to give:
\begin{equation}
\vec{H} \cdot \curlop \vec{E} = -\vec{H} \cdot \frac{\partial \vec{B}}{\partial t}.
\label{poynting1}
\end{equation}
Then, we take the scalar product of (\ref{eq:maxwell4}) with the electric field $\vec{E}$ on
both sides to give:
\begin{equation}
\vec{E} \cdot \curlop \vec{H} = \vec{E} \cdot \vec{J} + \vec{E} \cdot \frac{\partial \vec{D}}{\partial t}.
\label{poynting2}
\end{equation}
Now we subtract Eq.~(\ref{poynting2}) from Eq.~(\ref{poynting1}) to give:
\begin{equation}
\vec{H} \cdot \curlop \vec{E} - \vec{E} \cdot \curlop \vec{H} = 
- \vec{E} \cdot \vec{J} - \vec{E} \cdot \frac{\partial \vec{D}}{\partial t}
- \vec{H} \cdot \frac{\partial \vec{B}}{\partial t}.
\end{equation}
This may be rewritten as:
\begin{equation}
\frac{\partial}{\partial t} \left( 
\frac{1}{2} \varepsilon \vec{E}^2 + \frac{1}{2} \mu \vec{H}^2
 \right) =
- \divop \left( \vec{E} \times \vec{H} \right) - \vec{E} \cdot \vec{J}.
\label{poyntingtheorem}
\end{equation}
Equation (\ref{poyntingtheorem}) is Poynting's theorem.  It does not appear immediately
to tell us much about the energy in an electromagnetic field; but the physical interpretation
becomes a little clearer if we convert it from differential form into integral form.  Integrating
each term on either side over a volume $V$, and changing the first term on the right hand
side into an integral over the closed surface $A$ bounding $V$, we write:
\begin{equation}
\frac{\partial}{\partial t} \int_V \left( U_E + U_H \right) \, dV =
- \oint_A \vec{S} \cdot d\vec{A} - \int_V \vec{E} \cdot \vec{J} \, dV,
\label{poyntingtheoremintegralform}
\end{equation}
where:
\begin{eqnarray}
U_E & = & \frac{1}{2} \varepsilon \vec{E}^2 \\
U_H & = & \frac{1}{2} \mu \vec{H}^2 \\
\vec{S} & = & \vec{E} \times \vec{H}. \label{poyntingvector}
\end{eqnarray}

The physical interpretation follows from the volume integral on the right hand side of
Eq.~(\ref{poyntingtheoremintegralform}): this represents the rate at which the
electric field does work on the charges contained within the volume $V$.  If the field
does work on the charges within the field, then there must be energy contained within
the field that decreases as a result of the field doing work. Each of the terms within
the integral on the left hand side of Eq.~(\ref{poyntingtheoremintegralform}) has the
dimensions of energy density (energy per unit volume).  Therefore, the integral has
the dimensions of energy; it is then natural to interpret the full expression on the left
hand side of Eq.~(\ref{poyntingtheoremintegralform}) as the rate of change of energy
in the electromagnetic field within the volume $V$.  The quantities $U_E$ and $U_H$
represent the energy per unit volume in the electric field and in the magnetic field
respectively.

Finally, there remains the interpretation of the first term on the right hand side of
Eq.~(\ref{poyntingtheoremintegralform}).  As well as the energy in the field decreasing
as a result of the field doing work on charges, the energy may change as a result of a
flow of energy purely within the field itself (i.e. even in the absence of any electric charge).
Since the first term on the right hand side of Eq.~(\ref{poyntingtheoremintegralform})
is a surface integral, it is natural to interpret the vector inside the integral as the
energy flux within the field, i.e. the energy crossing unit area (perpendicular to the
vector) per unit time.  The vector $\vec{S}$ defined by Eq.~(\ref{poyntingvector}) is
called the \emph{Poynting vector}.

\subsection{Energy in an electromagnetic wave}

As an application of Poynting's theorem (or rather, of the expressions for energy density
and energy flux that arise from it), let us consider the energy in a plane electromagnetic
wave in free space.  As we noted above, if we use complex notation for the fields, then
we should take the real part to find the physical fields before using expressions involving
the products of fields (such as the expressions for the energy density and energy flux).

The electric field in a plane wave in free space is given by:
\begin{equation}
\vec{E} = \vec{E}_0 \cos \! \left( \vec{k} \cdot \vec{r} - \omega t + \phi_0 \right).
\end{equation}
Thus, the energy density in the electric field is:
\begin{equation}
U_E = \frac{1}{2} \varepsilon_0 \vec{E}^2
= \frac{1}{2} \varepsilon_0 \vec{E}^2_0 \cos^2 \! \left( \vec{k} \cdot \vec{r} - \omega t + \phi_0 \right).
\end{equation}
If we take the average over time at any point in space (or, the average over space at any
point in time), we find that the average energy density is:
\begin{equation}
\langle U_E \rangle = \frac{1}{4} \varepsilon_0 \vec{E}^2_0.
\end{equation}

The magnetic field in a plane wave in free space is given by:
\begin{equation}
\vec{B} = \vec{B}_0 \cos \! \left( \vec{k} \cdot \vec{r} - \omega t + \phi_0 \right),
\end{equation}
where:
\begin{equation}
| \vec{B}_0 | = \frac{| \vec{E}_0 |}{c}.
\label{beratioemwave}
\end{equation}
Since $\vec{B} = \mu_0 \vec{H}$, the energy density in the magnetic field is:
\begin{equation}
U_H = \frac{1}{2} \mu_0 \vec{H}^2
= \frac{1}{2\mu_0} \vec{B}^2_0 \cos^2 \! \left( \vec{k} \cdot \vec{r} - \omega t + \phi_0 \right).
\end{equation}
If we take the average over time at any point in space (or, the average over space at any
point in time), we find that the average energy density is:
\begin{equation}
\langle U_H \rangle = \frac{1}{4\mu_0} \vec{B}^2_0.
\end{equation}
Using the relationship (\ref{beratioemwave}) between the electric and magnetic fields
in a plane wave, this can be written:
\begin{equation}
\langle U_H \rangle = \frac{1}{4\mu_0} \frac{\vec{E}^2_0}{c^2}.
\end{equation}
Then, using Eq.~(\ref{speedoflight}):
\begin{equation}
\langle U_H \rangle = \frac{1}{4} \varepsilon_0 \vec{E}^2_0.
\end{equation}

We see that in a plane electromagnetic wave in free space, the energy is shared equally
between the electric field and the magnetic field, with the energy density averaged over time
(or, over space) given by:
\begin{equation}
\langle U \rangle = \frac{1}{2} \varepsilon_0 \vec{E}^2_0.
\label{energydensityplanewave}
\end{equation}

Finally, let us calculate the energy flux in the wave.  For this, we use the Poynting vector
(\ref{poyntingvector}):
\begin{equation}
\vec{S} = \vec{E} \times \vec{H}
= \hat{k} \frac{1}{\mu_0 c} \vec{E}_0^2
 \cos^2 \! \left( \vec{k} \cdot \vec{r} - \omega t + \phi_0 \right),
\end{equation}
where $\hat{k}$ is a unit vector in the direction of the wave vector.
The average value (over time at a particular point in space, or over space at a particular
time) is then given by:
\begin{equation}
\langle \vec{S} \rangle
= \frac{1}{2} \frac{1}{\mu_0 c} \vec{E}_0^2 \hat{k}
= \frac{1}{2} \varepsilon_0 c \vec{E}_0^2 \hat{k}
= \frac{\vec{E}^2_0}{2Z_0} \hat{k},
\end{equation}
where $Z_0$ is the impedance of free space, defined by:
\begin{equation}
Z_0 = \sqrt{\frac{\mu_0}{\varepsilon_0}}.
\end{equation}
$Z_0$ is a physical constant, with value $Z_0 \approx 376.73\,\Omega$.
Using Eq.~(\ref{energydensityplanewave}) we find the relation between energy flux
and energy density in a plane electromagnetic wave in free space:
\begin{equation}
\langle \vec{S} \rangle
= \langle U \rangle c \hat{k}.
\end{equation}
This is the relationship that we might expect in this case: the mean energy flux is given
simply by the mean
energy density moving at the speed of the wave in the direction of the wave.
But note that this is not the general case.  More generally, the energy in a wave
propagates at the group velocity, which may be different from the phase velocity.
For a plane electromagnetic wave in free space, the group velocity happens to be
equal to the phase velocity, $c$.  We shall discuss this further when we consider
energy propagation in waveguides.

\section{Electromagnetic potentials}

We have seen that to find non-trivial solutions for Maxwell's electromagnetic field
equations in free space, it is helpful to take additional derivatives of the equations,
since this allows us to construct separate equations for the electric and magnetic
fields.  The same technique can be used to find solutions for the fields when
sources (charge densities and currents) are present.  Such situations are important,
since they arise in the generation of electromagnetic waves.
However, it turns out that in systems where charges and currents are present, it
is often simpler to work with the \emph{electromagnetic potentials}, from which the
fields may be obtained by differentiation, than with the fields directly.

\subsection{Relationships between the potentials and the fields}

The scalar potential $\phi$ and vector potential $\vec{A}$ are defined so that the
electric and magnetic fields are obtained using the relations:
\begin{eqnarray}
\vec{B} & = & \curlop \vec{A}, \label{magneticpotential} \\
\vec{E} & = & -\gradop \phi - \frac{\partial \vec{A}}{\partial t}. \label{electricpotential}
\end{eqnarray}
We shall show below that as long as $\phi$ and $\vec{A}$ satisfy appropriate equations,
then the fields $\vec{B}$ and $\vec{E}$ derived from them using
Eqs.~(\ref{magneticpotential}) and (\ref{electricpotential}) satisfy Maxwell's equations.
But first, note that there is a many-to-one relationship between the potentials and the
fields.  That is, there are many different potentials that can give the same fields.
For example, we could add any uniform (independent of position) value to the scalar
potential $\phi$, and leave the electric field $\vec{E}$ unchanged, since the gradient of
a constant is zero.  Similarly, we could add any vector function with vanishing curl to the
vector potential $\vec{A}$; and if this function is independent of time, then again the
electric and magnetic fields are unchanged.  This property of the potentials is known as
\emph{gauge invariance}, and is of considerable practical value, as we shall see below.

\subsection{Equations for the potentials}

The fact that there is a relationship between the potentials and the fields implies
that the potentials that are allowed in physics have to satisfy certain equations,
corresponding to Maxwell's equations.  This is, of course, the case.  In this section.
we shall derive the equations that must be satisfied by the potentials, if the fields
that are derived from them are to satisfy Maxwell's equations.

However, to begin with, we show that two of Maxwell's equations (the ones
independent of the sources) are in fact satisfied if the fields are derived from 
\emph{any} potentials $\phi$ and $\vec{A}$ using Eqs.~(\ref{magneticpotential}) 
and (\ref{electricpotential}).
First, since the divergence of the curl of any differentiable vector field is always zero:
\begin{equation}
\divop \curlop \vec{A} \equiv 0,
\end{equation}
it follows that Maxwell's equation (\ref{eq:maxwell2}) is satisfied for any vector potential
$\vec{A}$.
Then, since the curl of the gradient of any differentiable scalar field is always zero:
\begin{equation}
\curlop \gradop \phi \equiv 0,
\end{equation}
Maxwell's equation (\ref{eq:maxwell3}) is satisfied for any scalar potential $\phi$
and vector potential $\vec{A}$ (as long as the magnetic field is obtained from the
vector potential by Eq.~(\ref{magneticpotential})).

Now let us consider the equations involving the source terms (the charge density $\rho$
and current density $\vec{J}$).
Differential equations for the potentials can be obtained by substituting from
Eqs.~(\ref{magneticpotential}) and (\ref{electricpotential}) into Maxwell's equations
(\ref{eq:maxwell1}) and (\ref{eq:maxwell4}).  We also need to use the constitutive
relations between the magnetic field $\vec{B}$ and the magnetic intensity $\vec{H}$,
and between the electric field $\vec{E}$ and the electric displacement $\vec{D}$.
For simplicity, let us assume a system of charges and currents in free space; then
the constitutive relations are:
\begin{equation}
\vec{D} = \varepsilon_0 \vec{E}, \qquad
\vec{B} = \mu_0 \vec{H}.
\end{equation}
Substituting from Eq.~(\ref{electricpotential}) into Maxwell's equation (\ref{eq:maxwell1})
gives:
\begin{equation}
\nabla^2 \phi + \frac{\partial}{\partial t} \divop \vec{A} = -\frac{\rho}{\varepsilon_0}.
\label{potentialsource1}
\end{equation}
Similarly, substituting from Eq.~(\ref{magneticpotential}) into Maxwell's equation
(\ref{eq:maxwell4}) gives (after some algebra):
\begin{equation}
\nabla^2 \vec{A} - \frac{1}{c^2} \frac{\partial^2 \vec{A}}{\partial t^2} = 
-\mu_0 \vec{J} + \gradop \left( \divop \vec{A} + \frac{1}{c^2} \frac{\partial \phi}{\partial t} \right).
\label{potentialsource2}
\end{equation}
Eqs.~(\ref{potentialsource1}) and (\ref{potentialsource2}) relate the electromagnetic
potentials to a charge density $\rho$ and current density $\vec{J}$ in free space.
Unfortunately, they are coupled equations (the scalar potential $\phi$ and vector potental
$\vec{J}$ each appear in both equations), and are rather complicated.
However, we noted above that the potentials for given electric and magnetic fields are not
unique: the potentials have the property of gauge invariance.  By imposing an additional
constraint on the potentials, known as a \emph{gauge condition}, it is possible to restrict
the choice of potentials.  With an appropriate choice of gauge, it is possible to
decouple the potentials, and furthermore, arrive at equations that have standard
solutions.  In particular, with the gauge condition:
\begin{equation}
\divop \vec{A} + \frac{1}{c^2} \frac{\partial \phi}{\partial t} = 0,
\label{lorenzgauge}
\end{equation}
then Eq.~(\ref{potentialsource1}) becomes:
\begin{equation}
\nabla^2 \phi - \frac{1}{c^2} \frac{\partial^2 \phi}{\partial t^2} = -\frac{\rho}{\varepsilon_0}.
\label{potentialsource2lorenz}
\end{equation}
and Eq.~(\ref{potentialsource2}) becomes:
\begin{equation}
\nabla^2 \vec{A} - \frac{1}{c^2} \frac{\partial^2 \vec{A}}{\partial t^2} = 
-\mu_0 \vec{J},
\label{potentialsource1lorenz}
\end{equation}
Eqs.~(\ref{potentialsource2lorenz}) and (\ref{potentialsource1lorenz}) have the form
of wave equations with source terms.  It is possible to write solutions in terms of integrals
over the sources: we shall do this shortly.  However, before we do so, it is important to note
that for any given potentials, it is possible to find new potentials that satisfy Eq.~(\ref{lorenzgauge}),
but give the same fields as the original potentials.  Eq.~(\ref{lorenzgauge}) is called the
\emph{Lorenz gauge}.  The proof proceeds as follows.

First we show that any function $\psi$ of position and time can be used to construct a gauge
transformation; that is, we can use $\psi$ to find new scalar and vector potentials (different
from the original potentials) that given the same electric and magnetic fields as the original
potentials.  Given the original potentials $\phi$ and $\vec{A}$, and a function $\psi$,
let us define new potentials $\phi^\prime$ and $\vec{A}^\prime$, as follows:
\begin{eqnarray}
\phi^\prime      & = & \phi + \frac{\partial \psi}{\partial t}, \label{gaugexform1} \\
\vec{A}^\prime & = & \vec{A} - \gradop \psi. \label{gaugexform2}
\end{eqnarray}
Eqs.~(\ref{gaugexform1}) and (\ref{gaugexform2}) represent a \emph{gauge transformation}.
If the original potentials give fields $\vec{E}$ and $\vec{B}$, then the magnetic field derived from
the new vector potential is:
\begin{equation}
\vec{B}^\prime = \curlop \vec{A}^\prime = \curlop \vec{A} = \vec{B},
\end{equation}
where we have used the fact that the curl of the gradient of any scalar function is zero.
The electric field derived from the new potentials is:
\begin{eqnarray}
\vec{E}^\prime & = & -\gradop \phi^\prime - \frac{\partial \vec{A}^\prime}{\partial t} \nonumber \\
                        & = & -\gradop \phi - \frac{\partial \vec{A}}{\partial t}
                                 -\gradop  \frac{\partial \psi}{\partial t}
                                + \frac{\partial}{\partial t} \gradop \psi \nonumber \\
                        & = & -\gradop \phi - \frac{\partial \vec{A}}{\partial t} \nonumber \\
                        & = & \vec{E}.
\end{eqnarray}
Here, we have made use of the fact that position and time are independent variables, so it
is possible to interchange the order of differentiation.  We see that for \emph{any} function
$\psi$, the fields derived from the new potentials are the same as the fields derived from the
original potentials.  We say that $\psi$ generates a gauge transformation: it gives us new
potentials, while leaving the fields unchanged.

Finally, we show how to choose a gauge transformation so that the new potentials satisfy
the Lorenz gauge condition.  In general, the new potentials satisfy the equation:
\begin{equation}
\divop \vec{A}^\prime + \frac{1}{c^2} \frac{\partial \phi^\prime}{\partial t}
= \divop \vec{A} + \frac{1}{c^2} \frac{\partial \phi}{\partial t}
- \nabla^2 \psi + \frac{1}{c^2} \frac{\partial^2 \psi}{\partial t^2}.
\end{equation}
Suppose we have potentials $\phi$ and $\vec{A}$ that satsify:
\begin{equation}
\divop \vec{A} + \frac{1}{c^2} \frac{\partial \phi}{\partial t} =  f,
\end{equation}
where $f$ is some function of position and time.  (If $f$ is non-zero, then the potentials 
$\phi$ and $\vec{A}$ do not satisfy the Lorenz gauge condition.)  
Therefore, if $\psi$ satisfies:
\begin{equation}
\nabla^2 \psi - \frac{1}{c^2} \frac{\partial^2 \psi}{\partial t^2} = f,
\label{generatingfunction}
\end{equation}
Then the new potentials $\phi^\prime$ and $\vec{A}^\prime$ satisfy the Lorenz gauge
condition:
\begin{equation}
\divop \vec{A^\prime} + \frac{1}{c^2} \frac{\partial \phi^\prime}{\partial t} = 0.
\end{equation}
Notice that Eq.~(\ref{generatingfunction}) again has the form of a wave equation, with a
source term.  Assuming that we can solve such an equation, then it is always possible to
find a gauge transformation such that, starting from some given original potentials, the new
potentials satisfy the Lorenz gauge condition.

\subsection{Solution of the wave equation with source term}

In the Lorenz gauge (\ref{lorenzgauge}):
\begin{equation}
\divop \vec{A} + \frac{1}{c^2} \frac{\partial \phi}{\partial t} =  0, \nonumber
\end{equation}
the vector potential $\vec{A}$ and the scalar potential $\phi$ satisfy the wave equations
(\ref{potentialsource1lorenz}) and (\ref{potentialsource2lorenz}):
\begin{eqnarray}
\nabla^2 \vec{A} - \frac{1}{c^2} \frac{\partial^2 \vec{A}}{\partial t^2} & = & 
-\mu_0 \vec{J}, \nonumber \\
\nonumber \\
\nabla^2 \phi - \frac{1}{c^2} \frac{\partial^2 \phi}{\partial t^2} & = & -\frac{\rho}{\varepsilon_0}.
\nonumber
\end{eqnarray}
Note that the wave equations have the form (for given charge density and current density)
of two \emph{uncoupled} second-order differential equations.  In many situations, it is
easier to solve these equations, than to solve Maxwell's equations for the fields (which
take the form of four first-order \emph{coupled} differential equations).

\begin{figure}[t]
\centering
\includegraphics[width=0.3\linewidth]{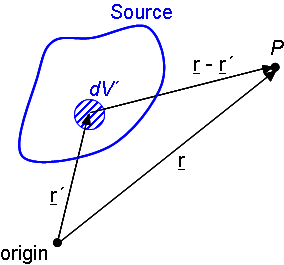}
\caption{Integration over a distributed source.
\label{fig:distributedsource}}
\end{figure}

We state (without giving a full proof) the solution to the wave equation with a source term.
The solution is:
\begin{equation}
\phi(\vec{r}, t) = \frac{1}{4\pi \varepsilon_0}
\int \frac{\rho(\vec{r}^\prime, t^\prime)}{|\vec{r}^\prime - \vec{r}|} \, dV^\prime,
\label{waveeqsourcesolution}
\end{equation}
where:
\begin{equation}
t^\prime = t - \frac{|\vec{r}^\prime - \vec{r}|}{c}.
\end{equation}
The integral extends over all space -- see Fig.\,\ref{fig:distributedsource}.  Note that the
source at each point in the integral has to
be evaluated at a time $t^\prime$, which depends on the distance between the source point
and the obeservation point (at which we are evaluating the potential).  From the equation
in the absence of any sources, we expect variations in the potential to propagate through
space at a speed $c$ (the speed of light).  The difference between $t^\prime$ and $t$ simply
accounts for the time taken for the effect of any change in the charge density at the source
point to propagate through space to the observation point.

We do not present a proof that Eq.~(\ref{waveeqsourcesolution}) represents a solution
to the wave equation (\ref{potentialsource2lorenz}).  However, we can at least see that in the
static case, Eq.~(\ref{potentialsource2lorenz}) reduces to the familiar form of Poisson's equation:
\begin{equation}
\nabla^2 \phi = -\frac{\rho}{\varepsilon_0}.
\label{poisson}
\end{equation}
For a point charge $q$ at a point $\vec{r}_0$, Eq.~(\ref{poisson}) has solution:
\begin{equation}
\phi(\vec{r}) = \frac{1}{4\pi \varepsilon_0}
\frac{q}{|\vec{r}_0 - \vec{r}|}.
\label{waveeqpointsourcesolution}
\end{equation}
This can be obtained directly from Eq.~(\ref{waveeqsourcesolution}) if the source is given by a Dirac delta function:
\begin{equation}
\rho(\vec{r}^\prime) = q\, \delta(\vec{r}^\prime - \vec{r}_0).
\end{equation}

The wave equation for the vector potential (\ref{potentialsource1lorenz}) has a solution that can
be expressed in a similar form to that for the scalar potential:
\begin{equation}
\vec{A}(\vec{r}, t) = \frac{\mu_0}{4\pi}
\int \frac{\vec{J}(\vec{r}^\prime, t^\prime)}{|\vec{r}^\prime - \vec{r}|} \, dV^\prime.
\label{waveeqsourcesolution1}
\end{equation}
Note that Eq.~(\ref{waveeqsourcesolution1}) can be expressed as three independent equations
(integrals), for the three components of the vector potential.

\subsection{Physical significance of the fields and potentials}

An electromagnetic field is really a way of describing the interaction between particles that have
electric charges.  Given a system of charged particles, one could, in principle, write down equations for
the evolution of the system purely in terms of the positions, velocities, and charges of the various
particles.  However, it is often convenient to carry out an intermediate step in which one computes
the fields generated by the particles, and then computes the effects of the fields on the motion of
the particles.  Maxwell's equations provide a prescription for computing the fields arising from
a given system of charges.  The effects of the fields on a charged particle are expressed by the
Lorentz force equation:
\begin{equation}
\vec{F} = q \left( \vec{E} + \vec{v} \times \vec{B} \right),
\label{lorentzforce}
\end{equation}
where $\vec{F}$ is the force on the particle, $q$ is the charge, and $\vec{v}$ is the velocity of
the particle.  The motion of the particle under the influence of a force $\vec{F}$ is then given by
Newton's second law of motion:
\begin{equation}
\frac{d}{dt} \gamma m \vec{v} = \vec{F}.
\label{newtonssecondlaw}
\end{equation}
Eqs.~(\ref{lorentzforce}) and (\ref{newtonssecondlaw}) make clear the physical significance of
the fields.  But what is the significance of the potentials?  At first, the feature of gauge invariance
appears to make it difficult to assign any definite physical significance to the potentials: in any
given system, we have a certain amount of freedom in changing the potentials without changing
the fields that are present.  However, let us consider first the case of a particle in a static
electric field.  In this case, the Lorentz force is given by:
\begin{equation}
\vec{F} = q \vec{E} = -q\, \gradop \phi.
\end{equation}
If the particle moves from position $\vec{r}_1$ to position $\vec{r_2}$ under the influence of the
Lorentz force, then the work done on the particle (by the field) is:
\begin{equation}
W = \int_{\vec{r}_1}^{\vec{r}_2} \vec{F} \cdot d\vec{\ell} = 
-q \, \int_{\vec{r}_1}^{\vec{r}_2} \gradop \phi \cdot d\vec{\ell} = 
-q \left[ \phi(\vec{r}_2) - \phi(\vec{r}_1) \right].
\end{equation}
Note that the work done by the field when the particle moves between two points depends on the
\emph{difference} in the potential at the two points; and that the work done is independent of
the path taken by the particle in moving between the two points.  This suggests that the scalar
potential $\phi$ is related to the energy of a particle in an \emph{electrostatic} field.  If a
time-dependent magnetic field is present, the analysis becomes more complicated.

A more complete understanding of the physical significance of the scalar and vector potentials
is probably best obtained in the context of Hamiltonian mechanics.  In this formalism, the equations
of motion of a particle are obtained from the Hamiltonian, $H(\vec{x},\vec{p};t)$; the Hamiltonian is
a function of the particle coordinates $\vec{x}$, the (canonical) momentum $\vec{p}$, and an
independent variable $t$ (often corresponding to the time).  Note that
the canonical momentum can (and generally does) differ from the usual mechanical momentum.
The Hamiltonian defines the dynamics of a system, in the same way that a force defines the
dynamics in Newtonian mechanics.  In Hamiltonian mechanics, the equations of motion of a particle
are given by Hamilton's equations:
\begin{eqnarray}
\frac{dx_i}{dt} & = & \frac{\partial H}{\partial p_i}, \label{hamilton1} \\
\frac{dp_i}{dt} & = & - \frac{\partial H}{\partial x_i}. \label{hamilton2}
\end{eqnarray}
In the case of a charged particle in an electromagnetic field, the Hamiltonian is given by:
\begin{equation}
H = c \sqrt{(\vec{p} - q\vec{A})^2 + m^2 c^2} + q\phi,  \label{hamilton3}
\end{equation}
where the canonical momentum is:
\begin{equation}
\vec{p} = \vec{\beta} \gamma mc + q \vec{A},  \label{hamilton4}
\end{equation}
where $\vec{\beta}$ is the normalised velocity of the particle, $\vec{\beta} = \vec{v}/c$.

Note that Eqs.~(\ref{hamilton1}) -- (\ref{hamilton4}) give the same dynamics as the Lorentz force
equation (\ref{lorentzforce}) together with Newton's second law of motion,
Eq.~(\ref{newtonssecondlaw}); they are just written in a different formalism.  The significant
point is that in the Hamiltonian formalism, the dynamics are expressed in terms of the potentials,
rather than the fields.  The Hamiltonian can be interpreted as the ``total energy'' of a particle,
${\cal E}$.
Combining equations (\ref{hamilton3}) and (\ref{hamilton4}), we find:
\begin{equation}
{\cal E} = \gamma mc^2 + q\phi.
\end{equation}
The first term gives the kinetic energy, and the second term gives the potential energy: this
is consistent with our interpretation above, but now it is more general.  Similarly, in 
Eq.~(\ref{hamilton4}) the ``total
momentum'' consists of a mechanical term, and a potential term:
\begin{equation}
\vec{p} = \vec{\beta} \gamma mc + q\vec{A}.
\end{equation}
The vector potential $\vec{A}$ contributes to the total momentum of the particle, in the same
way that the scalar potential $\phi$ contributes to the total energy of the particle.  Gauge
invariance allows us to find new potentials that leave the fields (and hence the dynamics) of
the system unchanged.  Since the fields are obtained by taking derivatives of the potentials,
this suggests that only changes in potentials between different positions and times are
significant for the dynamics of charged particles.  This in turn implies that only changes in
(total) energy and (total) momentum are significant for the dynamics.

\section{Generation of electromagnetic waves}

As an example of the practical application of the potentials in a physical system, let us consider
the generation of electromagnetic waves from an oscillating, infinitesimal electric dipole.
Although idealised, such a system provides a building block for constructing more realistic
sources of radiation (such as the half-wave antenna), and is therefore of real interest.  An
infinitesimal electric dipole oscillating at a single frequency is known as an \emph{Hertzian dipole}.

\begin{figure}[t]
\centering
\includegraphics[width=0.4\linewidth]{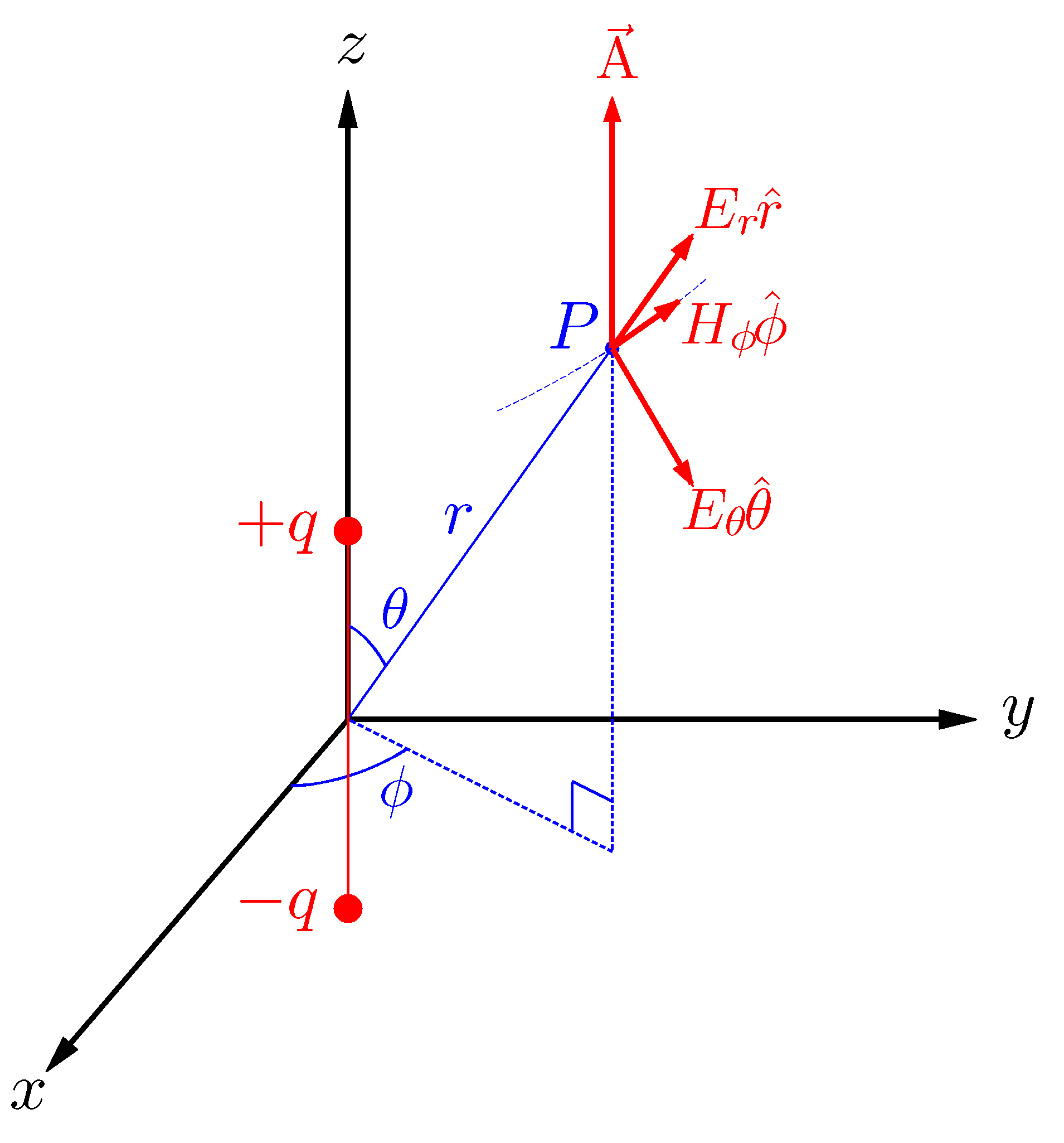}
\caption{Hertzian dipole: the charges oscillate around the origin along the $z$ axis with
infinitesimal amplitude.  The vector potential at any point is parallel to the $z$ axis, and
oscillates at the same frequency as the dipole, with a phase difference and amplitude
depending on the distance from the origin.
\label{fig:hertziandipole}}
\end{figure}

Consider two point-like particles located on the $z$ axis, close to and on opposite sides
of the origin.  Suppose that electric charge flows between the particles, so that the
charge on each particle oscillates, with the charge on one particle being:
\begin{equation}
q_1 = +q_0 e^{-i\omega t},
\end{equation}
and the charge on the other particle being:
\begin{equation}
q_2 = -q_0 e^{-i\omega t}.
\end{equation}
The situation is illustrated in Fig.\,\ref{fig:hertziandipole}.  The current at any point between
the charges is:
\begin{equation}
\vec{I} = \frac{dq_1}{dt} \hat{z} = -i\omega q_0 e^{-i\omega t} \hat{z}.
\end{equation}
In the limit that the distance between the charges approaches zero, the charge density vanishes; however, there remains a non-zero electric current at the origin, oscillating at frequency $\omega$
and with amplitude $I_0$, where:
\begin{equation}
I_0 = -i\omega q_0.
\end{equation}

Since the current is located only at a single point in space (the origin), it is straightforward
to perform the integral in Eq.~(\ref{waveeqsourcesolution1}), to find the vector potential at
any point away from the origin:
\begin{equation}
\vec{A} (\vec{r}, t) = \frac{\mu_0}{4\pi} (I_0 \ell)
\frac{e^{i(kr - \omega t)}}{r}\,\hat{z},
\label{hertziandipolevectorpotential}
\end{equation}
where:
\begin{equation}
k = \frac{\omega}{c},
\end{equation}
and $\ell$ is the length of the current: strictly speaking, we take the limit $\ell \to 0$, but
we do so, we increase the current amplitude $I_0$, so that the produce $I_0 \ell$ remains
non-zero and finite.

Notice that, with Eq.~(\ref{hertziandipolevectorpotential}), we have quickly found a relatively
simple expression for the vector potential around an Hertzian dipole.  From the vector potential
we can find the magnetic field; and from the magnetic field we can find the electric field.  By
working with the potentials rather than with the fields, we have greatly simplified the finding
of the solution in what might otherwise have been quite a complex problem.

The magnetic field is given, as usual, by $\vec{B} = \curlop \vec{A}$.  It is convenient to work
in spherical polar coordinates, in which case the curl is written as:
\begin{equation}
\nabla \times \vec{A} \equiv
\frac{1}{r^2 \sin \theta}
\left| \begin{array}{ccc}
\hat{r} & r\hat{\theta} & r\sin \theta \, \hat{\phi} \\
\frac{\partial}{\partial r} & \frac{\partial}{\partial \theta} & \frac{\partial}{\partial \phi} \\
A_r & r A_\theta & r\sin \theta A_\phi
\end{array} \right| .
\end{equation}
Evaluating the curl for the vector potential given by Eq.~(\ref{hertziandipolevectorpotential}) we find:
\begin{eqnarray}
B_r      & = & 0, \\
B_\theta & = & 0, \\
B_\phi   & = & \frac{\mu_0}{4\pi} (I_0\ell) k\, \sin \theta
\left( \frac{1}{kr} - i \right) \frac{e^{i(kr - \omega t)}}{r}.
\end{eqnarray}
The electric field can be obtained from $\nabla \times \vec{B} = \frac{1}{c^2}
\frac{\partial \vec{E}}{\partial t}$ (which follows from Maxwells equation (\ref{eq:maxwell4})
in free space).  The result is:
\begin{eqnarray}
E_r      & = & \frac{1}{4\pi \varepsilon_0} \frac{2}{c} (I_0\ell)
\left( 1 + \frac{i}{kr} \right)\, \frac{e^{i(kr - \omega t)}}{r^2}, \\
E_\theta & = & \frac{1}{4\pi \varepsilon_0} (I_0\ell) \frac{k}{c}
\, \sin \theta \left( \frac{i}{k^2r^2} + \frac{1}{kr} - i \right)\,
\frac{e^{i(kr - \omega t)}}{r}, \\
E_\phi   & = & 0.
\end{eqnarray}
Notice that the expressions for the fields are considerably more complicated than the expression
for the vector potential, and would be difficult to obtain by directly solving Maxwell's equations.

The expressions for the fields all involve a phase factor $e^{i(kr - \omega t)}$, with additional
factors giving the detailed dependence of the phase and amplitude on distance and angle from the
dipole.  The phase factor $e^{i(kr - \omega t)}$ means that the fields propagate as waves
in the radial direction, with frequency $\omega$ (equal to the frequency of the dipole), and
wavelength $\lambda$ given by:
\begin{equation}
\lambda = \frac{2\pi}{k} = \frac{2\pi c}{\omega}.
\end{equation}

If we make some approximations, we can simplify the expressions for the fields.  In fact, we can
identify two different regimes.  The \emph{near field} regime is defined by the condition $kr \ll 1$.
In this case, the fields are observed at a distance from the dipole much less than the wavelength
of the radiation emitted by the dipole.  The dominant field components are then:
\begin{eqnarray}
B_\phi   & \approx & \frac{\mu_0}{4\pi} (I_0\ell) \, \sin \theta \frac{e^{i(kr - \omega t)}}{r^2}, \\
E_r      & \approx & \frac{1}{4\pi \varepsilon_0} \frac{2i}{c} (I_0\ell)
\, \frac{e^{i(kr - \omega t)}}{k r^3}, \\
E_\theta & \approx & \frac{1}{4\pi \varepsilon_0} (I_0\ell) \frac{i}{c}
\, \sin \theta \, \frac{e^{i(kr - \omega t)}}{kr^3}.
\end{eqnarray}

The \emph{far field} regime is defined by the condition $kr \gg 1$.  In this regime, the fields
are observed at distances from the dipole that are large compared to the wavelength of the
radiation emitted by the dipole.  The dominant field components are:
\begin{eqnarray}
B_\phi   & \approx & -i \frac{\mu_0}{4\pi} (I_0\ell) \, k \sin \theta \frac{e^{i(kr - \omega t)}}{r}, \\
E_\theta & \approx & -i \frac{1}{4\pi \varepsilon_0} (I_0\ell) \frac{k}{c}
\, \sin \theta \, \frac{e^{i(kr - \omega t)}}{r}.
\end{eqnarray}
The following features of the fields in this regime are worth noting:
\begin{itemize}
\item The electric and magnetic field components are perpendicular to each other, and to the
(radial) direction in which the wave is propagating.
\item At any position and time, the electric and magnetic fields are in phase with each other.
\item The ratio between the magnitudes of the fields at any given position is
$| E_\theta | / | B_\phi | \approx c$.
\end{itemize}
These are all properties associated with plane waves in free space.  Furthermore, the amplitudes
of the fields falls off as $1/r$: at sufficiently large distance from the oscillating dipole, the
amplitudes decrease slowly with increasing distance.  At a large distance from an oscillating dipole,
the electromagnetic waves produced by the dipole make a good approximation to plane waves in
free space.

It is also worth noting the dependence of the field amplitudes on the polar angle $\theta$: the
amplitudes vanish for $\theta = 0$ and $\theta = \pi$, i.e. in the direction of oscillation of the
charges in the dipole.  However, the amplitudes reach a maximum for $\theta = \pi/2$, i.e. in a
plane through the dipole, and perpendicular to the direction of oscillation of the dipole.

\begin{figure}[t]
\centering
\includegraphics[width=0.4\linewidth]{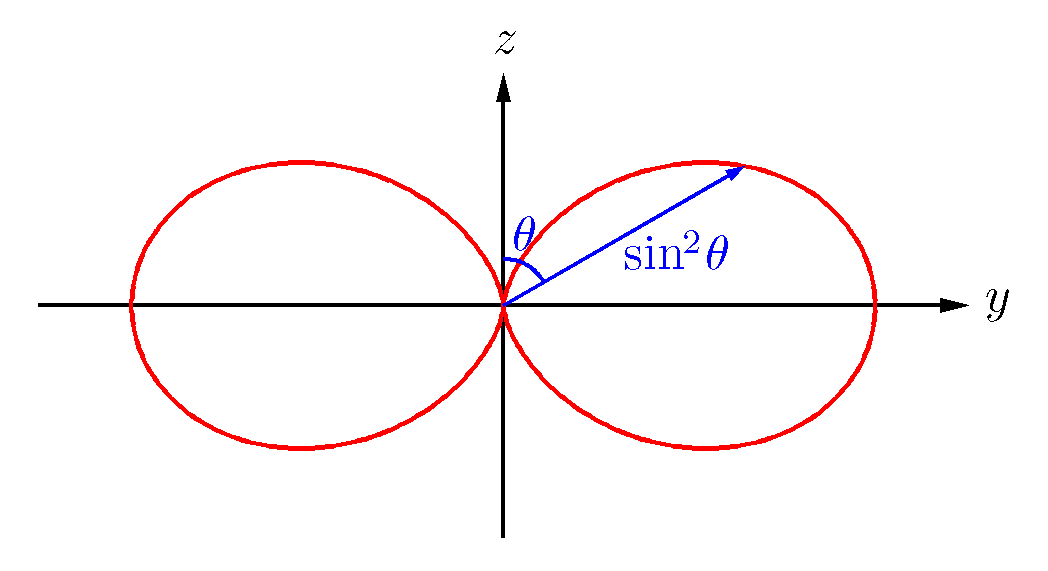}
\caption{Distribution of radiation power from an Hertzian dipole.  The current in the dipole
is oriented along the $z$ axis.  The distance of a point on the curve from the origin indicates
the relative power density in the direction from the origin to the point on the curve.
\label{fig:dipoleradiationpowerdistribution}}
\end{figure}

We have seen that electromagnetic waves carry energy.  This suggests that the Hertzian dipole
radiates energy, and that some energy ``input'' will be required to maintain the amplitude of
oscillation of the dipole.  This is indeed the case.  Let us calculate the rate at which the dipole
will radiate energy.  As usual, we use the expression:
\begin{equation}
\vec{S} = \vec{E} \times \vec{H},
\end{equation}
where the Poynting vector $\vec{S}$ gives the amount of energy in an electromagnetic field
crossing unit area (perpendicular to $\vec{S}$) per unit time.
Before taking the vector product, we need to take the real parts of the expressions for the fields:
\begin{eqnarray}
B_\phi   & = & \frac{\mu_0}{4\pi} (I_0\ell) k \frac{\sin \theta}{r}
\left( \frac{\cos (kr - \omega t)}{kr} + \sin (kr - \omega t) \right), \\
E_r      & = & \frac{1}{4\pi \varepsilon_0} \frac{2}{c} (I_0\ell) \frac{1}{r^2}
\left( \cos (kr - \omega t) - \frac{ \sin (kr - \omega t)}{kr} \right), \\
E_\theta & = & \frac{1}{4\pi \varepsilon_0} (I_0\ell) \frac{k}{c}
\, \frac{\sin \theta}{r}
\left(- \frac{\sin (kr - \omega t)}{k^2r^2} + \frac{\cos (kr - \omega t)}{kr} +
\sin (kr - \omega t) \right).
\end{eqnarray}
The full expression for the Poynting vector will clearly be rather complicated; but if we take
the average over time (or position), then we find that most terms vanish, and we are left with:
\begin{equation}
\langle \vec{S} \rangle = \frac{(I_0 \ell)^2 k^2}{32\pi^2 \varepsilon_0 c}  \frac{\sin^2 \theta}{r^2}
\, \hat{r}.
\label{hertziandipoleenergyflux}
\end{equation}
As expected, the radiation is directional, with most of the power emitted in a plane through
the dipole, and perpendicular to its direction of oscillation; no power is emitted in the
direction in which the dipole oscillates.  The power distribution is illustrated in
Fig.\,\ref{fig:dipoleradiationpowerdistribution}.
The power per unit area falls off with the square
of the distance from the dipole.  This is expected, from conservation of energy.

The total power emitted by the dipole is found by integrating the power per unit area given by
Eq.~(\ref{hertziandipoleenergyflux}) over a surface enclosing the dipole.  For simplicity, let us
take a sphere of radius $r$.  Then, the total (time averaged) power emitted by the dipole is:
\begin{equation}
\langle P \rangle = \int_{\theta = 0}^{\pi} \int_{\phi = 0}^{2\pi}
| \langle \vec{S} \rangle | \, r^2 \sin\theta \, d\theta \, d\phi.
\end{equation}
Using the result:
\begin{equation}
\int_{\theta = 0}^{\pi} \sin^3 \theta \, d\theta =  \frac{4}{3},
\end{equation}
we find:
\begin{equation}
\langle P \rangle = \frac{(I_0 \ell)^2 k^2}{12\pi \varepsilon_0 c} = 
\frac{(I_0 \ell)^2 \omega^2}{12\pi \varepsilon_0 c^3}.
\label{rayleighscattering}
\end{equation}
Notice that, for a given amplitude of oscillation, the total power radiated varies as the square
of the frequency of the oscillation.  The consequences of this fact are familiar in an everyday
observation.  Gas molecules in the Earth's atmosphere behave as small oscillating dipoles
when the electric charges within them respond to the electric field in the sunlight passing through
the atmosphere.  The dipoles re-radiate the energy they absorb, a phenomenon known as
\emph{Rayleigh scattering}.  The energy from the oscillating dipoles is radiated over a range of
directions; after many scattering ``events'', it appears to an observer that the light comes from
all directions, not just the direction of the original source (the sun).  Eq.~(\ref{rayleighscattering})
tells us that shorter wavelength (higher frequency) light is scattered much more strongly than
longer wavelength (lower frequency) light.  Thus, the sky appears blue.

\section{Boundary conditions}

Gauss' theorem and Stokes' theorem can be applied to Maxwell's equations
to derive constraints on the behaviour of electromagnetic fields at
boundaries between different materials.  For RF systems in particle accelerators, the boundary
conditions at the surfaces of highly-conductive materials are of particular significance.

\subsection{General boundary conditions}

\begin{figure}[t]
\centering
\includegraphics[width=0.35\linewidth]{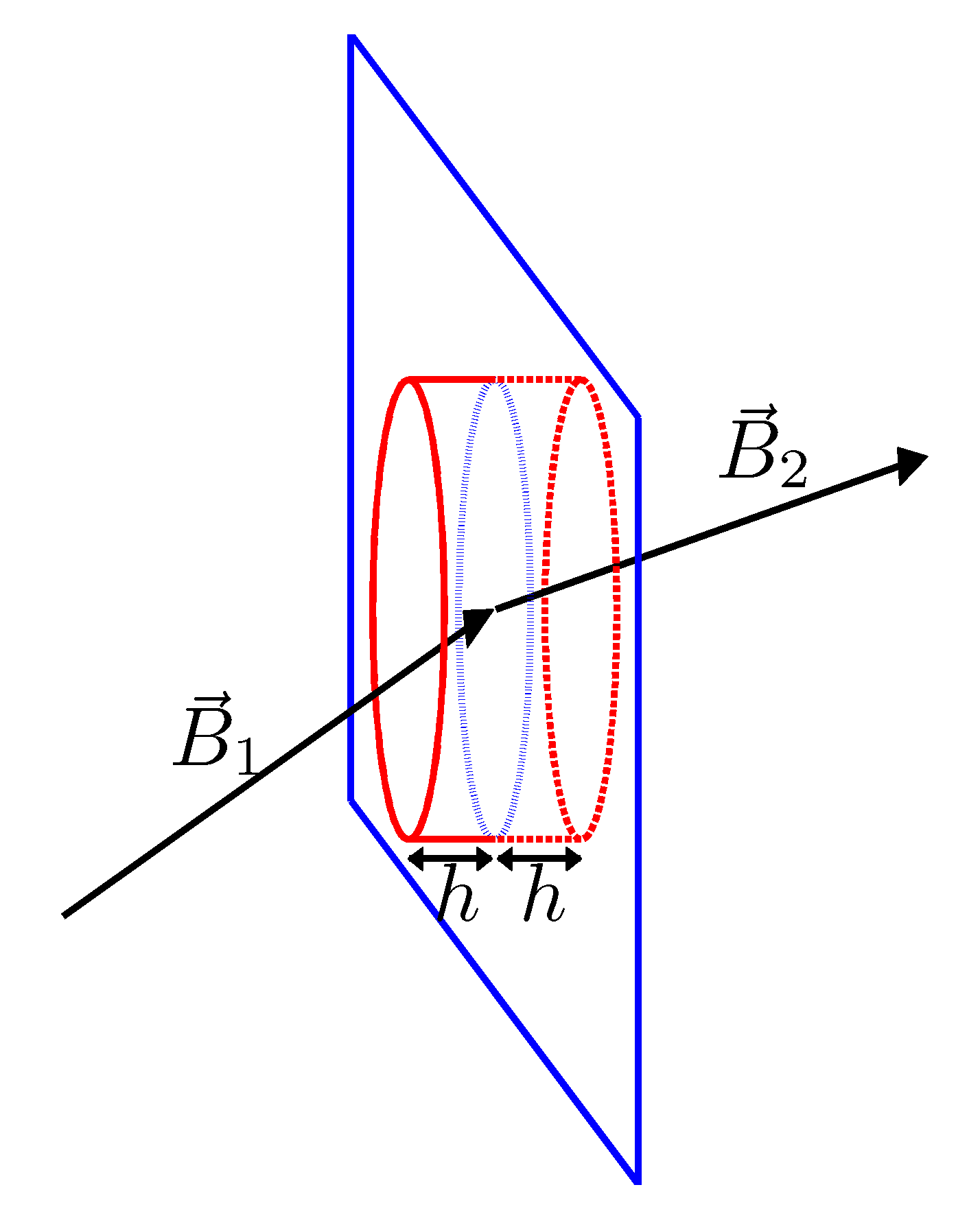}
\hspace{0.06\linewidth}
\includegraphics[width=0.4\linewidth]{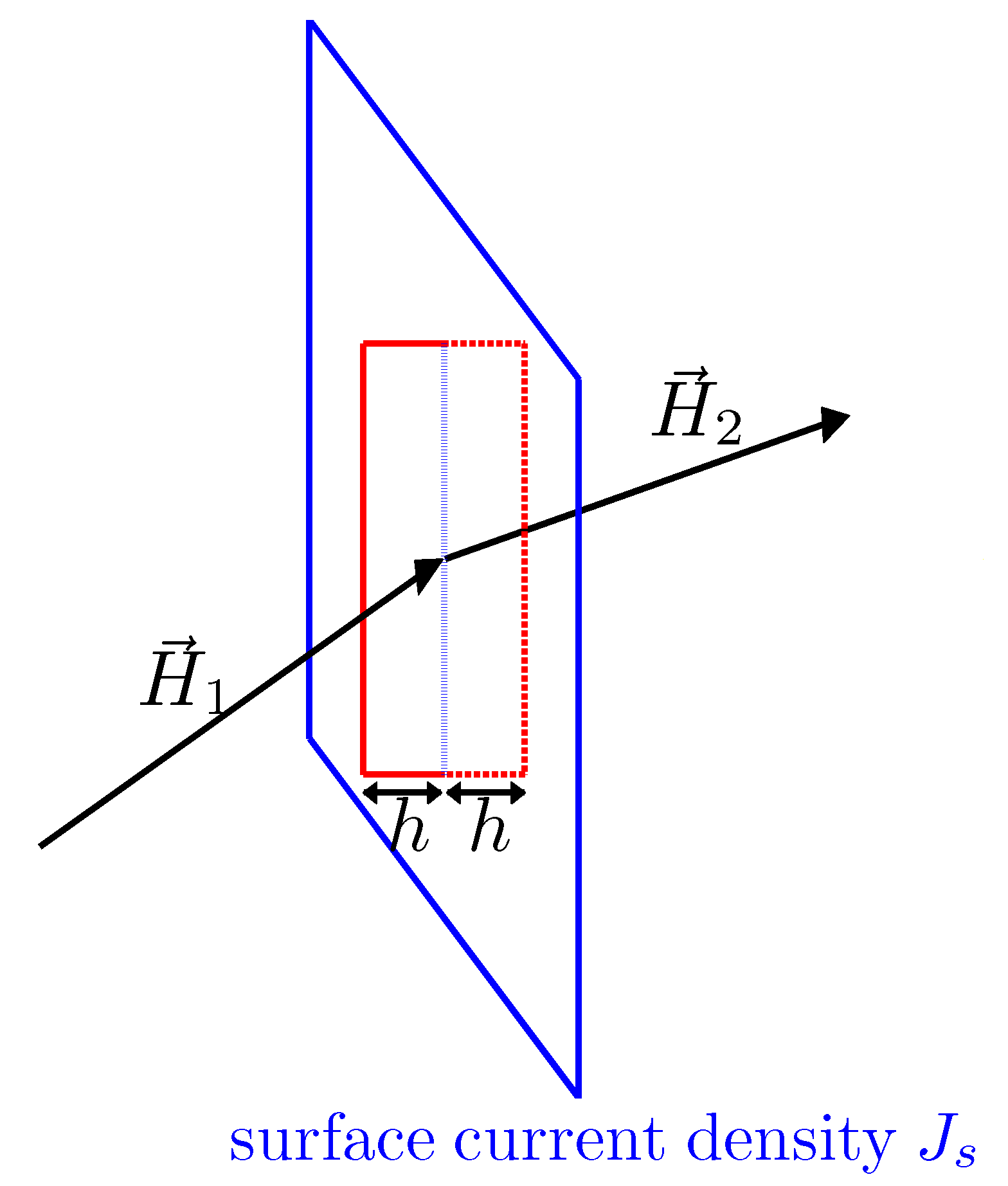}
\caption{(a) Left: ``Pill box'' surface for derivation of the boundary
conditions on the normal component of the magnetic flux density at the
interface between two media.  (b) Right: Geometry for derivation of the
boundary conditions on the tangential component of the magnetic
intensity at the interface between two media.
\label{fig:boundaryconditionb}}
\end{figure}

Consider first a short cylinder or ``pill box'' that crosses the boundary
between two media, with the flat ends of the cylinder parallel to the
boundary, see Fig.\,\ref{fig:boundaryconditionb}\,(a).  Applying Gauss'
theorem to Maxwell's equation (\ref{eq:maxwell2}) gives:
\begin{equation}
\int_V \divop \vec{B}\, dV = \oint_{\partial V} \vec{B} \cdot d\vec{S} = 0,
\nonumber
\end{equation}
where the boundary $\partial V$ encloses the volume $V$ within the cylinder.
If we take the limit where the length of the cylinder ($2h$ --
see Fig.\,~\ref{fig:boundaryconditionb}\,(a)) approaches zero,
then the only contributions to the surface integral come from the flat ends;
if these have infinitesimal area $dS$, then since the orientations of these
surfaces are in opposite directions on opposite sides of the boundary, and
parallel to the normal component of the magnetic field, we find:
\begin{equation}
- B_{1\perp}\, dS + B_{2\perp}\, dS = 0,
\nonumber
\end{equation}
where $B_{1\perp}$ and $B_{2\perp}$ are the normal components of the
magnetic flux density on either side of the boundary.  Hence:
\begin{equation}
B_{1\perp} = B_{2\perp}.
\label{eq:boundarynormalb}
\end{equation}
In other words, the normal component of the magnetic flux density is
continuous across a boundary.

Applying the same argument, but starting from Maxwell's equation
(\ref{eq:maxwell1}), we find:
\begin{equation}
D_{2\perp} - D_{1\perp} = \rho_s,
\label{eq:boundarynormald}
\end{equation}
where $D_\perp$ is the normal component of the electric displacement, and
$\rho_s$ is the \emph{surface} charge density (i.e. the charge per unit area, existing
purely on the boundary).

A third boundary condition, this time on the component of the magnetic
field parallel to a boundary, can be obtained by applying Stokes' theorem
to Maxwell's equation (\ref{eq:maxwell4}).
In particular, we consider a surface $S$ bounded by a loop $\partial S$ that
crosses the boundary of the material, see
Fig.~\ref{fig:boundaryconditionb}\,(b).  If we integrate both sides of 
Eq.~(\ref{eq:maxwell4}) over that surface, and apply Stokes' theorem 
(\ref{eq:stokestheorem}), we find:
\begin{equation}
\int_S \curlop \vec{H} \cdot d\vec{S}
 = \oint_{\partial S} \vec{H} \cdot d\vec{l}
 = \int_S \vec{J} \cdot d\vec{S} + \frac{\partial}{\partial t} \int_S \vec{D} \cdot d\vec{S},
\label{integralcurlh}
\end{equation}
where $I$ is the total current flowing through the surface $S$.
Now, let the surface $S$ take the form of a thin strip, with the short
ends perpendicular to the boundary, and the long ends parallel to the
boundary.  In the limit that the length of the short ends goes to
zero, the area of $S$ goes to zero: the electric displacement integrated
over $S$ becomes zero.  In principle, there may be some ``surface current'',
with density (i.e. current per unit length) $\vec{J}_s$: this contribution to the right hand side
of Eq.~(\ref{integralcurlh}) remains non-zero in the limit that the lengths of the short sides
of the loop go to zero.  In particular, note that we are interested in the component of
$\vec{J}_s$ that is perpendicular to the component of $\vec{H}$ parallel to the surface.  We
denote this component of the surface current density $J_{s\perp}$.  Then, we find from
Eq.~(\ref{integralcurlh}) (taking the limit of zero length for the short sides of the loop):
\begin{equation}
H_{2\parallel} - H_{1\parallel} = -J_{s\perp}, \label{eq:boundaryparallelh}
\end{equation}
where $H_{1\parallel}$ is the component of the magnetic intensity
parallel to the boundary at a point on one side of the boundary,
and $H_{2\parallel}$ is the component of the magnetic intensity
parallel to the boundary at a nearby point on the other side of the
boundary.

A final boundary condition can be obtained using the same argument that led to
Eq.~(\ref{eq:boundaryparallelh}), but starting from Maxwell's equation (\ref{eq:maxwell4}).
The result is:
\begin{equation}
E_{2\parallel} = E_{1\parallel}, \label{eq:boundaryparallele}
\end{equation}
that is, the tangential component of the electric field $\vec{E}$ is continuous across any
boundary.

\subsection{Electromagnetic waves on boundaries}

The boundary conditions (\ref{eq:boundarynormalb}), (\ref{eq:boundarynormald}), 
(\ref{eq:boundaryparallelh}), and (\ref{eq:boundaryparallele}) must be satisfied for the fields
in an electromagnetic wave incident on the boundary between two media.  This requirement
leads to the familiar phenomena of reflection and refraction: the laws of reflection and
refraction, and the amplitudes of the reflected and refracted waves can be derived from 
the boundary conditions, as we shall now show.

\begin{figure}[t]
\centering
\includegraphics[width=0.5\linewidth]{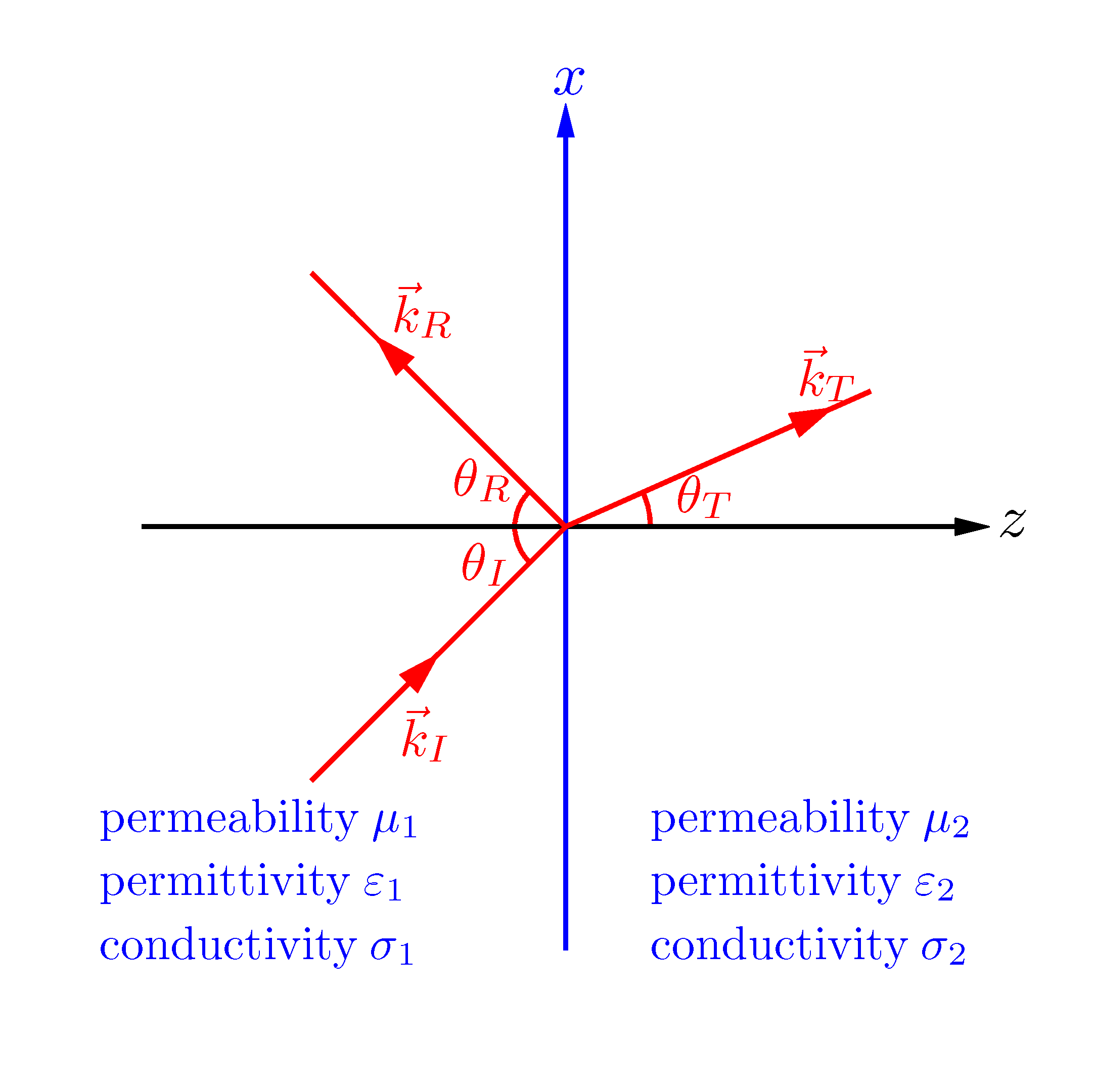}
\caption{Incident, reflected, and transmitted waves on a boundary between two media.
\label{fig:reflectionrefraction1}}
\end{figure}

Consider a plane boundary between two media (see Fig.\,\ref{fig:reflectionrefraction1}).
We choose the coordinate system so that the
boundary lies in the $x$-$y$ plane, with the $z$ axis pointing from medium 1 into medium 2.
We write a general expression for the electric field in a plane wave incident on the boundary
from medium 1:
\begin{equation}
\vec{E}_I = \vec{E}_{0I} e^{i(\vec{k}_I \cdot \vec{r} - \omega_I t)}.
\end{equation}
In order to satisfy the boundary conditions, there must be a wave present on the far side of
the boundary, i.e. a transmitted wave in medium 2:
\begin{equation}
\vec{E}_T = \vec{E}_{0T} e^{i(\vec{k}_T \cdot \vec{r} - \omega_T t)}.
\end{equation}
Let us assume that there is also an additional (reflected) wave in medium 1, i.e. on the
incident side of the boundary.  It will turn out that such a wave will be required by the boundary
conditions.  The electric field in this wave can be written:
\begin{equation}
\vec{E}_R = \vec{E}_{0R} e^{i(\vec{k}_R \cdot \vec{r} - \omega_R t)}.
\end{equation}

First of all, the boundary conditions must be satisfied at all times.  This is only possible if all
waves are oscillating with the same frequency:
\begin{equation}
\omega_I = \omega_T = \omega_R = \omega.
\end{equation}
Also, the boundary conditions must be satisfied for all points on the boundary.  This is only
possible if the phases of all the waves vary in the same way across the boundary.  Therefore,
if $\vec{p}$ is \emph{any} point on the boundary:
\begin{equation}
\vec{k}_I \cdot \vec{p} = \vec{k}_T \cdot \vec{p} = \vec{k}_R \cdot \vec{p}.
\label{wavevectorsdotp}
\end{equation}
Let us further specify our coordinate system so that $\vec{k}_I$ lies in the $x$-$z$ plane, i.e.
the $y$ component of $\vec{k}_I$ is zero.  Then, if we choose $\vec{p}$ to lie on the $y$ axis,
we see from Eq.~(\ref{wavevectorsdotp}) that:
\begin{equation}
k_{Ty} = k_{Ry} = k_{Iy} = 0.
\end{equation}
Therefore, the transmitted and reflected waves also lie in the $x$-$z$ plane.

Now let us choose $\vec{p}$ to lie on the $x$ axis.  Then, again using Eq.~(\ref{wavevectorsdotp}),
we find that:
\begin{equation}
k_{Tx} = k_{Rx} = k_{Ix}.
\label{reflectionrefraction1}
\end{equation}
If we define the angle $\theta_I$ as the angle between $\vec{k}_I$ and the $z$ axis (the
normal to the boundary), and similarly for $\theta_T$ and $\theta_R$, then
Eq.~(\ref{reflectionrefraction1}) can be expressed:
\begin{equation}
k_T \sin \theta_T = k_R  \sin \theta_R = k_I  \sin \theta_I.
\label{reflectionrefraction2}
\end{equation}
Since the incident and reflected waves are travelling in the same medium, and have the
same frequency, the magnitudes of their wave vectors must be the same, i.e. $k_R = k_I$.
Therefore, we have the law of reflection:
\begin{equation}
\sin \theta_R = \sin \theta_I.
\label{lawofreflection}
\end{equation}
The angles of the transmitted and incident waves must be related by:
\begin{equation}
\frac{\sin \theta_I}{\sin \theta_T} = \frac{k_T}{k_I} = \frac{v_1}{v_2},
\label{refraction1}
\end{equation}
where $v_1$ and $v_2$ are the phase velocities in the media 1 and 2 respectively, and
we have used the dispersion relation $v = \omega/k$.  If we define the refractive index $n$ of
a medium as the ratio between the speed of light in vacuum to the speed of light in the medium:
\begin{equation}
n = \frac{c}{v},
\label{refractiveindex}
\end{equation}
then Eq.~(\ref{refraction1}) can be expressed:
\begin{equation}
\frac{\sin \theta_I}{\sin \theta_T} = \frac{n_2}{n_1}.
\label{snellslaw}
\end{equation}
This is the familiar form of the law of refraction, Snell's law.

So far, we have derived expressions for the relative directions of the incident, reflected, and
transmitted waves.  To do this, we have only used the fact that boundary conditions on
the fields in the wave exits.  Now, we shall derive expressions for the relative amplitudes
of the waves: for this, we shall need to apply the boundary conditions themselves.

\begin{figure}[t]
\centering
\includegraphics[width=0.4\linewidth]{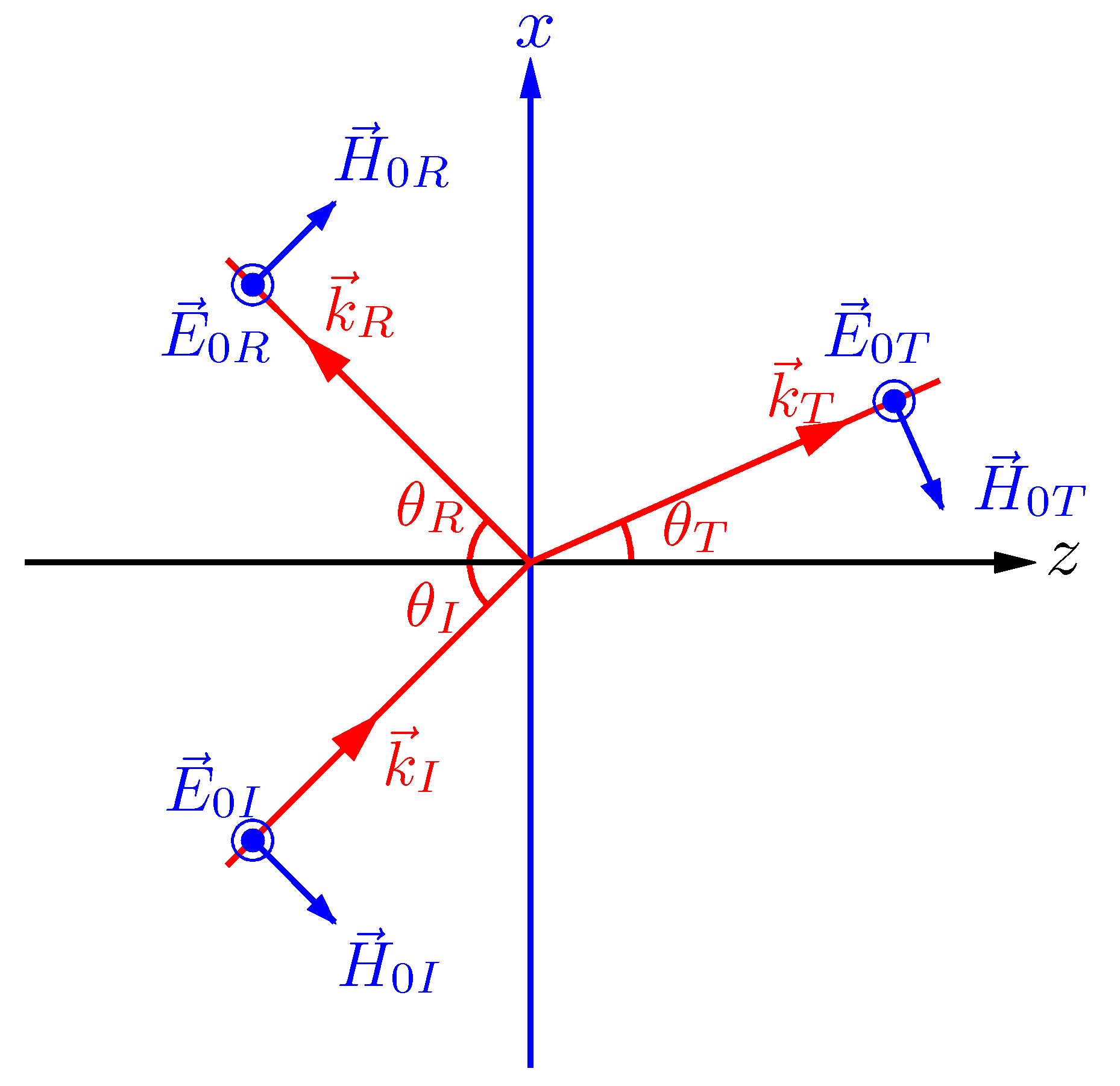}
\includegraphics[width=0.4\linewidth]{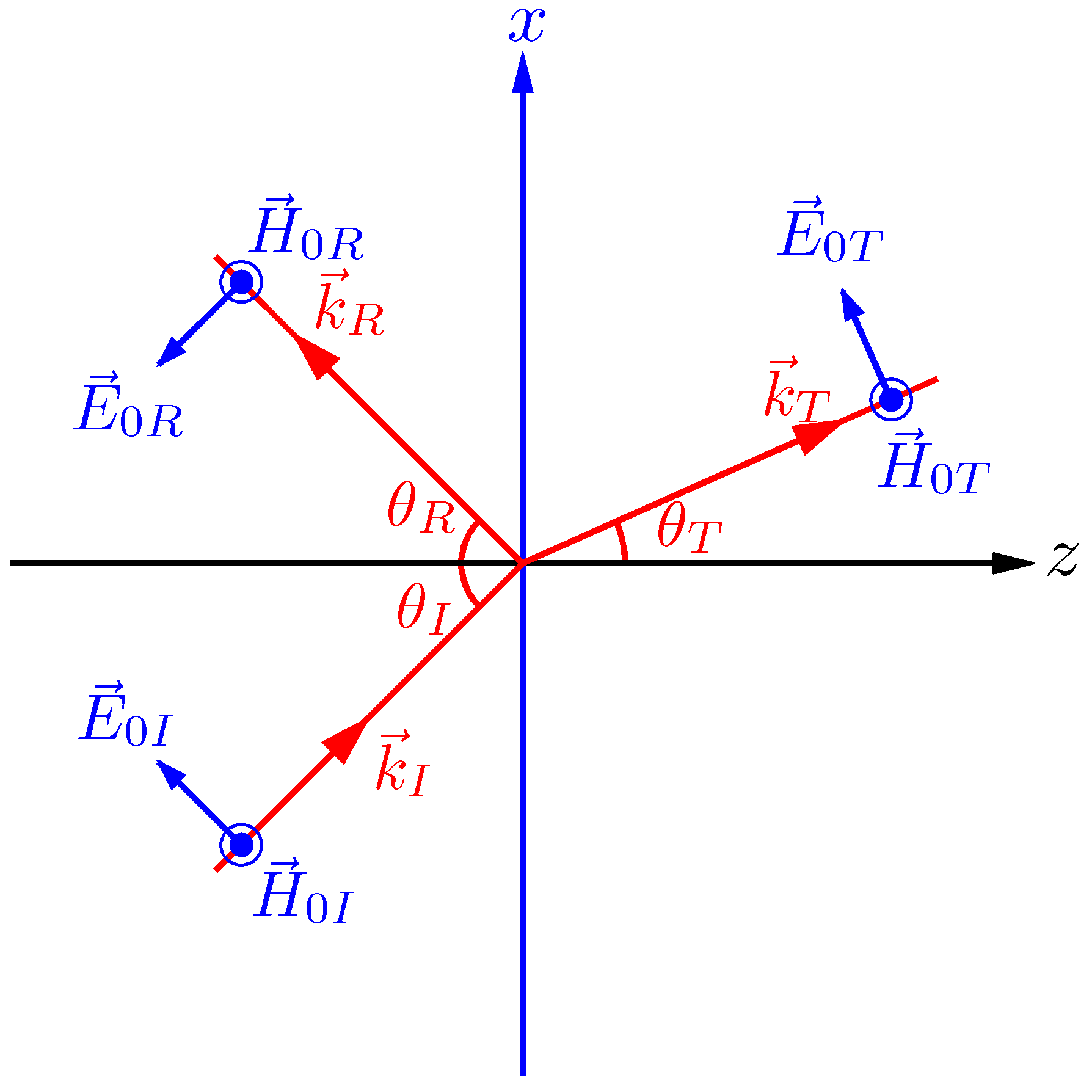}
\caption{Electric and magnetic fields in the incident, reflected, and transmitted waves
on a boundary between two media.  Left: The incident wave is $N$ polarised, i.e. with the
electric field normal to the plane of incidence.  Right: The incident wave is $P$ polarised,
i.e. with the electric field parallel to the plane of incidence.
\label{fig:reflectionrefraction2}}
\end{figure}

It turns out that there are different relationships between the amplitudes of the waves, depending
on the orientation of the electric field with respect to the plane of incidence (that is, the plane
defined by the normal to the boundary and the wave vector of the incident wave).  Let us first
consider the case that the electric field is normal to the plane of incidence, i.e. ``$N$ polarisation'',
see Fig.\,\ref{fig:reflectionrefraction2}, left.
Then, the electric field must be tangential to the boundary.  Using the boundary condition 
(\ref{eq:boundaryparallele}), the tangential component of the electric field is continuous
across the boundary, and so:
\begin{equation}
\vec{E}_{0I} + \vec{E}_{0R} = \vec{E}_{0T}.
\label{fresnel1}
\end{equation}
Using the boundary condition (\ref{eq:boundaryparallelh}), the tangential component of the 
magnetic intensity $\vec{H}$ is also continuous across the boundary.  However, because the
magnetic field in a plane wave is perpendicular to the electric field, the magnetic intensity in
each wave must lie in the plane of incidence, and at an angle to the boundary.  Taking the
directions of the wave vectors into account:
\begin{equation}
\vec{H}_{0I} \cos \theta_I - \vec{H}_{0R} \cos \theta_I = \vec{H}_{0T} \cos \theta_T.
\label{fresnel2}
\end{equation}
Using the definition of the impedance $Z$ of a medium as the ratio between the amplitude
of the electric field and the amplitude of the magnetic intensity:
\begin{equation}
Z = \frac{E_0}{H_0},
\label{impedancedefn}
\end{equation}
we can solve Eqs.~(\ref{fresnel1}) and (\ref{fresnel2}) to give:
\begin{eqnarray}
\left( \frac{E_{0R}}{E_{0I}} \right)_{\!\!N} & = & 
\frac{Z_2 \cos \theta_I - Z_1 \cos \theta_T}{Z_2 \cos \theta_I + Z_1 \cos \theta_T}, \label{fresnel3} \\
\left( \frac{E_{0T}}{E_{0I}} \right)_{\!\!N} & = & 
\frac{2Z_2 \cos \theta_I}{Z_2 \cos \theta_I + Z_1 \cos \theta_T}. \label{fresnel4}
\end{eqnarray}

Following a similar procedure for the case that the electric field is oriented so that it
is parallel to the plane of incidence (``$P$ polarisation'', see Fig.\,\ref{fig:reflectionrefraction2},
right), we find:
\begin{eqnarray}
\left( \frac{E_{0R}}{E_{0I}} \right)_{\!\!P} & = & 
\frac{Z_2 \cos \theta_T - Z_1 \cos \theta_I}{Z_2 \cos \theta_T + Z_1 \cos \theta_I}, \label{fresnel5} \\
\left( \frac{E_{0T}}{E_{0I}} \right)_{\!\!P} & = & 
\frac{2Z_2 \cos \theta_I}{Z_2 \cos \theta_T + Z_1 \cos \theta_I}. \label{fresnel6}
\end{eqnarray}

Equations (\ref{fresnel3}), (\ref{fresnel4}), (\ref{fresnel5}) and (\ref{fresnel6}) are known as
Fresnel's equations: they give the amplitudes of the reflected
and transmitted waves relative to the amplitude of the incident wave, in terms of the properties of the
media (specifically, the impedance) on either side of the boundary, and the angle of the incident wave.
Many important phenomena, including total internal reflection, and polarisation by reflection, follow
from Fresnel's equations.  However, we shall focus on the consequences for a wave incident on a
good conductor.

First, note that for a dielectric with permittivity $\varepsilon$ and permeability $\mu$, the impedance
is given by:
\begin{equation}
Z = \sqrt{\frac{\mu}{\varepsilon}}.
\end{equation}
This follows from Eq.~(\ref{impedancedefn}), using the constitutive relation $\vec{B} = \mu \vec{H}$,
and the relation between the field amplitudes in an electromagnetic wave $E_0 / B_0 = v$, where the
phase velocity $v = 1/\sqrt{\mu \varepsilon}$. 

Now let us consider what happens when a wave is incident on the surface of a conductor.
Using Maxwell's equation (\ref{eq:maxwell3}), the impedance can be written (in general) for a
plane wave:
\begin{equation}
Z = \frac{E_0}{H_0} = \frac{\mu \omega}{k},
\end{equation}
In a good conductor (for which the conductivity $\sigma \gg \omega \varepsilon$), the wave vector
is complex; and this means that the impedance will also be complex.  This implies a phase difference
between the electric and magnetic fields, which does indeed occur in a conductor.  In the context of
Fresnel's equations, complex impedances will describe the phase relationships between the incident,
reflected, and transmitted waves. From Eq.~(\ref{wavevectorgoodconductor}), the wave vector
in a good conductor is given (approximately) by:
\begin{equation}
k \approx (1 + i)\sqrt{\frac{\omega \sigma \mu}{2}}.
\end{equation}
Therefore, we can write:
\begin{equation}
Z \approx (1 - i) \sqrt{\frac{\omega \mu}{2\sigma}} =
(1 - i) \sqrt{\frac{\omega \varepsilon}{2\sigma}} \sqrt{\frac{\mu}{\varepsilon}}.
\end{equation}
Consider a wave incident on a good conductor from a dielectric.  If the permittivity and permeability
of the conductor are similar to those in the dielectric, then, since $\sigma \gg \omega \varepsilon$
(by definition, for a good conductor), the impedance of the conductor will be much less than the
impedance of the dielectric.  Fresnel's equations become (for both $N$ and $P$ polarisation):
\begin{equation}
\frac{E_{0R}}{E_{0I}}  \approx 
-1, \qquad
 \frac{E_{0T}}{E_{0I}} \approx 
0.
\end{equation}
There is (almost) perfect reflection of the wave (with a change of phase); and very little of the wave
penetrates into the conductor.

At optical frequencies and below, most metals are good conductors.  In practice, as we expect
from the above discussion, most metals have highly reflective surfaces.  This is of considerable
importance for RF systems in particle accelerators, as we shall see when we consider cavities
and waveguides, in the following sections.

\subsection{Fields on the boundary of an ideal conductor}

We have seen that a good conductor will reflect most of the energy in a wave incident on its
surface.  We shall define an \emph{ideal conductor} as a material that reflects all the energy
in an electromagnetic wave incident on its surface\footnote{It is tempting to identify
superconductors with ideal conductors; however, superconductors are rather complicated
materials, that show sometimes surprising behaviour not always consistent with our
definition of an ideal conductor.}.  In that case, the fields at any point inside the ideal conductor will
be zero at all times.  From the boundary conditions (\ref{eq:boundarynormalb}) and
(\ref{eq:boundaryparallele}), this implies that, at the surface of the conductor:
\begin{equation}
B_\perp = 0, \qquad
E_\parallel = 0.
\end{equation}
That is, the normal component of the magnetic field, and the tangential component of the electric
field must vanish at the boundary.  These conditions impose strict constraints on the patterns of electromagnetic field that can persist in RF cavities, or that can propagate along waveguides.

The remaining boundary conditions, (\ref{eq:boundarynormald}) and (\ref{eq:boundaryparallelh}),
allow for discontinuities in the normal component of the electric field, and the tangential component
of the magnetic field, depending on the presence of surface charge and surface current.  In an
ideal conductor, both surface charge and surface current can be present: this allows the field to
take non-zero values at the boundary of (and within a cavity enclosed by) an ideal conductor.

\section{Fields in cavities}

In the previous section, we saw that most of the energy in an electromagnetic wave is reflected
from the surface of a good conductor.  This provides the possibility of storing electromagnetic
energy in the form of standing waves in a cavity; the situation will be analogous to a standing
mechanical wave on, say, a violin string.  We also saw in the previous section that there are
constraints on the fields on the surface of a good conductor: in particular, at the surface of an
ideal conductor, the normal component of the magnetic field and the tangential component of
the electric field must both vanish.  As a result, the possible field patterns (and frequencies) of
the standing waves that can persist within the cavity are determined by the shape of the cavity.
This is one of the most important practical aspects for RF cavities in particle accelerators.
Usually, the energy stored in a cavity is needed to manipulate a charged particle beam in a
particular way (for example, to accelerate or deflect the beam).  The effect on the beam is
determined by the field pattern.  Therefore, it is important to design the shape of the cavity, so
that the fields in the cavity interact with the beam in the desired way; and that undesirable
interactions (which always occur to some extent) are minimised.  The relationship between
the shape of the cavity and the different field patterns (or \emph{modes}) that can persist
within the cavity will be the main topic of the present section.  Other practical issues (for
example, how the electromagnetic waves enter the cavity) are beyond our scope.

\subsection{Modes in a rectangular cavity}

\begin{figure}
\centering
\includegraphics[width=0.6\textwidth]{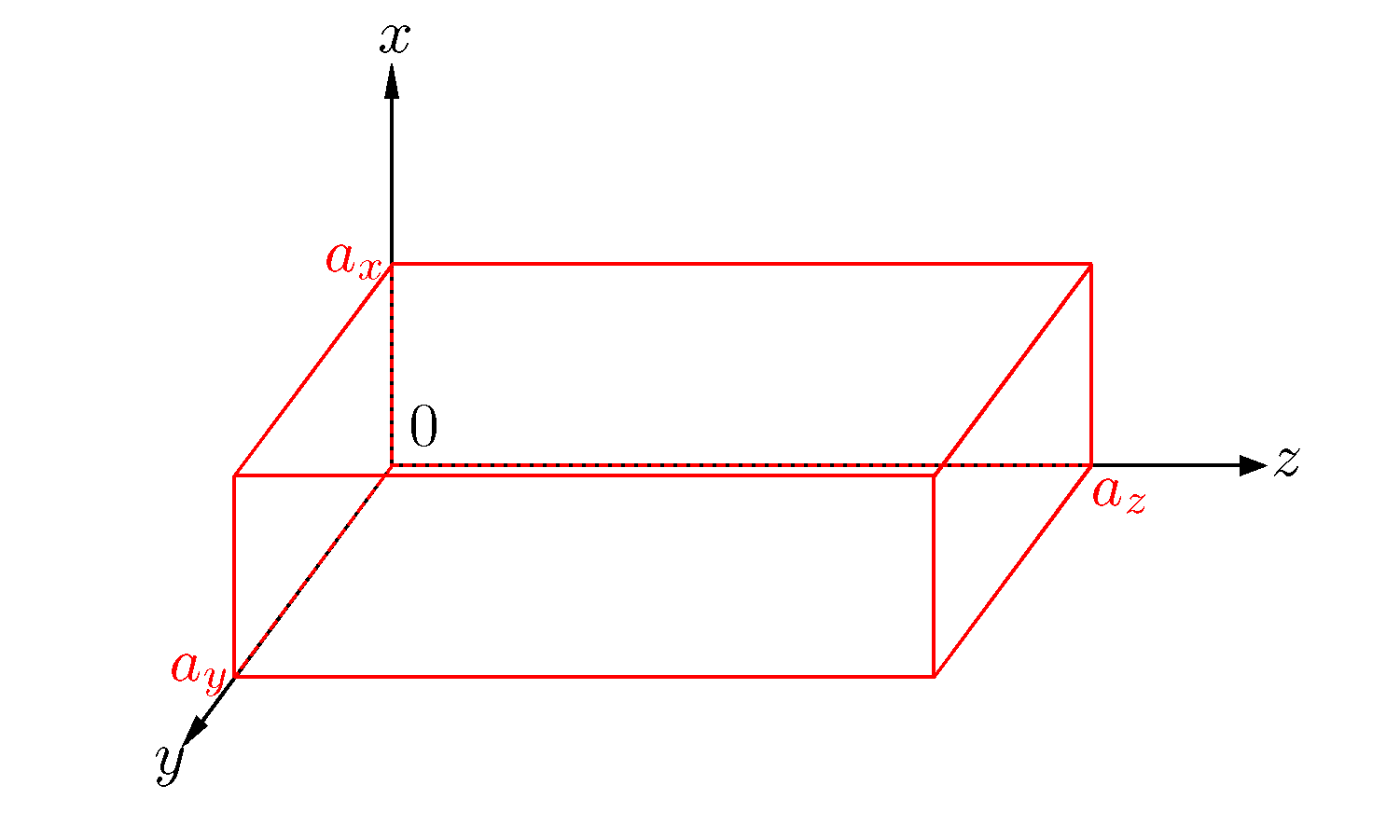}
\caption{Rectangular cavity.\label{fig:rectangularcavity}}
\end{figure}

We consider first a rectangular cavity with perfectly conducting walls, containing a perfect
vacuum (see Fig.\,\ref{fig:rectangularcavity}).  The wave equation for the electric field inside
the cavity is:
\begin{equation}
\nabla^2 \vec{E} - \frac{1}{c^2} \frac{\partial^2 \vec{E}}{\partial t^2} = 0,
\end{equation}
where $c$ is the speed of light in a vacuum.  There is a similar equation for the
magnetic field $\vec{B}$.  We look for solutions to the wave equations for $\vec{E}$ and
$\vec{B}$ that also satisfy Maxwell's equations, and also satisfy the boundary conditions
for the fields at the walls of the cavity.  If the walls of the cavity are ideal conductors, then the
boundary conditions are:
\begin{eqnarray}
E_\parallel & = & 0, \\
B_\perp & = & 0,
\end{eqnarray}
where $E_\parallel$ is the component of the electric field tangential to the wall, and
$B_\perp$ is the component of the magnetic field normal to the wall.

Free-space plane wave solutions will not satisfy the boundary conditions.  However, we can
look for solutions of the form:
\begin{equation}
\vec{E}(x,y,z,t) = \vec{E}_{\vec{r}} e^{-i\omega t},
\end{equation}
where $\vec{E}_{\vec{r}} = \vec{E}_{\vec{r}} (x,y,z)$ is a vector function of position (independent
of time).
Substituting into the wave equation, we find that the spatial dependence satisfies:
\begin{equation}
\nabla^2 \vec{E}_{\vec{r}} + \frac{\omega^2}{c^2} \vec{E}_{\vec{r}} = 0.
\label{waveequation1}
\end{equation}
The full solution can be derived using the method of separation of variables (in fact,
we have begun the process by separating the time from the spatial variables).
However, it is sufficient to quote the result, and verify the solution simply by substitution into
the wave equation.  The components of the electric field in the rectangular cavity are given by:
\begin{eqnarray}
E_x & = & E_{x0} \cos k_x x \, \sin k_y y\, \sin k_z z\, e^{-i\omega t}, \\
E_y & = & E_{y0} \sin k_x x \, \cos k_y y\, \sin k_z z\, e^{-i\omega t}, \\
E_z & = & E_{z0} \sin k_x x \, \sin k_y y\, \cos k_z z\, e^{-i\omega t}.
\end{eqnarray}
To satisfy the wave equation, we require:
\begin{equation}
k_x^2 + k_y^2 + k_z^2 = \frac{\omega^2}{c^2}.
\end{equation}
Maxwell's equation $\nabla \cdot \vec{E} = 0$, imposes a constraint on the components
of the wave vector and the amplitudes of the field components:
\begin{equation}
k_x E_{x0} + k_y E_{y0} + k_z E_{z0} = 0. \label{rectangularcavitydive}
\end{equation}

\begin{figure}
\centering
\includegraphics[width=0.4\textwidth]{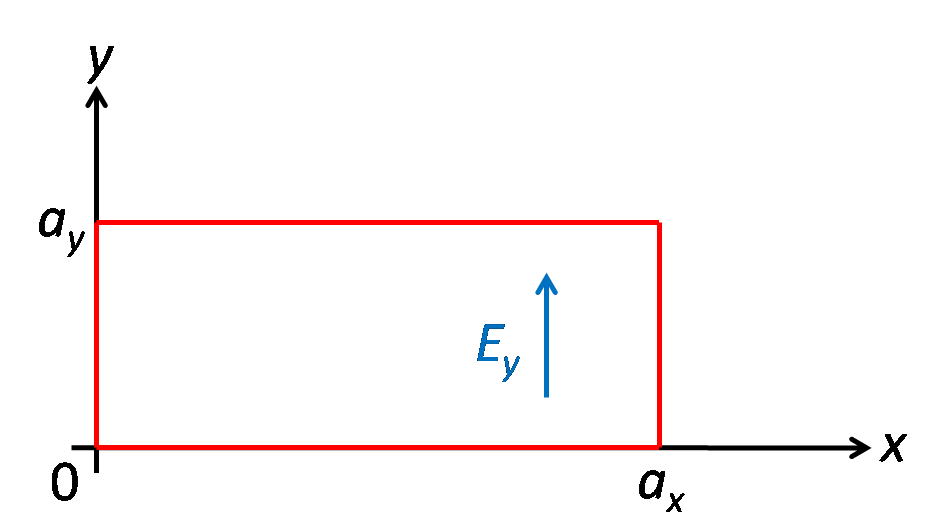}
\caption{Boundary conditions in a rectangular cavity.  The field component $E_y$ is
parallel to the walls of the cavity (and must therefore vanish) at $x = 0$ and $x = a_x$.\label{fig:rectangularcavityboundaryconditions}}
\end{figure}

We also need to satisfy the boundary conditions, in particular that the tangential
component of the electric field vanishes at the walls of the cavity.  This imposes
additional constraints on $k_x$, $k_y$ and $k_z$.
Consider:
\begin{equation}
E_y = E_{y0} \sin k_x x \, \cos k_y y\, \sin k_z z\, e^{-i\omega t}.
\end{equation}
The boundary conditions require that $E_y = 0$ at $x = 0$ and $x = a_x$, for all
$y$, $z$, and $t$ (see Fig.\,\ref{fig:rectangularcavityboundaryconditions}).
These conditions are satisfied if $k_x a_x = m_x \pi$, where $m_x$ is an integer.
To satisfy \emph{all} the boundary conditions, we require:
\begin{equation}
k_x = \frac{m_x \pi}{a_x}, \qquad
k_y = \frac{m_y \pi}{a_y}, \qquad
k_z = \frac{m_z \pi}{a_z},
\end{equation}
where $m_x$, $m_y$ and $m_z$ are integers.  Note that these integers play a large
part in determining the shape of the electric field (though even when these numbers
are specified, there is still some freedom in choosing the relative amplitudes of the
field components).  The quantities $m_x$, $m_y$, and $m_z$, are called the
\emph{mode numbers}.  The frequency of oscillation is determined completely by
the mode numbers, for a given size and shape of cavity.

The magnetic field can be obtained from the electric field, using Maxwell's equation:
\begin{equation}
\nabla \times \vec{E} = -\frac{\partial \vec{B}}{\partial t}.
\end{equation}
This gives:
\begin{eqnarray}
B_x & = & \frac{i}{\omega} (E_{y0}k_z - E_{z0}k_y) \sin k_x x \, \cos k_y y\, \cos k_z z\, e^{-i\omega t}, \\
B_y & = & \frac{i}{\omega} (E_{z0}k_x - E_{x0}k_z) \cos k_x x \, \sin k_y y\, \cos k_z z\, e^{-i\omega t}, \\
B_z & = & \frac{i}{\omega} (E_{x0}k_y - E_{y0}k_x) \cos k_x x \, \cos k_y y\, \sin k_z z\, e^{-i\omega t}. \label{rectangularcavitybz}
\end{eqnarray}
It is left as an exercise for the reader to show that these fields satisfy the
boundary condition on the magnetic field at the walls of the cavity, and also satisfy
the remaining Maxwell's equations:
\begin{equation}
\nabla \cdot \vec{B} = 0, \quad \textrm{and} \quad \nabla \times \vec{B} = \frac{1}{c^2} \frac{\partial \vec{E}}{\partial t}.
\end{equation}

\begin{figure}
\centering
\includegraphics[width=1.0\textwidth]{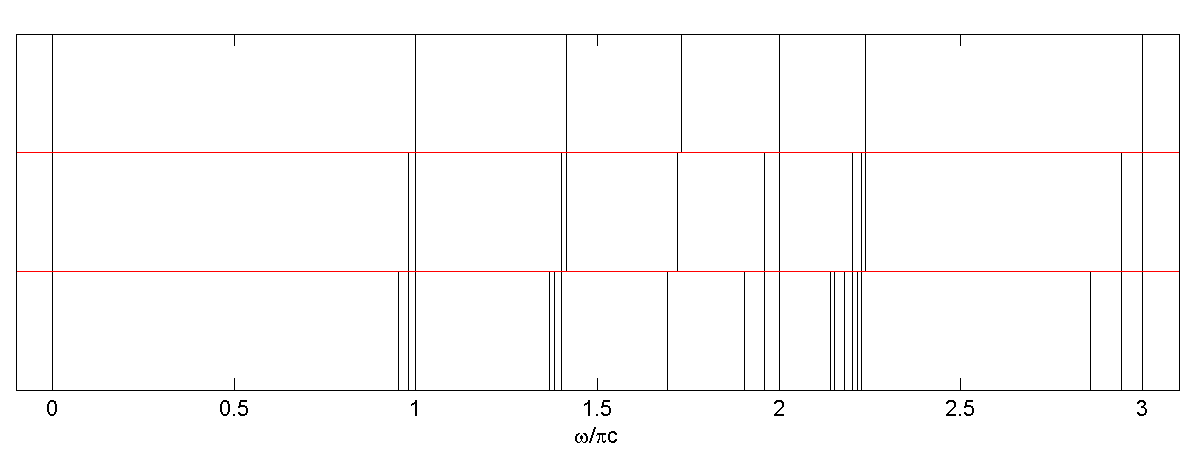}
\caption{Mode spectra in rectangular cavities.  Top: all side lengths equal. Middle: two side
lengths equal. Bottom: all side lengths different.  Note that we show all modes, including those
with two (or three) mode numbers equal to zero, even though such modes will have zero amplitude.
\label{fig:rectangularcavitymodespectra}}
\end{figure}

Note that the frequency of oscillation of the wave in the cavity is determined
by the mode numbers $m_x$, $m_y$ and $m_z$:
\begin{equation}
\omega = \pi c \sqrt{
\left( \frac{m_x}{a_x} \right)^2 +
\left( \frac{m_y}{a_y} \right)^2 +
\left( \frac{m_z}{a_z} \right)^2.
}
\end{equation}
For a cubic cavity ($a_x = a_y = a_z$), there will be a degree of degeneracy,
i.e. there will generally be several different sets of mode numbers leading to
different field patterns, but all with the same frequency of oscillation.
The degeneracy can be broken by making the side lengths different: see
Fig.\,\ref{fig:rectangularcavitymodespectra}.

\begin{figure}
\centering
\includegraphics[width=0.8\textwidth]{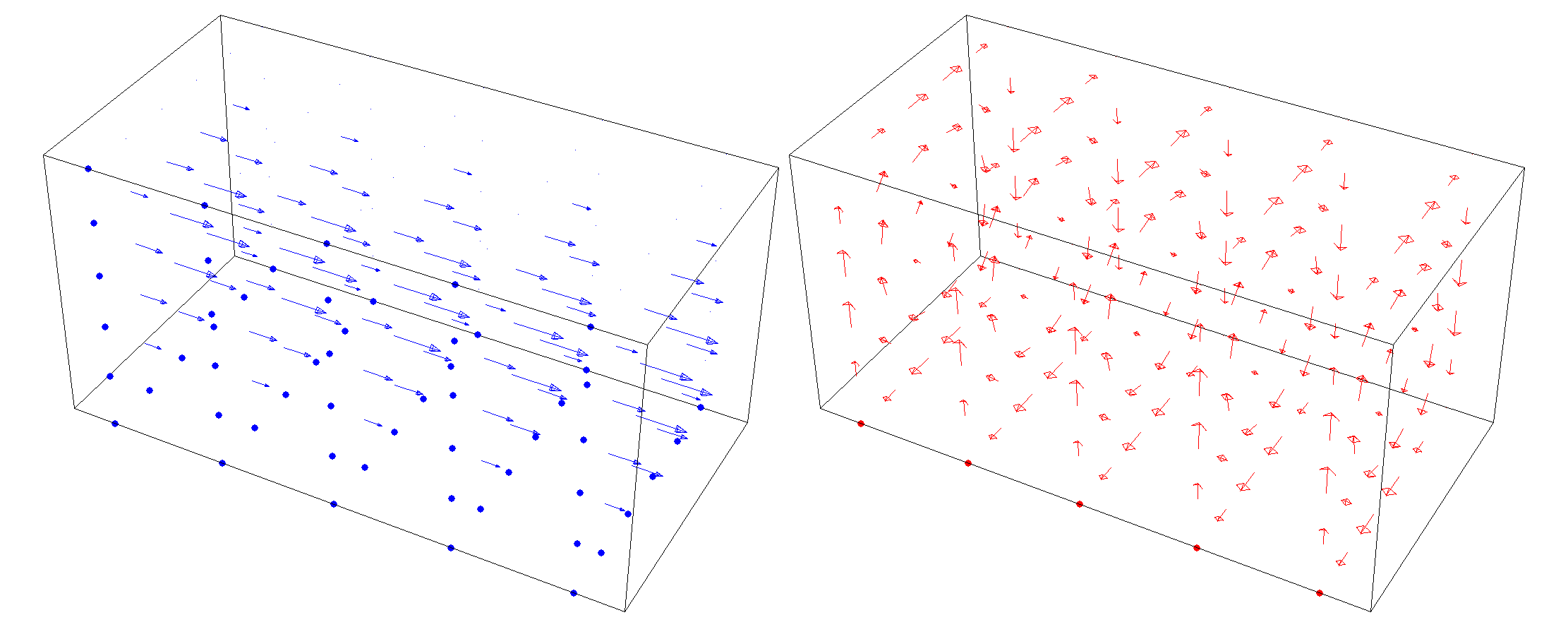}
\includegraphics[width=0.8\textwidth]{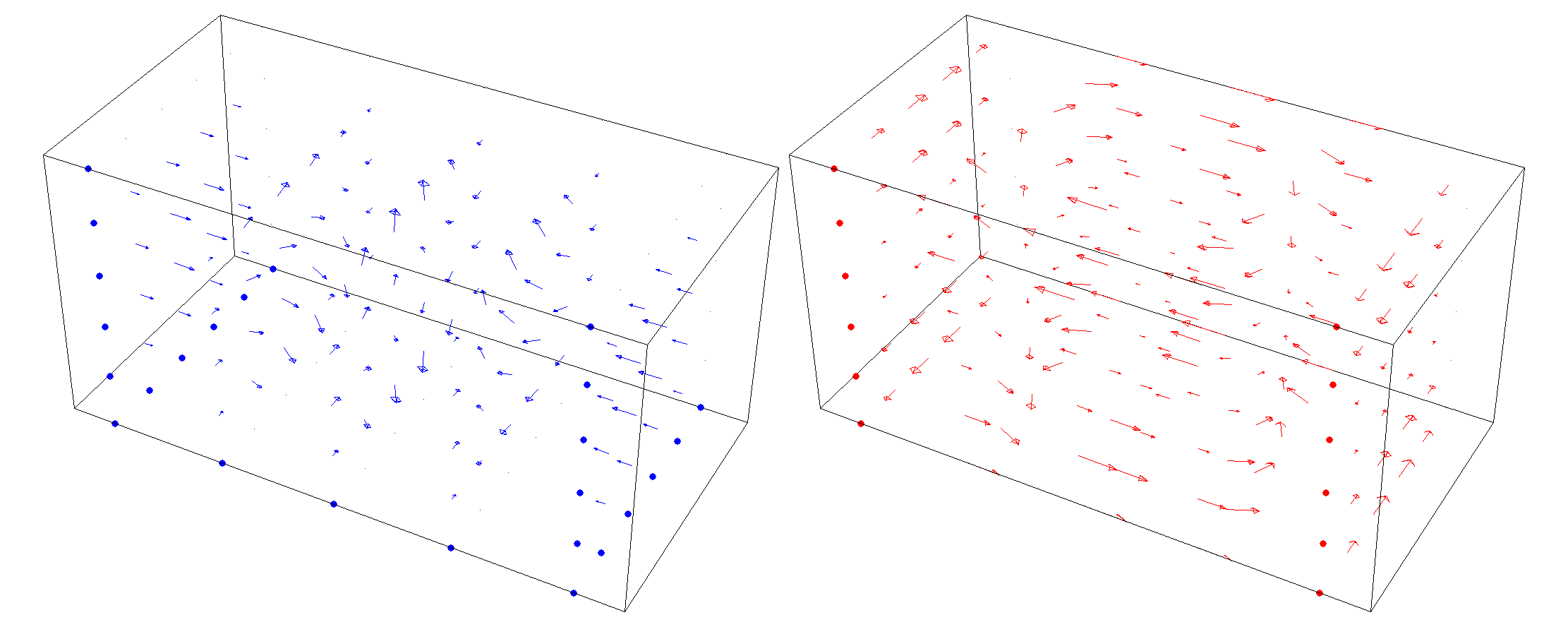}
\includegraphics[width=0.8\textwidth]{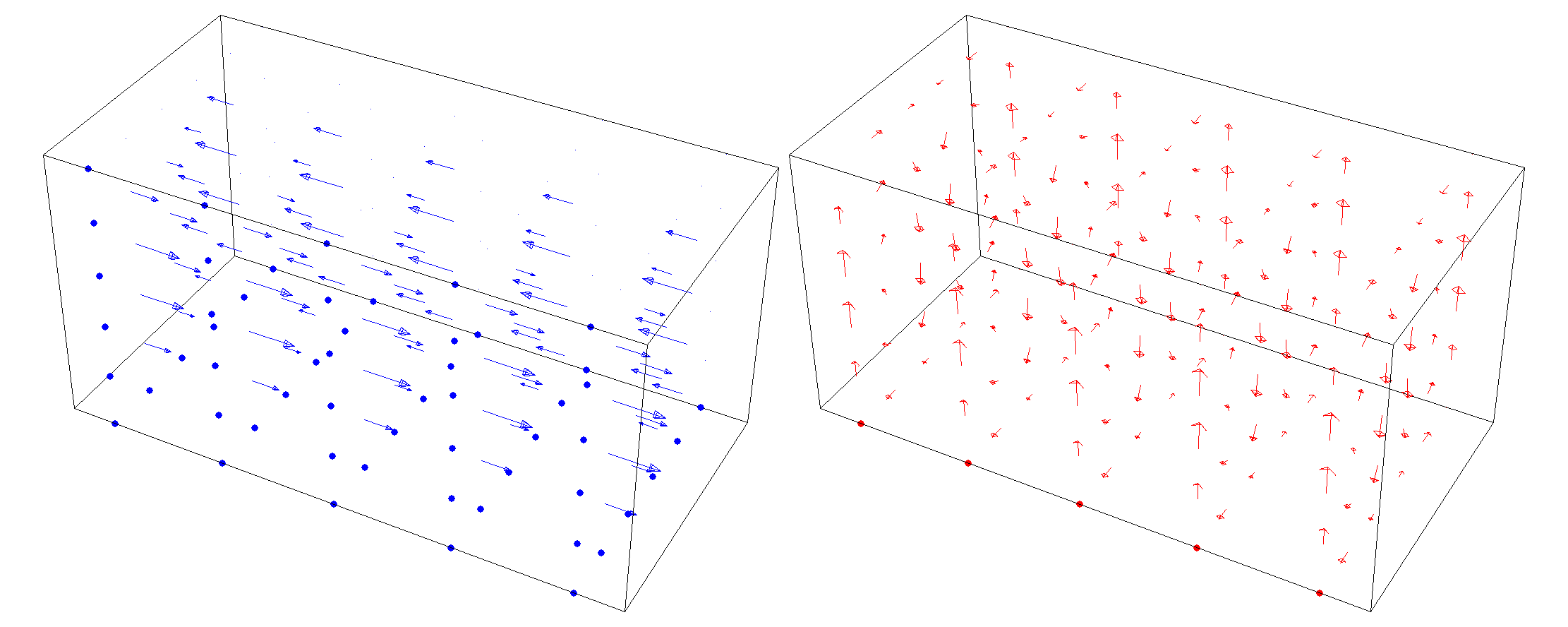}
\includegraphics[width=0.8\textwidth]{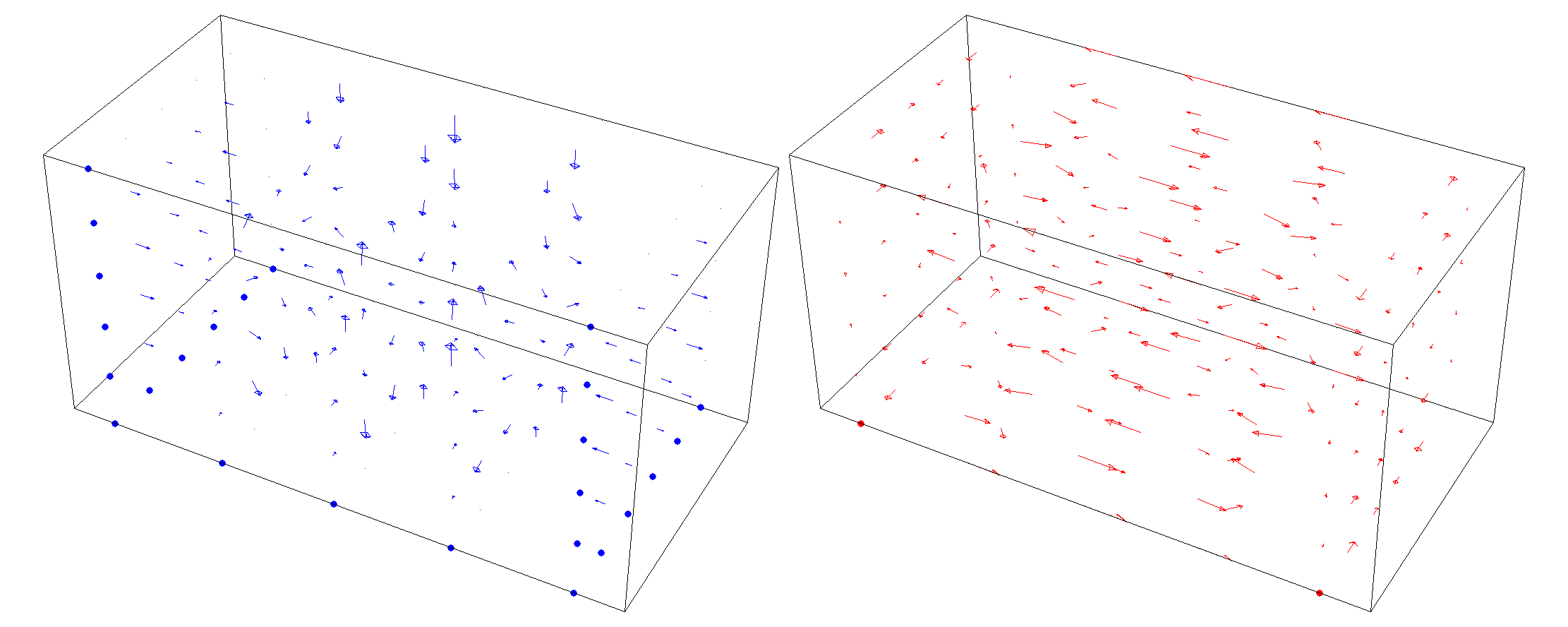}
\caption{Examples of modes in rectangular cavity.  From top to bottom:
(110), (111), (210), and (211).
\label{fig:rectangularcavitymodeexamples}}
\end{figure}

Some examples of field patterns in different modes of a rectangular cavity are shown
in Fig.\,\ref{fig:rectangularcavitymodeexamples}.

\subsection{Quality factor}

Note that the standing wave solution represents an oscillation that will continue
indefinitely: there is no mechanism for dissipating the energy.  This is because we have
assumed that the walls of the cavity are made from an ideal conductor, and the energy
incident upon a wall is completely reflected.  These assumptions are implicit in the boundary
conditions we have imposed, that the tangential component of the electric field and the
normal component of the magnetic field vanish at the boundary.  

In practice, the walls of the cavity will not be perfectly conducting, and the boundary
conditions will vary slightly from those we have assumed.  The electric and (oscillating)
magnetic fields on the walls will induce currents, which will dissipate the energy.  The rate
of energy dissipation is usually quantified by the ``quality factor'', $Q$.  The equation of
motion for a damped harmonic oscillator:
\begin{equation}
\frac{d^2 u}{dt^2} + \frac{\omega}{Q} \frac{du}{dt} + \omega^2 u = 0
\end{equation}
has the solution:
\begin{equation}
u = u_0 \, e^{-\frac{\omega t}{2Q}} \, \cos (\omega^\prime t - \phi),
\end{equation}
where:
\begin{equation}
\omega^\prime = \omega \sqrt{\frac{4Q^2 - 1}{4Q^2}}.
\end{equation}
The quality factor $Q$ is (for $Q \gg 1$) the number of oscillations over which the
energy in the oscillator (proportional to the square of the amplitude of $u$) falls by
a factor $1/e$.  More precisely, the rate of energy dissipation (the dissipated power,
$P_d$) is given by:
\begin{equation}
P_d = - \frac{d {\cal E}}{dt} = \frac{\omega}{Q} {\cal E}.
\label{dissipatedpowerqfactor}
\end{equation}

If a field is generated in a rectangular cavity corresponding to one of the modes we have
calculated, the fields generating currents in the walls will be small, and the dissipation will
be slow: such modes (with integer values of $m_x$, $m_y$ and $m_z$) will have a high
quality factor, compared to other field patterns inside the cavity.  A mode with a high quality
factor is called a resonant mode.  In a rectangular cavity, the modes corresponding to
integer values of the mode numbers are resonant modes.

\subsection{Energy stored in a rectangular cavity}

It is often useful to know the energy stored in a cavity.  For a rectangular cavity, it is
relatively straightforward to calculate the energy stored in a particular mode, given the
amplitude of the fields.  The energy density in an electric field is:
\begin{equation}
U_E = \frac{1}{2} \vec{D} \cdot \vec{E}.
\end{equation}
Therefore, the total energy stored in the electric field in a cavity is:
\begin{equation}
{\mathcal E}_E = \frac{1}{2} \varepsilon_0 \int \vec{E}^2 \, dV,
\end{equation}
where the volume integral extends over the entire volume of the cavity.
In a resonant mode, we have:
\begin{equation}
\int_0^{a_x} \cos^2 k_x x \, dx =
\int_0^{a_x} \sin^2 k_x x \, dx = \frac{1}{2},
\end{equation}
where $k_x = m_x \pi/a_x$, and $m_x$ is a non-zero integer.
We have similar results for the $y$ and $z$ directions,
so we find (for $m_x$, $m_y$ and $m_z$ all non-zero integers):
\begin{equation}
{\mathcal E}_E = \frac{1}{16} \varepsilon_0 (E_{x0}^2 + E_{y0}^2 + E_{z0}^2) \cos^2 \omega t.
\end{equation}
The energy varies as the square of the field amplitude, and oscillates
sinusoidally in time.

Now let us calculate the energy in the magnetic field.  The energy density is:
\begin{equation}
U_B = \frac{1}{2} \vec{B} \cdot \vec{H}.
\end{equation}
Using:
\begin{equation}
k_x^2 + k_y ^2 + k_z^2 = \frac{\omega^2}{c^2}, \quad \textrm{and} \quad
E_{x0} k_x + E_{y0} k_y + E_{z0} k_z  = 0, 
\end{equation}
we find, after some algebraic manipulation (and noting that the magnetic field is 90$^\circ$
out of phase with the electric field):
\begin{equation}
{\mathcal E}_B = \frac{1}{16} \frac{1}{\mu_0 c^2} (E_{x0}^2 + E_{y0}^2 + E_{z0}^2) \sin^2 \omega t.
\end{equation}
As in the case of the electric field, the numerical factor is correct if
the mode numbers $m_x$, $m_y$ and $m_z$ are non-zero integers.

Finally, using $1/c^2 = \mu_0 \varepsilon_0$, we have (for $m_x$, $m_y$ and $m_z$
non-zero integers):
\begin{equation}
{\mathcal E}_E + {\mathcal E}_B = \frac{1}{16} \varepsilon_0 (E_{x0}^2 + E_{y0}^2 + E_{z0}^2).
\end{equation}
The total energy in the cavity is constant over time, although the energy ``oscillates''
between the electric field and the magnetic field.

The power flux in the electromagnetic field is given by the Poynting vector:
\begin{equation}
\vec{S} = \vec{E} \times \vec{H}.
\end{equation}
Since the electric and magnetic fields in the cavity are 90$^\circ$ out of phase
(if the electric field varies as $\cos \omega t$, then the magnetic field varies
as $\sin \omega t$), averaging the Poynting vector over time at any point in the
cavity gives zero: this is again consistent with conservation of energy.

\subsection{Shunt impedance}

We have seen that in practice, some of the energy stored in a cavity will be dissipated in
the walls, and that the rate of energy dissipation for a given mode is measured by the
quality factor, $Q$ (Eq.\,(\ref{dissipatedpowerqfactor})):
\begin{equation}
P_d = -\frac{d {\mathcal E}}{dt} = \frac{\omega}{Q}{\mathcal E}. \nonumber
\end{equation}
For a mode with a longitudinal electric field component $E_{z0} = V_0/L$ (where $L$ is
the length of the cavity), we define the \emph{shunt impedance}, $R_s$:
\begin{equation}
R_s = \frac{V_0^2}{2P_d}.
\label{shuntimpedance}
\end{equation}
(Note that different definitions of the shunt impedance are used, depending on the
context: some definitions omit the factor 1/2).

Combining Eqs.\,(\ref{dissipatedpowerqfactor}) and (\ref{shuntimpedance}), we see that:
\begin{equation}
\frac{R_s}{Q} = \frac{V_0^2}{2P_d}\cdot \frac{P_d}{\omega {\mathcal E}}
= \frac{V_0^2}{2\omega {\mathcal E}}. \label{roverq}
\end{equation}

Consider a mode with $B_z = 0$.  Such modes have only transverse components of the
magnetic field, and are called TM modes.
Using equation (\ref{rectangularcavitybz}), we see that the electric field in TM
modes obeys:
\begin{equation}
k_y E_{x0} = k_x E_{y0}.
\end{equation}
We also have, from (\ref{rectangularcavitydive}):
\begin{equation}
k_x E_{x0} + k_y E_{y0} + k_z E_{z0} = 0.
\end{equation}
These relations allow us to write the energy stored in the cavity purely in terms
of the mode numbers and the peak longitudinal electric field:
\begin{equation}
{\mathcal E} = \frac{\varepsilon_0}{8}
\left( \frac{k_x^2 +k_y^2 + k_z^2}{k_x^2 +k_y^2} \right) E_{z0}^2.
\label{tmenergy}
\end{equation}

Combining equations (\ref{roverq}) and (\ref{tmenergy}), we see that:
\begin{equation}
\frac{R_s}{Q} = \frac{16}{\varepsilon_0}
\left(
\frac{k_x^2 +k_y^2}{k_x^2 +k_y^2 + k_z^2}
\right) \frac{L^2}{\omega}.
\end{equation}

For a TM mode in a rectangular cavity, the quantity $R_s/Q$ depends only on the
length of the cavity and the mode numbers.  It is independent of such factors as the
material of the walls.  In fact, this result generalises: for TM modes, $R_s/Q$
depends only on the geometry of the cavity, and the mode numbers.  This is of practical
significance since, to optimise the design of a cavity for accelerating a beam, the goal
is to maximise $R_s/Q$ for the accelerating mode, and minimise this quantity for all other
modes.  Since $R_s/Q$ is independent of such quantities as the material from which
the cavity is made, it is possible while designing the cavity to focus on optimising the
geometry to achieve the highest $R_s/Q$ for the desired mode (and minimise $R_s/Q$)
for the other modes.  Properties of the material from which the cavity will be made can
safely be neglected at this stage of the design process.

\subsection{Cylindrical cavities}

Most cavities in accelerators are closer to a cylindrical than a rectangular geometry.
It is worth considering the solutions to Maxwell's equations, subject to the usual
boundary conditions, for a cylinder with perfectly conducting walls.
We can find the modes in just the same way as we did for a rectangular cavity:
that is, we find solutions to the wave equations for the electric and magnetic fields
using separation of variables; then we find the ``allowed'' solutions by imposing the
boundary conditions.
The algebra is more complicated this time, because we have to work in cylindrical
polar coordinates.  We will not go through the derivation in detail: the solutions
for the fields can be checked by taking the appropriate derivatives.

\begin{figure}
\centering
\includegraphics[width=0.7\textwidth]{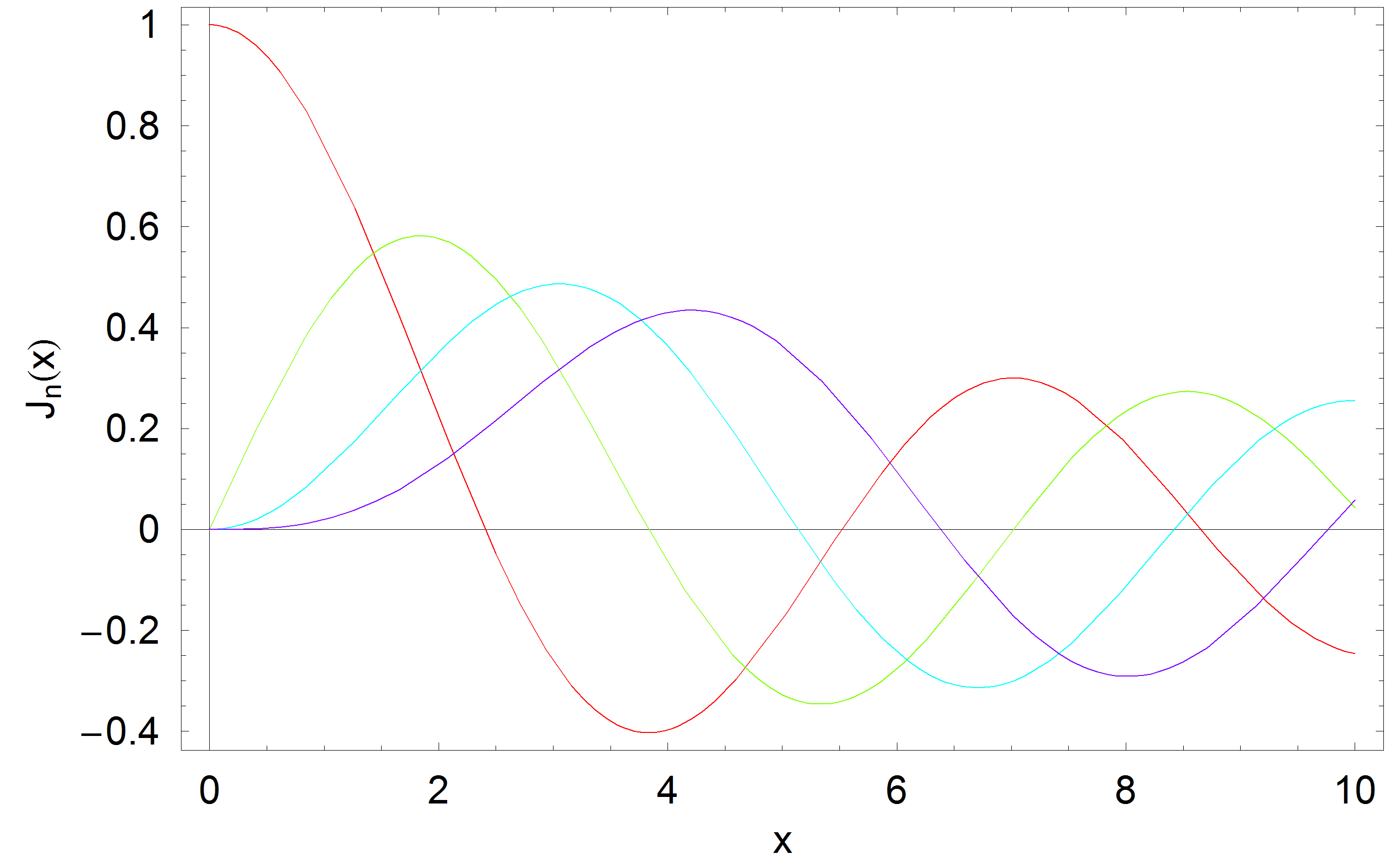}
\caption{Bessel functions. The red, green, blue and purple lines show (respectively) the first, second, third and fourth order Bessel functions $J_n(x)$. \label{fig:besselfunctions}}
\end{figure}

One set of modes (not the most general solution) we can write down is as follows:
\begin{eqnarray}
E_r      & = & -i B_0 \frac{n\omega}{k_r^2 r} J_n(k_r r) \, \sin n\theta \, \sin k_z z \, e^{-i\omega t}, \\
         &   & \nonumber \\
E_\theta & = & -i B_0 \frac{\omega}{k_r} J^\prime_n(k_r r)\, \cos n\theta \, \sin k_z z \, e^{-i\omega t}, \\
         &   & \nonumber \\
E_z      & = &  0, \\
         &   & \nonumber \\
B_r      & = &  B_0 \frac{k_z}{k_r} J^\prime_n(k_r r)\, \cos n\theta \, \cos k_z z \, e^{-i\omega t}, \\
         &   & \nonumber \\
B_\theta & = & -B_0 \frac{n k_z}{k_r^2 r} J_n(k_r r) \, \sin n\theta \, \cos k_z z \, e^{-i\omega t}, \\
         &   & \nonumber \\
B_z      & = &  B_0 J_n(k_r r)\, \cos n\theta \, \sin k_z z \, e^{-i\omega t}.
\end{eqnarray}
Note that $J_n(x)$ is a Bessel function of order $n$, and $J^\prime_n(x)$ is the
derivative of $J_n(x)$.
The Bessel functions (Fig.\,\ref{fig:besselfunctions}) are solutions of the differential equation:
\begin{equation}
y^{\prime\prime} + \frac{y^\prime}{x} + \left( 1 - \frac{n^2}{x^2} \right) y = 0.
\end{equation}
This equation appears when we separate variables in finding a solution to the wave equation
in cylindrical polar coordinates.

Because of the dependence of the fields on the azimuthal angle $\theta$, we require
that $n$ is an integer: $n$ provides an azimuthal index in specifying a mode.

From the boundary conditions, $E_\theta$ and $B_r$ must both vanish on the curved
wall of the cavity, i.e. when $r = a$, where $a$ is the radius of the cylinder.
Therefore, we have a constraint on $k_r$:
\begin{equation}
J^\prime_n(k_r a) = 0,
\end{equation}
or:
\begin{equation}
k_r = \frac{p^\prime_{nm}}{a},
\end{equation}
where $p^\prime_{nm}$ is the $m$th zero of the derivative of the
$n$th order Bessel function.
This equation is analogous to the conditions we had for the rectangular cavity,
e.g. $k_x = m_x\pi / a_x$.
We can use the integer $m$ as a radial index in specifying a mode.

We also need to have $B_z = E_r = E_\theta = 0$ on the flat ends of the cavity.
Assuming the flat ends of the cavity are at $z = 0$ and $z = L$, these boundary conditions
are satisfied if:
\begin{equation}
\sin k_z L = 0, \qquad \textrm{therefore} \quad
k_z = \frac{\ell \pi}{L}, 
\end{equation}
where $\ell$ is an integer.
$\ell$ provides a longitudinal index in specifying a mode.

Also, we find that:
\begin{equation}
\nabla^2 \vec{E} = -(k_r^2 + k_z^2) \vec{E},
\end{equation}
so from the wave equation:
\begin{equation}
\nabla^2 \vec{E} - \frac{1}{c^2}\frac{\partial^2 \vec{E}}{\partial t^2} = 0,
\end{equation}
we must have:
\begin{equation}
k_r^2 + k_z^2 = \frac{\omega^2}{c^2}.
\end{equation}
Similar equations hold for the magnetic field, $\vec{B}$.

\begin{figure}
\centering
\includegraphics[width=0.6\textwidth]{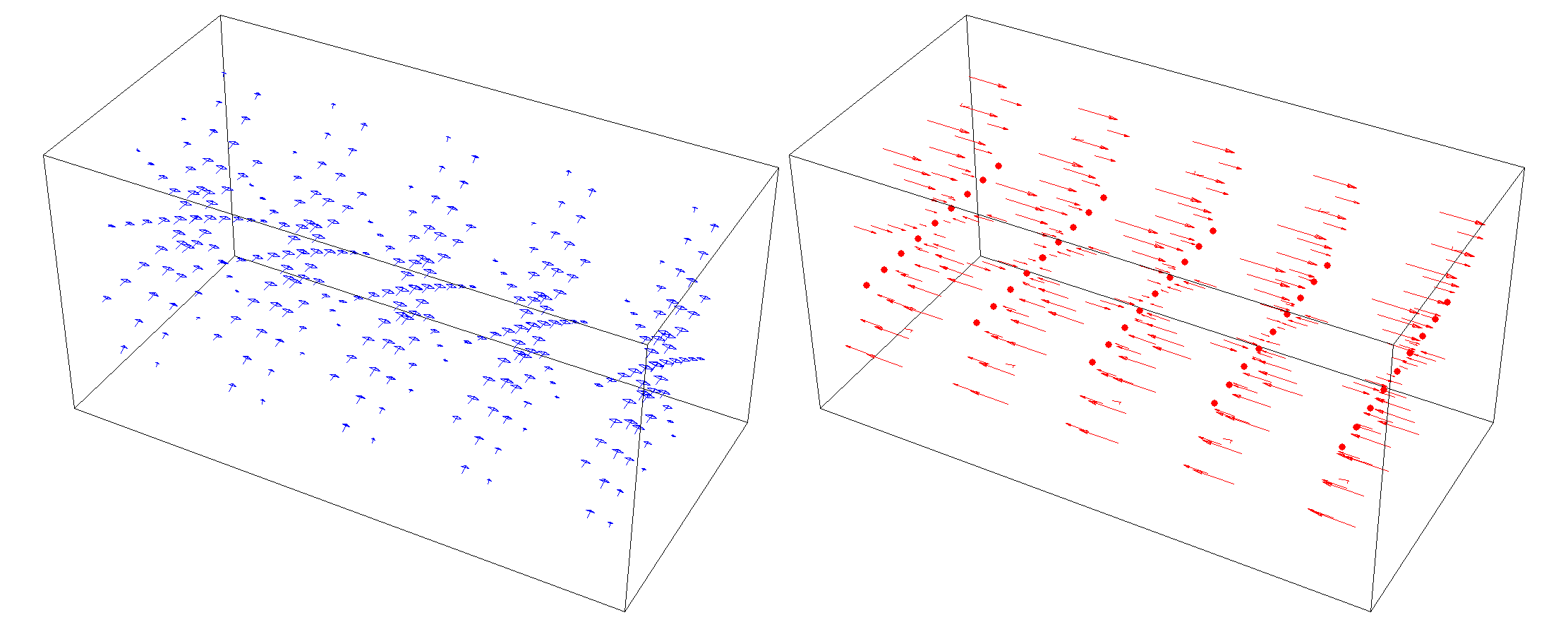} \\
\includegraphics[width=0.6\textwidth]{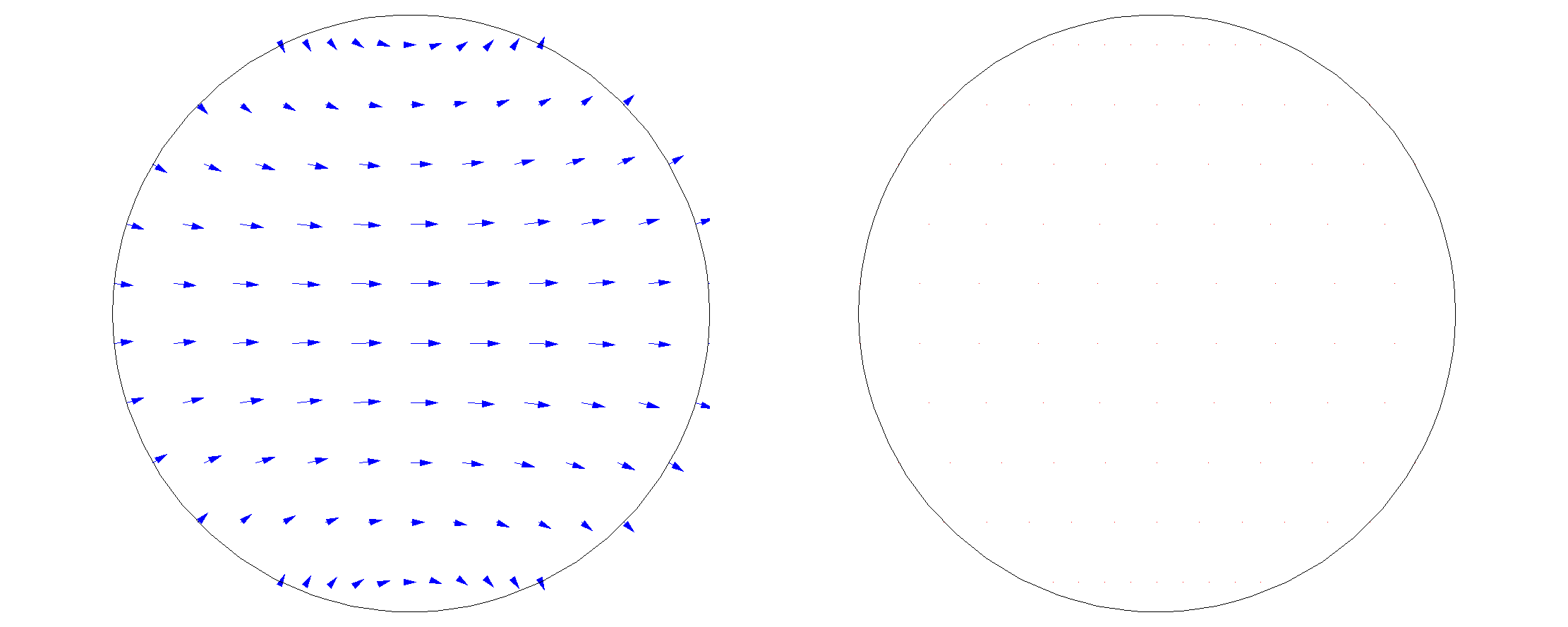}
\caption{TE$_{110}$ mode in a cylindrical cavity. Left: Electric field.  Right: Magnetic field.  Top: 3-dimensional view.  Bottom: Cross-sectional view.  \label{fig:cylindricalte110}}
\end{figure}

In the modes that we are considering, the longitudinal component of the electric
field $E_z = 0$, i.e. the electric field is purely transverse.
These modes are known as ``TE'' modes.  A specific mode of this type, with indices
$n$ (azimuthal), $m$ (radial), and $\ell$ (longitudinal) is commonly written as
TE$_{nm\ell}$.
The frequency of mode TE$_{nm\ell}$ depends on the dimensions of the cavity, and is given by:
\begin{equation}
\omega_{nm\ell} = c \sqrt{ k_r^2 + k_z^2 }
 = c\sqrt{\left( \frac{p^\prime_{nm}}{a} \right)^2 + \left( \frac{\ell \pi}{L} \right)^2}.
\end{equation}
The fields in the TE$_{110}$ mode in a cylindrical cavity are shown in Fig.\,\ref{fig:cylindricalte110}.

TE modes are useful for giving a transverse deflection to a beam in an accelerator,
but are not much use for providing acceleration.
Fortunately, cylindrical cavities allow another set of modes that have non-zero
longitudinal electric field:
\begin{eqnarray}
E_r      & = & -E_0 \frac{k_z}{k_r} J^\prime_n(k_r r) \, \cos n\theta \, \sin k_z z \, e^{-i\omega t}, \\
         &   & \nonumber \\
E_\theta & = &  E_0 \frac{n k_z}{k_r^2 r} J_n(k_r r)\, \sin n\theta \, \sin k_z z \, e^{-i\omega t}, \\
         &   & \nonumber \\
E_z      & = &  E_0 J_n(k_r r)\, \cos n\theta \, \cos k_z z \, e^{-i\omega t}, \\
         &   & \nonumber \\
B_r      & = & i E_0 \frac{n \omega}{c^2 k_r^2 r} J_n(k_r r)\, \sin n\theta \, \cos k_z z \, e^{-i\omega t}, \\
         &   & \nonumber \\
B_\theta & = & i E_0 \frac{\omega}{c^2 k_r} J^\prime_n(k_r r) \, \cos n\theta \, \cos k_z z \, e^{-i\omega t}, \\
         &   & \nonumber \\
B_z      & = &  0.
\end{eqnarray}
In these modes, the magnetic field is purely transverse (zero longitudinal component);
therefore, they are referred to as ``TM'' modes.
As before, for physical fields, $n$ must be an integer.
The boundary conditions on the fields give:
\begin{equation}
k_r = \frac{p_{nm}}{a}, \qquad \textrm{and} \quad
k_z = \frac{\ell \pi}{L},
\end{equation}
where $p_{nm}$ is the $m$th zero of the $n$th order Bessel function $J_n(x)$.
The frequency of a mode TM$_{nm\ell}$ is given by:
\begin{equation}
\omega_{nm\ell} = c \sqrt{ k_r^2 + k_z^2 }
 = c\sqrt{\left( \frac{p_{nm}}{a} \right)^2 + \left( \frac{\ell \pi}{L} \right)^2}.
\end{equation}

\begin{figure}
\centering
\includegraphics[width=0.8\textwidth]{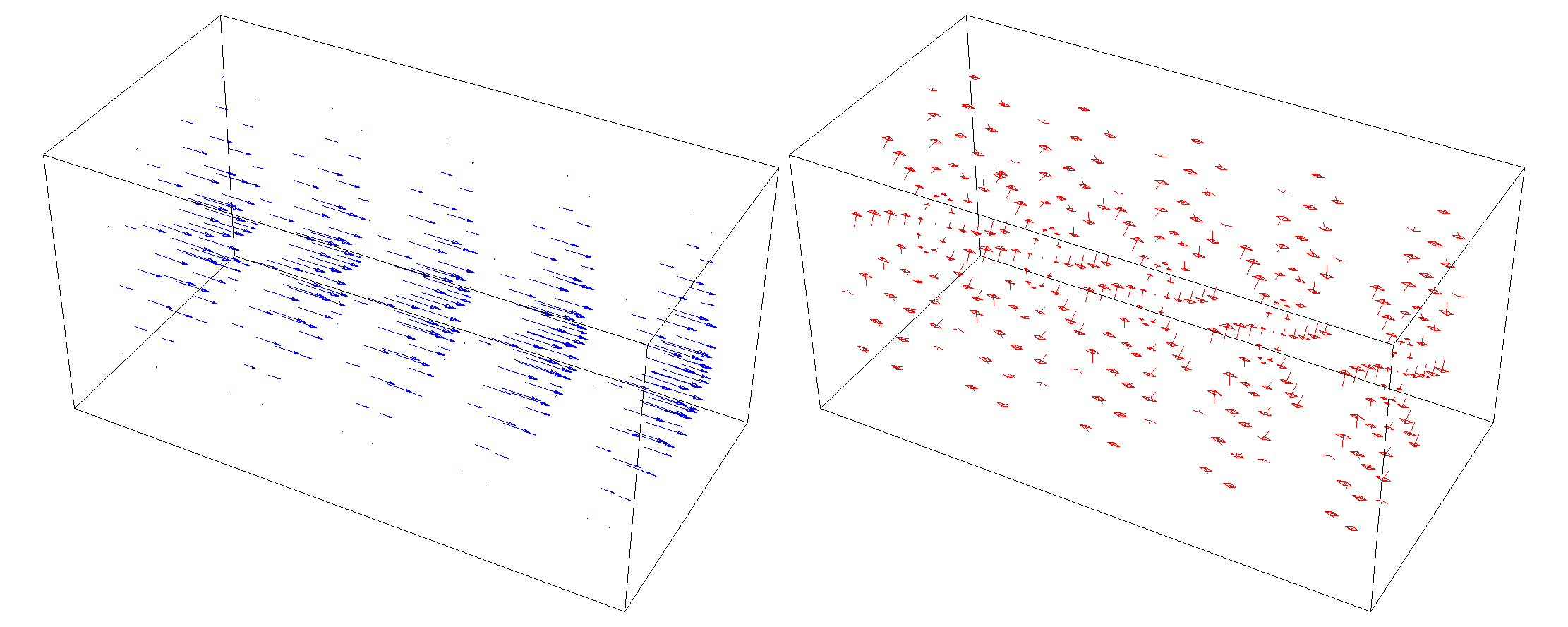}
\caption{Fields in the TM$_{010}$ mode in a cylindrical cavity. Left: Electric field.
Right: Magnetic field. \label{fig:cylindricaltm010}}
\end{figure}

The lowest frequency accelerating mode in a cylindrical cavity is the TM$_{010}$
mode ($n = 0$, $m = 1$, $\ell = 0$).
The fields in the TM$_{010}$ (see Fig.\,\ref{fig:cylindricaltm010}) mode are given by:
\begin{eqnarray}
E_r      & = & 0, \\
         &   & \nonumber \\
E_\theta & = & 0, \\
         &   & \nonumber \\
E_z      & = &  E_0 J_0\!\left( p_{01}\frac{r}{a} \right) \, e^{-i\omega t}, \\
         &   & \nonumber \\
B_r      & = & 0, \\
         &   & \nonumber \\
B_\theta & = & -i \frac{E_0}{c} J_1\!\left( p_{01}\frac{r}{a} \right) e^{-i\omega t}, \\
         &   & \nonumber \\
B_z      & = &  0.
\end{eqnarray}

The frequency of the TM$_{010}$ mode is determined by the radius of the cavity,
and is independent of its length:
\begin{equation}
\omega_{010} = p_{01}\frac{c}{a}.
\end{equation}
(Note that $p_{01} \approx 2.40483$.)
However, to get the maximum acceleration from the cavity,
the time taken for a particle to pass through the cavity should be one half of
the RF period, i.e. $\pi/\omega$.
Therefore, for best efficiency, the length of the cavity should
be $L = \pi v/\omega$, where $v$ is the velocity of the particle.  For ultra-relativistic particles,
the length of the cavity should be $L = \lambda/2$, where $\lambda$ is the wavelength of an
electromagnetic wave with angular frequency $\omega$ in free space.

\section{Waveguides}

Cavities are useful for storing energy in electromagnetic fields, but it is also necessary to
transfer electromagnetic energy between different locations, e.g. from an RF power source
such as a klystron, to an RF cavity.  Waveguides are generally used for carrying large amounts
of energy (high power RF).  For low power RF signals (e.g. for timing or control systems),
transmission lines are generally used.  Although the basic physics in waveguides and transmission
lines is the same -- both involve electromagnetic waves propagating through bounded regions --
different formalisms are used for their analysis, depending on the geometry of the boundaries.
In this section, we consider waveguides; we consider transmission lines in the following section.

As was the case for cavities, the patterns of the fields in the resonant modes are determined
by the geometry of the boundary.  The resonant modes in a waveguide with circular cross-section
will look somewhat different from the modes in a waveguide with rectangular cross-section.
The physical principles and general methods are the same in both cases; however, the algebra is
somewhat easier for a rectangular waveguide.  Therefore, we shall consider here only waveguides
with rectangular cross-section.

\subsection{Modes in a rectangular waveguide}

\begin{figure}
\centering
\includegraphics[width=0.5\textwidth]{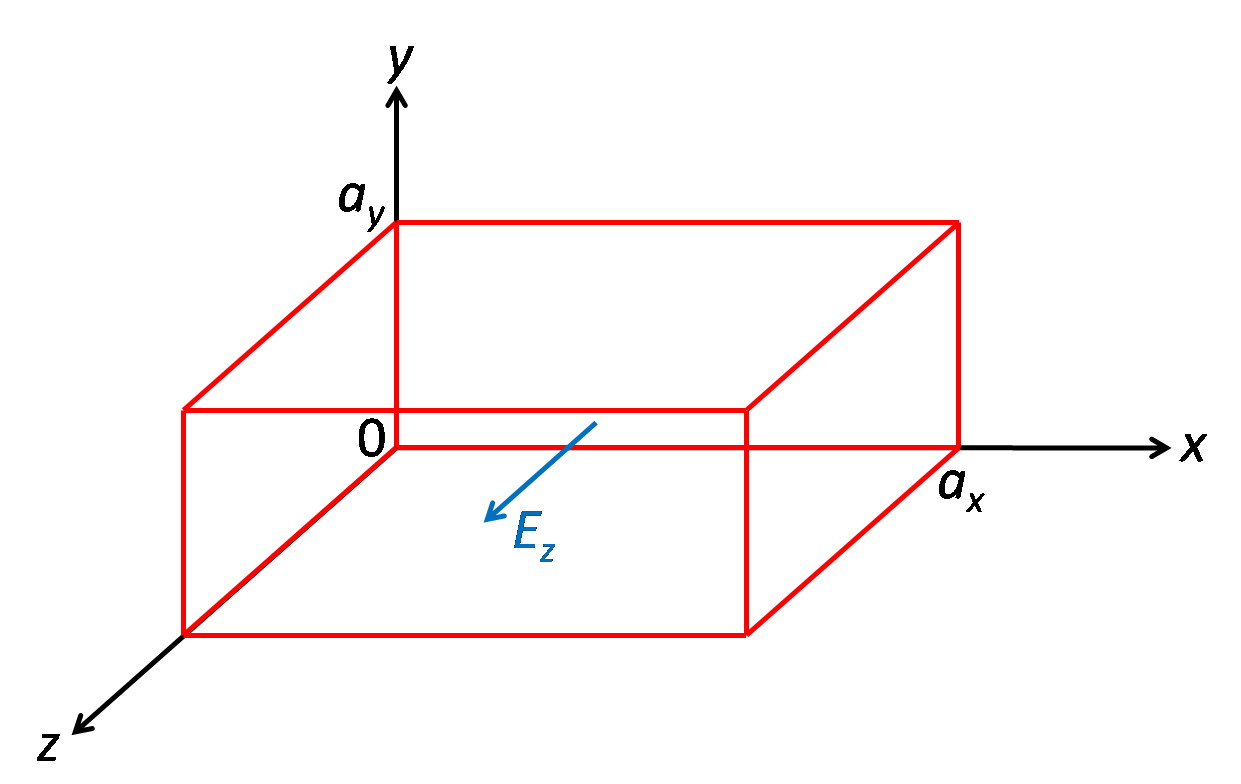}
\caption{Rectangular waveguide. \label{fig:rectangularwaveguide}}
\end{figure}

Consider an ideal conducting tube with rectangular cross-section, of width $a_x$ and
height $a_y$ (Fig.\,\ref{fig:rectangularwaveguide}).  This is essentially a cavity resonator
with length $a_z \to \infty$.
The electric field must solve the wave equation:
\begin{equation}
\nabla^2 \vec{E} - \frac{1}{c^2} \frac{\partial^2 \vec{E}}{\partial t^2} = 0,
\end{equation}
together (as usual) with Maxwell's equations.
By comparison with the rectangular cavity case, we expect to find standing waves
in $x$ and $y$, with plane wave solution in $z$.  Therefore, we try a solution
of the form:
\begin{eqnarray}
E_x & = & E_{x0} \cos k_x x \, \sin k_y y \, e^{i(k_z z - \omega t)}, \label{waveguideex} \\
    &   & \nonumber \\
E_y & = & E_{y0} \sin k_x x \, \cos k_y y \, e^{i(k_z z - \omega t)}, \label{waveguideey} \\
    &   & \nonumber \\
E_z & = & i E_{z0} \sin k_x x \, \sin k_y y \, e^{i(k_z z - \omega t)}. \label{waveguideez}
\end{eqnarray}
To satisfy the wave equation, we must have:
\begin{equation}
k_x^2 + k_y^2 + k_z^2 = \frac{\omega^2}{c^2}.
\label{ksquared}
\end{equation}
We must of course also satisfy Maxwell's equation:
\begin{equation}
\nabla \cdot \vec{E} = 0,
\end{equation}
at all $x$, $y$ and $z$.  This leads to the condition:
\begin{equation}
k_x E_{x0} + k_y E_{y0} + k_z E_{z0} = 0, \label{divergencee}
\end{equation}
which is the same as we found for a rectangular cavity.

Now we apply the boundary conditions.  In particular, the tangential component of
the electric field must vanish at all boundaries.
For example, we must have $E_z = 0$ for all $t$ and all $z$, at $x = 0$ and $x = a_x$,
and at $y = 0$ and $y = a_y$ (see Fig.\,\ref{fig:rectangularwaveguide}).
Therefore, $\sin k_x a_x = 0$, and $\sin k_y a_y = 0$, and hence:
\begin{eqnarray}
k_x & = & \frac{m_x \pi}{a_x}, \qquad m_x = 0,1,2\dots \label{boundaryconditionx} \\
    &   & \nonumber \\
k_y & = & \frac{m_y \pi}{a_y}, \qquad m_y = 0,1,2\dots \label{boundaryconditiony} 
\end{eqnarray}
These conditions also ensure that $E_x = 0$ at $y = 0$ and at $y = a_y$, and that
$E_y = 0$ at $x = 0$ and at $x = a_x$.

We can find the magnetic field in the waveguide from:
\begin{equation}
\nabla \times \vec{E} = -\frac{\partial \vec{B}}{\partial t}.
\end{equation}
The result is:
\begin{eqnarray}
B_x & = & \frac{1}{\omega}(k_y E_{z0} - k_z E_{y0}) \sin k_x x \, \cos k_y y \, e^{i(k_z z - \omega t)}, \label{waveguidebx} \\
    &   & \nonumber \\
B_y & = & \frac{1}{\omega}(k_z E_{x0} - k_x E_{z0}) \cos k_x x \, \sin k_y y \, e^{i(k_z z - \omega t)}, \label{waveguideby} \\
    &   & \nonumber \\
B_z & = & \frac{1}{i\omega}(k_x E_{y0} - k_y E_{x0}) \cos k_x x \, \cos k_y y \, e^{i(k_z z - \omega t)}. \label{waveguidebz}
\end{eqnarray}
The conditions (\ref{boundaryconditionx}) and (\ref{boundaryconditiony}):
\begin{eqnarray}
k_x & = & \frac{m_x \pi}{a_x}, \qquad m_x = 0,1,2\dots \nonumber \\
    &   & \nonumber \\
k_y & = & \frac{m_y \pi}{a_y}, \qquad m_y = 0,1,2\dots \nonumber
\end{eqnarray}
ensure that the boundary conditions on the magnetic field are satisfied.
Also, the conditions (\ref{ksquared}):
\begin{equation}
k_x^2 + k_y^2 + k_z^2 = \frac{\omega^2}{c^2}, \nonumber
\end{equation}
and (\ref{divergencee}):
\begin{equation}
k_x E_{x0} + k_y E_{y0} + k_z E_{z0} = 0, \nonumber
\end{equation}
ensure that Maxwell's equation:
\begin{equation}
\nabla \times \vec{B} = \frac{1}{c^2} \frac{\partial \vec{E}}{\partial t}
\end{equation}
is satisfied.

In a rectangular cavity, the mode could be specified by a set of three numbers,
$m_x$, $m_y$, and $m_z$, that determined the wave numbers and the frequency of 
oscillation of the field.
Because $m_x$, $m_y$ and $m_z$ had to be integers, there was a discrete spectrum
of allowed frequencies of oscillation in a cavity.
In a waveguide, we only need two integer mode numbers, $m_x$ and $m_y$, corresponding
to the wave numbers in the transverse dimensions.
The third wave number, in the longitudinal direction, is allowed to take values
over a continuous range: this is because of the absence of boundaries in the
longitudinal direction.  Therefore, there is a continuous range of frequencies
also allowed.

When we discussed cylindrical cavities, we wrote down two sets of modes for the
fields: TE modes ($E_z = 0$), and TM modes ($B_z = 0$).
We can also define TE and TM modes in rectangular cavities, and rectangular
waveguides.
For a TE mode in a waveguide, the longitudinal component
of the electric field is zero at all positions and times, which implies that:
\begin{equation}
E_{z0} = 0.
\end{equation}
From (\ref{divergencee}), it follows that in a TE mode, the wave numbers and
amplitudes must be related by:
\begin{equation}
k_x E_{x0} + k_y E_{y0} = 0.
\end{equation}
Similarly, for a TM mode, we have (using \ref{waveguidebz}):
\begin{equation}
B_{z0} = 0, \quad k_x E_{y0} - k_y E_{x0} = 0.
\end{equation}

\subsection{Cut-off frequency in a waveguide}

We have seen that in a waveguide, specifying the mode does not uniquely specify the
frequency.  This is a consequence of the fact that there are no boundaries in the
longitudinal direction: there are only two mode numbers required to specify the transverse
dependence of the field; the longitudinal dependence is not constrained by boundaries.
The frequency of oscillation of fields in the waveguide depends on the longitudinal
variation of the field, as well as on the transverse variation.  However, if we want a wave
to propagate along the waveguide, there is a minimum frequency of oscillation that is
needed to achieve this, as we shall now show.

Let us write equation (\ref{ksquared}) in the form:
\begin{equation}
k_z^2 = \frac{\omega^2}{c^2} - k_x^2 - k_y^2.
\end{equation}
If the wave is to propagate with no attentuation (i.e. propagate without any loss in amplitude),
then $k_z$ must be real.  We therefore require that:
\begin{equation}
\frac{\omega^2}{c^2} > k_x^2 + k_y^2.
\end{equation}
This can be written:
\begin{equation}
\frac{\omega}{c} > \pi \sqrt{\left( \frac{m_x}{a_x} \right)^2 + \left( \frac{m_y}{a_y} \right)^2}.
\label{waveguidemodefrequency}
\end{equation}
We define the \emph{cut-off} frequency $\omega_{\textrm{\small co}}$:
\begin{equation}
\omega_{\textrm{\small co}} =  \pi c \sqrt{\left( \frac{m_x}{a_x} \right)^2 + \left( \frac{m_y}{a_y} \right)^2}.
\end{equation}
Above the cut-off frequency, the wave will propagate without loss of amplitude (in a
waveguide with ideal conducting walls).
At the cut-off frequency, the group velocity of the wave is zero: the wave is effectively a
standing wave.  Below the cut-off frequency, oscillating fields can still exist within the
waveguide, but they take the form of attenuated (evanescent) waves, rather than
propagating waves: the longitudinal wave number is purely imaginary.

Note that at least one of the mode numbers $m_x$ and $m_y$ must be non-zero, for
a field to be present at all. Therefore, the minimum cut-off frequency of all the possible
modes is given by:
\begin{equation}
\omega_{\textrm{\small co}} =  \frac{\pi}{a}c,
\end{equation}
where $c$ is the speed of light, and $a$ is the side length of the \emph{longest}
side of the waveguide cross-section.

\subsection{Phase velocity, dispersion, and group velocity}

\begin{figure}
\centering
\includegraphics[width=0.5\textwidth]{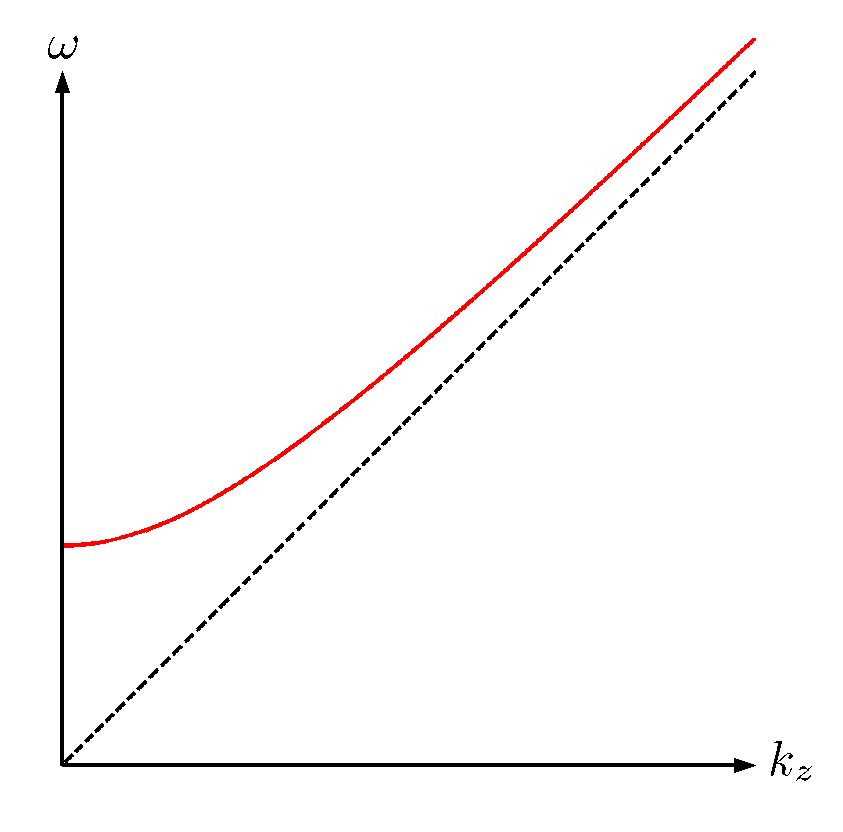}
\caption{Dispersion curve (in red) for waves in a waveguide. The broken black line
shows phase velocity $\omega/k_z = c$. \label{fig:dispersioncurvewaveguide}}
\end{figure}

Consider an electromagnetic wave that propagates down a waveguide, with longitudinal
wave number $k_z$ and frequency $\omega$.
The phase velocity $v_p$ is given by:
\begin{equation}
v_p = \frac{\omega}{k_z} = \frac{\sqrt{k_x^2 + k_y^2 + k_z^2}}{k_z} c.
\end{equation}
This is the speed at which one would have to travel along the waveguide to stay
at a constant phase of the electromagnetic wave (i.e. $k_z z - \omega t = $ constant).
However, since $k_x$, $k_y$ and $k_z$ are all real, it follows that the phase
velocity is greater than the speed of light:
\begin{equation}
v_p > c.
\end{equation}
The dispersion curve (Fig.\,\ref{fig:dispersioncurvewaveguide}) shows how the
frequency $\omega$ varies with the longitudinal wave number $k_z$.

Although the phase velocity of a wave in a waveguide is greater than the speed of light,
this does not violate special relativity, since energy and information
signals travel down the waveguide at the group velocity, rather than the phase
velocity.
The group velocity $v_g$ is given by:
\begin{equation}
v_g = \frac{d \omega}{dk_z} = \frac{k_z}{\sqrt{k_x^2 + k_y^2 + k_z^2}} c
 = \frac{k_z}{\omega} c^2,
\label{groupvelocity}
\end{equation}
so we have:
\begin{equation}
v_g < c.
\end{equation}
Note that for a rectangular waveguide, the phase and group velocities are related by:
\begin{equation}
v_p v_g = c^2.
\end{equation}

\begin{figure}
\centering
\includegraphics[width=0.6\linewidth]{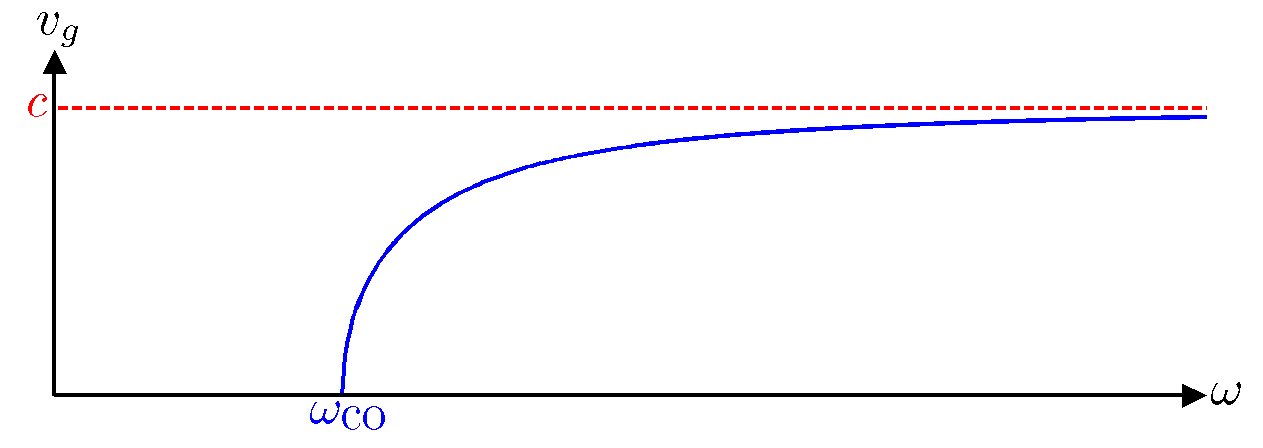}
\caption{Group velocity in a waveguide. \label{waveguidedispersion}}
\end{figure}

Using equation (\ref{ksquared}):
\begin{equation}
k_x^2 + k_y^2 + k_z^2 = \frac{\omega^2}{c^2}, \nonumber
\end{equation}
and the boundary conditions (\ref{boundaryconditionx}) and (\ref{boundaryconditiony}):
\begin{equation}
k_x = \frac{m_x \pi}{a_x} \qquad \textrm{and} \quad
k_y = \frac{m_y \pi}{a_y}, \nonumber
\end{equation}
we can write the group velocity (\ref{groupvelocity}) for a given mode $(m_x,m_y)$ in
terms of the frequency $\omega$:
\begin{equation}
v_g = c \sqrt{1 - \frac{\pi^2 c^2}{\omega^2} \left( \frac{m_x^2}{a_x^2} + \frac{m_y^2}{a_y^2} \right)}.
\end{equation}
Energy in the wave propagates along the waveguide only for
$\omega > \omega_{\textrm{\small co}}$
(where $\omega_{\textrm{\small co}}$ is the cut-off frequency).
Note the limiting behaviour of the group velocity, for
$\omega > \omega_{\textrm{\small co}}$:
\begin{eqnarray}
\lim_{\omega \to \infty} v_g & = & c, \\
 & & \nonumber \\
\lim_{\omega \to \omega_{\textrm{\small co}}} v_g & = & 0.
\end{eqnarray}
The variation of group velocity with frequency is illustrated in Fig.\,\ref{waveguidedispersion}.

\subsection{Energy density and energy flow in a waveguide}

The main purpose of a waveguide is to carry electromagnetic energy from one place
to another: therefore, it is of interest to know  the energy density and the energy flow in
a waveguide, for a given mode and field amplitude.

The energy density in the waveguide can be calculated from:
\begin{equation}
U = \frac{1}{2} \varepsilon_0 \vec{E}^2 + \frac{1}{2} \frac{\vec{B}^2}{\mu_0}.
\end{equation}
Using the expressions for the fields (\ref{waveguideex}) -- (\ref{waveguideez}),
and assuming that both $k_x$ and $k_y$ are non-zero, we see that the
\emph{time-average energy per unit length} in the electric field is:
\begin{equation}
\int\!\!\int \langle U_E \rangle \, dx \, dy = \frac{1}{16} \varepsilon_0 A 
          \left( E_{x0}^2 + E_{y0}^2 +E_{z0}^2 \right),
\end{equation}
where $A$ is the cross-sectional area of the waveguide.

The fields in a waveguide are just a superposition of plane waves in free space.
We know that in such waves, the energy is shared equally between the electric and
magnetic fields.
Using the expressions for the magnetic field (\ref{waveguidebx}) -- (\ref{waveguidebz}),
we can indeed show that the magnetic energy is equal (on average) to the electric
energy - though the algebra is rather complicated.
The total time-average energy per unit length in the waveguide is then given by:
\begin{equation}
\int\!\!\int \langle U \rangle \, dx \, dy = \frac{1}{8} \varepsilon_0 A 
          \left( E_{x0}^2 + E_{y0}^2 +E_{z0}^2 \right).
\end{equation}

The energy flow along the waveguide is given by the Poynting vector,
$\vec{S} = \vec{E} \times \vec{H}$.
Taking the real parts of the field components, we find for the time-average
transverse components:
\begin{equation}
\langle S_x \rangle = \langle S_y \rangle = 0.
\end{equation}
The longitudinal component is non-zero; taking the time average and integrating
over the cross section gives the time average energy flow along the waveguide
\begin{equation}
\int\!\!\int \langle S_z \rangle \,dx\,dy = \frac{1}{8\mu_0} A ( E_{x0}^2 + E_{y0}^2 + E_{z0}^2 )
\frac{k_z}{\omega}.
\end{equation}
Using equation (\ref{groupvelocity}) for the group velocity $v_g$, the time-average
energy flow along the waveguide can be written in the form:
\begin{equation}
\int\!\!\int \langle S_z \rangle \,dx\,dy = \frac{1}{8} \varepsilon_0 A ( E_{x0}^2 + E_{y0}^2 + E_{z0}^2 ) v_g.
\end{equation}
Thus, the energy density per unit length and the rate of energy flow are related by:
\begin{equation}
\int\!\!\int \langle S_z \rangle \,dx\,dy = v_g \int\!\!\int \langle U \rangle \,dx\,dy.
\end{equation}
This is consistent with our interpretation of the group velocity as the speed at
which energy is carried by a wave.

The boundary conditions that we have applied implicitly assume ideal conducting walls.  In that
case, there will be no energy losses to the walls, and electromagnetic waves can propagate
indefinitely along the waveguide without any reduction in amplitude.  In practice, there will be
some attenuation, because of the finite conductivity of the walls of the wavguide.  However,
using a metal with high conductivity (such as copper or aluminium), attenuation lengths can
be quite long, of order many metres.  This makes waveguides very suitable for high-power
RF applications.

\section{Transmission lines}

Transmission lines are used (as are waveguides) to guide electromagnetic waves
from one place to another.  A coaxial cable (used, for example, to connect a
radio or television to an aerial) is an example of a transmission line.
Transmission lines may be less bulky and less expensive than waveguides; but
they generally have higher losses, so are more appropriate for carrying
low-power signals over short distances.

\subsection{$LC$ model of a transmission line}

\begin{figure}
\centering
\includegraphics[width=0.7\textwidth]{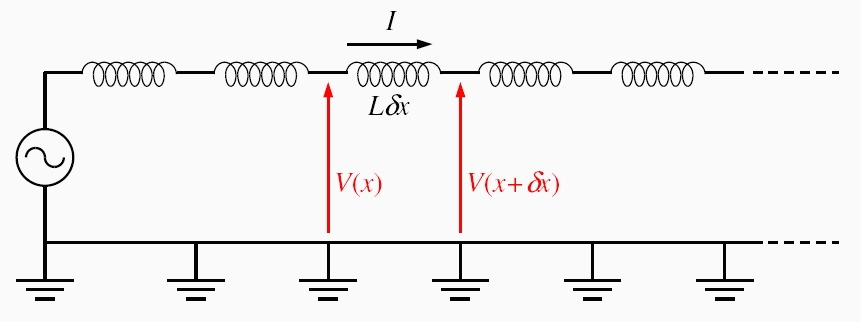}
\caption{Inductance in a transmission line. \label{fig:lcmodelinductance}}
\end{figure}

Consider an infinitely long, parallel wire with zero resistance.  In general, the wire
will have some inductance per unit length, $L$, which means that when an alternating
current $I$ flows in the wire, there will be a potential difference between
different points along the wire (Fig.\,\ref{fig:lcmodelinductance}).
If $V$ is the potential at some point along the
wire with respect to earth, then the potential difference between two points
along the wire is given by:
\begin{equation}
\Delta V = \frac{\partial V}{\partial x} \delta x = -L \delta x \frac{\partial I}{\partial t}.
\label{dVdx1}
\end{equation}

\begin{figure}
\centering
\includegraphics[width=0.7\textwidth]{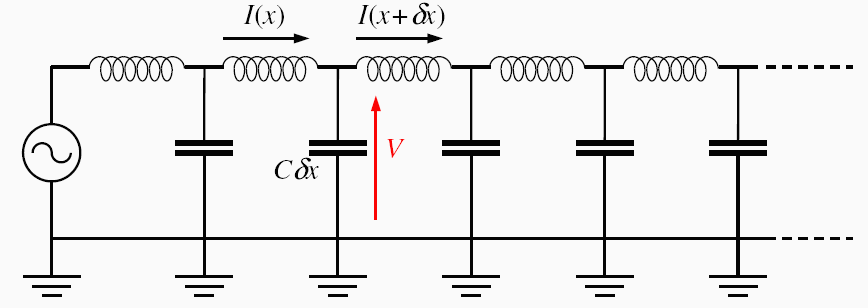}
\caption{Capacitance in a transmission line. \label{fig:lcmodelcapacitance}}
\end{figure}

In general, as well as the inductance, there will also be some capacitance per
unit length, $C$, between the wire and earth (Fig.\,\ref{fig:lcmodelcapacitance}).
This means that the current in the wire can vary with position:
\begin{equation}
\frac{\partial I}{\partial x} \delta x = -C \delta x \frac{\partial V}{\partial t}.
\label{dIdx1}
\end{equation}
The ``complete'' model, including both inductance
and capacitance, is referred to as the $LC$ model for a transmission line.

We wish to solve the equations for the current and voltage in the transmission line,
as functions of position and time.
Let us take Equations (\ref{dVdx1}) and (\ref{dIdx1}) above:
\begin{eqnarray}
\frac{\partial V}{\partial x} & = & -L \frac{\partial I}{\partial t}, \label{dvdx} \\
 & & \nonumber \\
\frac{\partial I}{\partial x} & = & -C \frac{\partial V}{\partial t}. \label{didx}
\end{eqnarray}
These equations are analogous to Maxwell's equations: they show how changes
(with respect to time and position) in current and voltage are related to each other.
They take the form of two, coupled first-order differential equations.  The fact that
they are coupled means that the solution is not obvious; however, we can take the
same approach that we did with Maxwell's equations in free space.  By taking
additional derivatives, we can arrive at two \emph{uncoupled} second-order
differential equations.  The solution to the uncoupled equations is more apparent than
the solution to the coupled equations.
Differentiate (\ref{dvdx}) with respect to $t$:
\begin{equation}
\frac{\partial^2 V}{\partial x \partial t} = -L \frac{\partial^2 I}{\partial t^2},
\end{equation}
and (\ref{didx}) with respect to $x$:
\begin{equation}
\frac{\partial^2 I}{\partial x^2} = -C \frac{\partial^2 V}{\partial t \partial x}.
\end{equation}
Hence:
\begin{equation}
\frac{\partial^2 I}{\partial x^2} = LC \frac{\partial^2 I}{\partial t^2}. \label{wavei}
\end{equation}
Similarly (by differentiating (\ref{dvdx}) with respect to $x$ and (\ref{didx})
with respect to $t$), we find:
\begin{equation}
\frac{\partial^2 V}{\partial x^2} = LC \frac{\partial^2 V}{\partial t^2} \label{wavev}.
\end{equation}

Equations (\ref{wavei}) and (\ref{wavev}) are wave equations for the current in the
wire, and the voltage between the wire and earth.  The waves travel with speed $v$,
given by:
\begin{equation}
v = \frac{1}{\sqrt{LC}}.
\end{equation}
The solutions to the wave equations may be written:
\begin{eqnarray}
V & = & V_0 e^{i\left( k x - \omega t \right)}, \\
 & & \nonumber \\
I & = & I_0 e^{i\left( k x - \omega t \right)},
\end{eqnarray}
where the phase velocity is:
\begin{equation}
v = \frac{\omega}{k} = \frac{1}{\sqrt{LC}}.
\end{equation}
Note that the inductance per unit length $L$ and the capacitance per unit
length $C$ are real and positive.  Therefore, if the frequency $\omega$ is
real, the wave number $k$ will also be real: this implies that waves
propagate along the transmission line with constant amplitude.  This
result is expected, given our assumption about the line having zero
resistance.

The solutions must also satisfy the first-order equations (\ref{dVdx1}) and
(\ref{dIdx1}).  Substituting the above solutions into these equations, we
find:
\begin{eqnarray}
k V_0 & = & \omega L I_0, \\
k I_0 & = & \omega C V_0.
\end{eqnarray}
Hence:
\begin{equation}
\frac{V_0}{I_0} = \sqrt{\frac{L}{C}} = Z.
\end{equation}
The ratio of the voltage to the current is called the \emph{characteristic
impedance}, $Z$, of the transmission line.  $Z$ is measured in ohms, $\Omega$.
Note that, since $L$ and $C$ are real and positive, the impedance is a real
number: this means that the voltage and current are in phase.  The characteristic
impedance of a transmission line is analogous to the impedance of a medium for
electromagnetic waves: the impedance of a transmission line gives the ratio of the
voltage amplitude to the current amplitude; the impedance of a medium for
electromagnetic waves gives the ratio of the electric field amplitude to the
magnetic field amplitude.

\subsection{Impedance matching}

\begin{figure}
\centering
\includegraphics[width=0.7\textwidth]{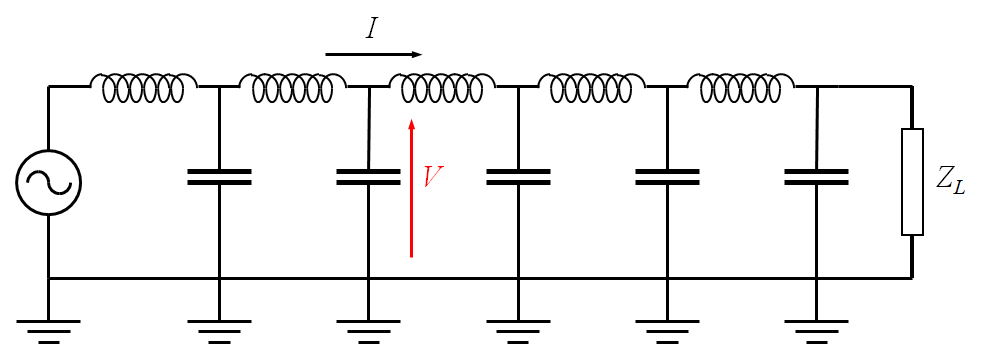}
\caption{Termination of a transmission line with an impedance $R$.
\label{fig:transmissionlinetermination}}
\end{figure}

So far, we have assumed that the transmission line has infinite length.
Obviously, this cannot be achieved in practice.  We can terminate the
transmission line using a ``load'' with impedance $Z_L$ that dissipates the energy in the wave
while maintaining the same ratio of voltage to current as exists all along
the transmission line -- see Fig.\,\ref{fig:transmissionlinetermination}.
In that case, our above analysis for the infinite
line will remain valid for the finite line, and we say that the impedances
of the line and the load are properly \emph{matched}.

What happens if the impedance of the load, $Z_L$, is not properly matched to
the characteristic impedance of the transmission line, $Z$?  In that case, we
need to consider a solution consisting of a superposition of waves travelling in
opposite directions:
\begin{equation}
V = V_0 e^{i \left( k x - \omega t \right)} + 
  K V_0 e^{i \left(-k x - \omega t \right)}.
\label{TerminatedTransmissionLineVoltage}
\end{equation}
The corresponding current is given by:
\begin{equation}
I = \frac{V_0}{Z} e^{i \left( k x - \omega t \right)} - 
  K \frac{V_0}{Z} e^{i \left(-k x - \omega t \right)}.
\end{equation}
Note the minus sign in the second term in the expression for the current:
this comes from equations (\ref{dvdx}) and (\ref{didx}).
Let us take the end of the transmission line, where the load is located, to be
at $x=0$.  At this position, we have:
\begin{eqnarray}
V & = & V_0 e^{-i \omega t} \left( 1 + K \right), \\
 & & \nonumber \\
I & = & \frac{V_0}{Z} e^{-i \omega t} \left( 1 - K \right).
\end{eqnarray}
If the impedance of the load is $Z_L$, then:
\begin{equation}
Z_L = \frac{V}{I} = Z \frac{1 + K}{1 - K}.
\end{equation}
Solving this equation for $K$ (which gives the relative amplitude and phase of
the ``reflected'' wave), we find:
\begin{equation}
K = \frac{Z_L / Z - 1}{Z_L / Z + 1} = \frac{Z_L - Z}{Z_L + Z}.
\label{ReflectionAmplitude}
\end{equation}
Note that if $Z_L = Z$, there is no reflected wave: the termination is correctly
matched to the characteristic impedance of the transmission line.

\subsection{``Lossy'' transmission lines}

\begin{figure}
\centering
\includegraphics[width=0.5\textwidth]{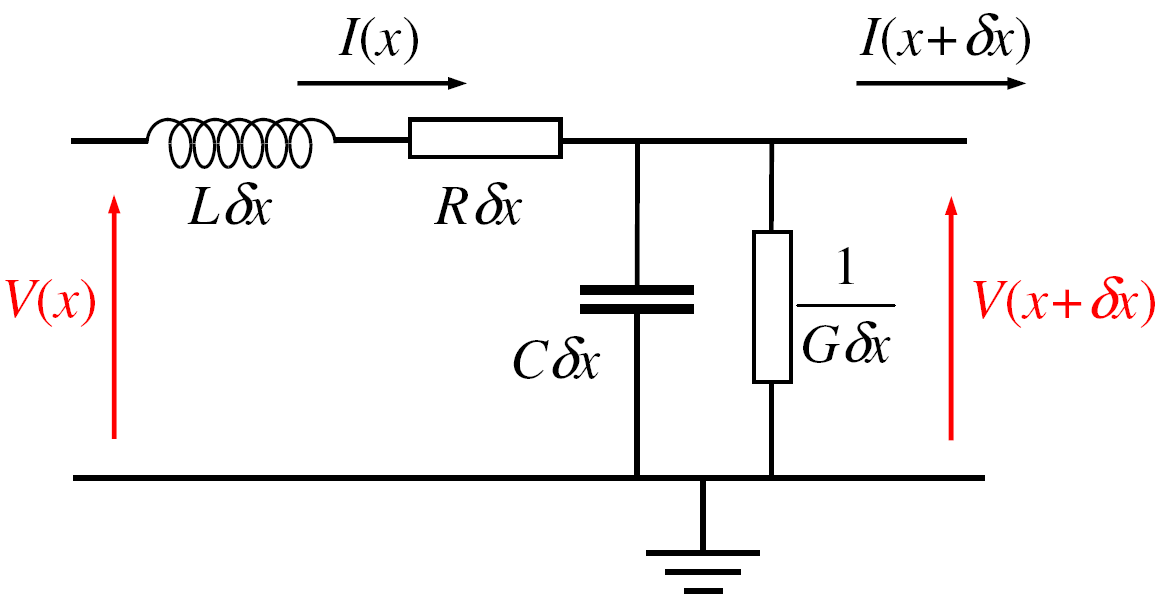}
\caption{A ``lossy'' transmission line. \label{fig:lossyline}}
\end{figure}

So far, we have assumed that the conductors in the transmission line have
zero resistance, and are separated by a perfect insulator.  Usually, though, the
conductors will have finite conductivity; and the insulator will have some
finite resistance.
To understand the impact that this has, we need to modify our transmission line
model to include:
\begin{itemize}
\item a resistance per unit length $R$ in series with the inductance;
\item a conductance per unit length $G$ in parallel with the capacitance.
\end{itemize}
The modified transmission line is illustrated in Fig.\,\ref{fig:lossyline}.

The equations for the current and voltage are then:
\begin{eqnarray}
\frac{\partial V}{\partial x} & = & -L \frac{\partial I}{\partial t} - RI,
\label{lossydvdx} \\
 & & \nonumber \\
\frac{\partial I}{\partial x} & = & -C \frac{\partial V}{\partial t} - GV.
\label{lossydidx}
\end{eqnarray}
We can find solutions to the equations (\ref{lossydvdx}) and (\ref{lossydidx})
for the voltage and current in the lossy transmission line by considering
the case that we propagate a wave with a single, well-defined frequency $\omega$.
In that case, we can replace each time derivative by a factor $-i\omega$.  The
equations become:
\begin{eqnarray}
\frac{\partial V}{\partial x} & = & i\omega L I - RI = -\tilde{L} \frac{\partial I}{\partial t},
\label{lossydvdx1} \\
 & & \nonumber \\
\frac{\partial I}{\partial x} & = & i\omega C V - GV = -\tilde{C} \frac{\partial V}{\partial t},
\label{lossydidx1}
\end{eqnarray}
where
\begin{equation}
\tilde{L} = L - \frac{R}{i\omega} \qquad \textrm{and} \qquad \tilde{C} = C - \frac{G}{i\omega}.
\end{equation}
The new equations (\ref{lossydvdx1}) and (\ref{lossydidx1}) for the lossy transmission
line look exactly like the original equations (\ref{dvdx}) and (\ref{didx}) for a
lossless transmission line, but with the capacitance $C$ and inductance $L$ replaced
by (complex) quantities $\tilde{C}$ and $\tilde{L}$.  The imaginary parts of 
$\tilde{C}$ and $\tilde{L}$ characterise the losses in the lossy transmission line.

Mathematically, we can solve the equations for a lossy transmission line in exactly
the same way as we did for the lossless line.  In particular, we find for the
phase velocity:
\begin{equation}
v = \frac{1}{\sqrt{\tilde{L}\tilde{C}}} =
\frac{1}{\sqrt{\left( L + i\frac{R}{\omega} \right) \left( C + i\frac{G}{\omega} \right)}},
\label{lossyphasevelocity}
\end{equation}
and for the impedance:
\begin{equation}
Z = \sqrt{\frac{\tilde{L}}{\tilde{C}}} = \sqrt{\frac{L + i\frac{R}{\omega}}{C  + i\frac{G}{\omega}}}.
\label{lossyimpedance}
\end{equation}
Since the impedance (\ref{lossyimpedance}) is now a complex number, there will be
a phase difference (given by the complex phase of the impedance) between the
current and voltage in the transmission line.
Note that the phase velocity (\ref{lossyphasevelocity}) depends explicitly on the
frequency.  That means that a lossy transmission line will exhibit dispersion:
waves of different frequencies will travel at different speeds, and a the shape
of a wave ``pulse'' composed of different frequencies will change as it travels
along the transmission line.
Dispersion is one reason why it is important to keep losses in a transmission line
as small as possible (for example, by using high-quality materials).  The other
reason is that in a lossy transmission line, the wave amplitude will
attenuate, much like an electromagnetic wave propagating in a conductor.

Recall that we can write the phase velocity:
\begin{equation}
v = \frac{\omega}{k},
\end{equation}
where $k$ is the wave number appearing in the solution to the wave equation:
\begin{equation}
V = V_0 e^{i \left( k x - \omega t \right)},
\end{equation}
and similarly for the current $I$.  Using Eq.\,(\ref{lossyphasevelocity})
for the phase velocity, we have:
\begin{equation}
k = \omega \sqrt{LC} \sqrt{\left( 1 + i \frac{R}{\omega L} \right) \left( 1 + i \frac{G}{\omega C} \right)}.
\end{equation}
Let us assume $R \ll \omega L$ (i.e. good conductivity along the transmission line)
and $G \ll \omega C$ (i.e. poor conductivity between the lines); then we can make
a Taylor series expansion, to find:
\begin{equation}
k \approx \omega \sqrt{LC} \left[ 1 + \frac{i}{2\omega} \left( \frac{R}{L} + \frac{G}{C} \right) \right] .
\label{lossyk}
\end{equation}
Finally, we write:
\begin{equation}
k = \alpha + i \beta,
\end{equation}
and equate real and imaginary parts in equation (\ref{lossyk}) to give:
\begin{equation}
\alpha \approx \omega \sqrt{LC},
\end{equation}
and:
\begin{equation}
\beta \approx \frac{1}{2} \left( \frac{R}{Z_0} + G Z_0 \right),
\end{equation}
where $Z_0 = \sqrt{L/C}$ is the impedance with $R = G = 0$ (not to be confused with
the impedance of free space).  Note that since:
\begin{equation}
V = V_0 e^{i \left( k x - \omega t \right)}
  = V_0 e^{-\beta x} e^{i \left( \alpha x - \omega t \right)},
\end{equation}
the value of $\alpha$ gives the wavelength $\lambda = 1/ \alpha$, and the value of
$\beta$ gives the attenuation length $\delta = 1/ \beta$.

\subsection{Example: a coaxial cable}

A lossless transmission line has two key properties: the phase velocity $v$,
and the characteristic impedance $Z$.  These are given in terms of the inductance
per unit length $L$, and the capacitance per unit length $C$:
\begin{equation}
v = \frac{1}{\sqrt{LC}}, \qquad Z = \sqrt{\frac{L}{C}}. \nonumber
\end{equation}
The problem, when designing or analysing a transmission line, is to calculate the
values of $L$ and $C$.  These are determined by the geometry of the transmission
line, and are calculated by solving Maxwell's equations.

\begin{figure}
\centering
\includegraphics[width=0.7\textwidth]{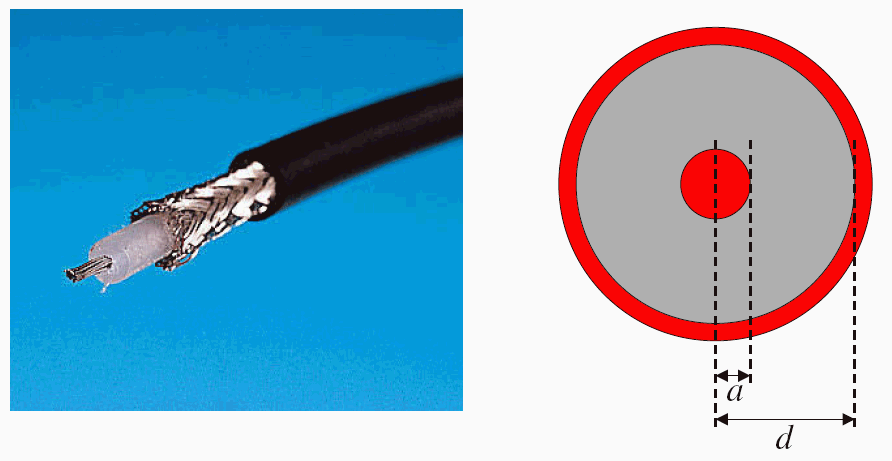}
\caption{Coaxial cable transmission line. \label{fig:coaxialcable}}
\end{figure}

As an example, we consider a coaxial cable transmission line, consisting
of a central wire of radius $a$, surrounded by a conducting ``sheath'' of
internal radius $d$ -- see Fig.\,\ref{fig:coaxialcable}.  The central wire and
surrounding sheath are separated
by a dielectric of permittivity $\varepsilon$ and permeability $\mu$.
Suppose that the central wire carries charge per unit length $+\lambda$, and the
surrounding sheath carries charge per unit length $-\lambda$ (so that the sheath
is at zero potential).  We can apply Maxwell's equation (\ref{eq:maxwell1}) with
Gauss' theorem, to find that
the electric field in the dielectric is given by:
\begin{equation}
|\vec{E}| = \frac{\lambda}{2\pi \varepsilon r},
\end{equation}
where $r$ is the radial distance from the axis.  The potential between the
conductors is given by:
\begin{equation}
V = \int_a^d \vec{E} \cdot d\vec{r}
  = \frac{\lambda}{2\pi \varepsilon} \ln \! \left( \frac{d}{a} \right).
\end{equation}
Hence the capacitance per unit length of the coaxial cable is:
\begin{equation}
C = \frac{\lambda}{V} = \frac{2\pi\varepsilon}{\ln \left( d/a \right)}.
\label{coaxialcapacitance}
\end{equation}

\begin{figure}
\centering
\includegraphics[width=0.4\textwidth]{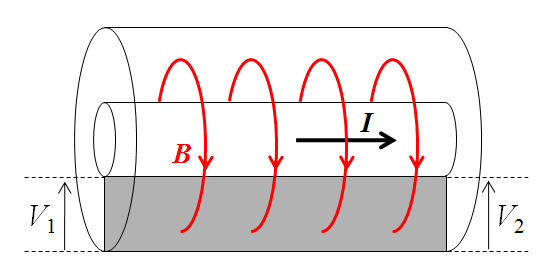}
\caption{Inductance in a coaxial cable. A change in the current $I$ flowing in the cable
will lead to a change in the magnetic flux through the shaded area; by Faraday's law, the
change in flux will induce an electromotive force, which results in a difference between
the voltages $V_2$ and $V_1$. \label{fig:coaxialinductance}}
\end{figure}

To find the inductance per unit length, we consider a length $l$ of the cable.
If the central wire carries a current $I$, then the magnetic field at a radius
$r$ from the axis is given by:
\begin{equation}
|\vec{B}| = \frac{\mu I}{2\pi r}.
\end{equation}
The flux through the shaded area shown in Fig.\,\ref{fig:coaxialinductance} is given by:
\begin{equation}
\Phi = l \int_a^d |\vec{B}| dr
     = \frac{\mu l I}{2\pi} \ln \! \left( \frac{d}{a} \right).
\end{equation}
A change in the flux through the shaded area will (by Faraday's law) induce an
electromotive force around the boundary of this area; assuming that the outer sheath
of the coaxial cable is earthed, the change in voltage between two points in the cable
separated by distance $l$ will be:
\begin{equation}
\Delta V = V_2 - V_1 = - \frac{d\Phi}{dt}.
\end{equation}
The change in the voltage can also be expressed in terms of the inductance per unit
length of the cable:
\begin{equation}
\Delta V = -lL\frac{dI}{dt},
\end{equation}
where $I$ is the current flowing in the cable.
Hence the inductance per unit length is given by:
\begin{equation}
L = \frac{\Phi}{l I} = \frac{\mu}{2 \pi} \ln \! \left( \frac{d}{a} \right).
\end{equation}
With the expression in Eq.\,(\ref{coaxialcapacitance}) for the capacitance per unit
length:
\begin{equation}
C = \frac{2\pi\varepsilon}{\ln \left( d/a \right)}, \nonumber
\end{equation}
the phase velocity of waves along the coaxial cable is given by:
\begin{equation}
v = \frac{1}{\sqrt{LC}} = \frac{1}{\sqrt{\mu \varepsilon}},
\end{equation}
and the characteristic impedance of the cable is given by:
\begin{equation}
Z = \sqrt{\frac{L}{C}}
  = \frac{1}{2\pi} \ln \! \left( \frac{d}{a} \right) \sqrt{\frac{\mu}{\varepsilon}}.
\end{equation}

\end{document}